\documentclass[aps,prx,amsmath,amsfonts,amssymb,notitlepage,twocolumn,superscriptaddress,footinbib,floatfix]{revtex4-2}

\usepackage{CJK}
\usepackage{bm}
\usepackage{soul}
\usepackage[caption=false,subrefformat=parens,labelformat=parens]{subfig}
\usepackage{color} 
\usepackage[table,dvipsnames]{xcolor}
\usepackage[colorlinks,linktocpage,bookmarks=false]{hyperref}
 
\usepackage{enumitem}
\usepackage{comment}

\usepackage{multirow}
\usepackage{xspace}

\newcommand\myshade{85}
\definecolor{myrulecolor}{RGB}{150,20,0}
\colorlet{mylinkcolor}{violet}
\colorlet{mycitecolor}{YellowOrange}
\colorlet{myurlcolor}{Aquamarine}
\hypersetup{
	linkcolor = myurlcolor!\myshade!black,
	citecolor = mycitecolor!\myshade!black,
	urlcolor = myrulecolor!\myshade!black,
}  

\usepackage{graphicx}
\graphicspath{{FIG/}}
\usepackage{multirow}
\usepackage{braket} 
\usepackage{booktabs}
\usepackage{ulem}

\AtBeginDocument{
	\heavyrulewidth=.08em
	\lightrulewidth=.05em
	\cmidrulewidth=.03em
	\belowrulesep=.65ex
	\belowbottomsep=0pt
	\aboverulesep=.4ex
	\abovetopsep=0pt
	\cmidrulesep=\doublerulesep
	\cmidrulekern=.5em
	\defaultaddspace=.5em
}

\newcommand{\beq}{\begin{equation}}
\newcommand{\eeq}{\end{equation}}
\newcommand{\bea}{\begin{eqnarray}}
\newcommand{\eea}{\end{eqnarray}}

\newcommand{\qq}{\mathbf{q}}

\DeclareMathAlphabet\mathbfcal{OMS}{cmsy}{b}{n}

\newcommand{\bfk}{\mathbf{k}}

\newcommand{\bfT}{\mathbf{T}}

\newcommand{\bfC}{\mathbf{C}}

\newcommand{\bfR}{\mathbf{R}}

\newcommand{\calC}{\mathcal{C}}

\newcommand\CP{\mathbb{C}P}
\newcommand\RP{\mathbb{R}P}

\newcommand{\bfq}{\mathbf{q}}
\newcommand{\bfr}{\mathbf{r}}

\renewcommand\[{\begin{equation}}
\renewcommand\]{\end{equation}}

\newcommand{\paper}{{\color{black} paper}\xspace}

\makeindex

\begin{document} 
	\begin{CJK*}{UTF8}{gbsn} 
\title{Classification of Classical Spin Liquids:  Detailed Formalism and Suite of Examples}

\author{Han Yan (闫寒)}
\affiliation{Department of Physics and Astronomy, Rice University, Houston, TX 77005, USA}
\affiliation{Smalley-Curl Institute, Rice University, Houston, TX 77005, USA}
\author{Owen Benton} 
\affiliation{Max Planck Institute for Physics of Complex Systems, N\"{o}thnitzer Str.\ 38, Dresden 01187, Germany}
 \author{Andriy H. Nevidomskyy}
\affiliation{Department of Physics and Astronomy, Rice University, Houston, TX 77005, USA}
\author{Roderich Moessner} 
\affiliation{Max Planck Institute for Physics of Complex Systems, N\"{o}thnitzer Str.\ 38, Dresden 01187, Germany}
\begin{abstract}
The hallmark of highly frustrated systems is the presence of many states close in energy to the ground state. Fluctuations between these states can preclude the emergence of any form of order and lead to the appearance of spin liquids.
Even on the classical level, spin liquids are not all alike:
they may have algebraic or exponential  correlation decay, and various
forms of long wavelength description, including vector or tensor gauge theories.
Here, we introduce a classification scheme, allowing us to 
fit the diversity of classical spin liquids (CSLs) into a general framework as well as predict and construct new kinds.
CSLs with either algebraic or exponential  correlation-decay can be classified via the properties of the bottom flat band(s) in their   soft-spin Hamiltonians. 
The classification of the former is based on the algebraic structures of gapless points in the  spectra, which relate directly to the emergent generalized Gauss's laws that control the low temperature physics. 
The second category of CSLs, meanwhile, are classified by the fragile topology of the gapped bottom band(s).
Utilizing the classification scheme we construct  new
models realizing exotic CSLs,
including one  with anisotropic generalized Gauss's laws and charges with subdimensional mobility, 
one with a network of pinch-line singularities in its
correlation functions, and a series of fragile topological CSLs connected by zero-temperature  transitions.
\end{abstract}
\maketitle
\end{CJK*}

\tableofcontents

\begin{figure*}[th!]
 \centering
 \includegraphics[width= \textwidth]{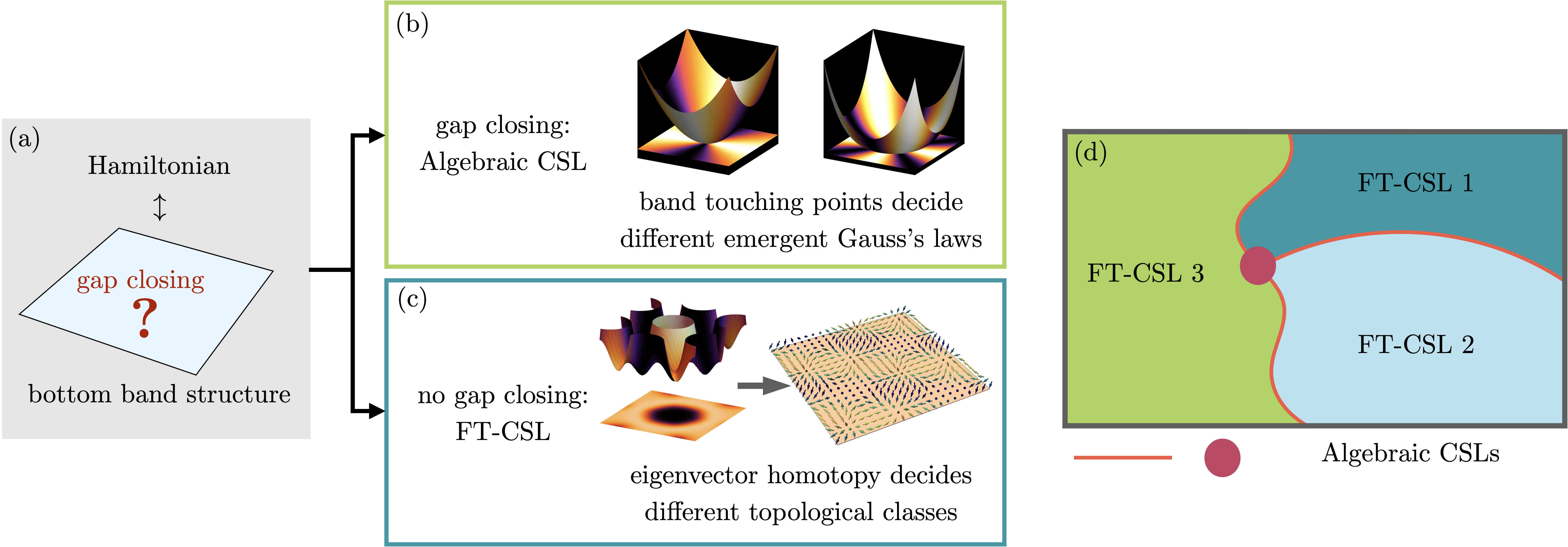} 
 \caption{(a)  The classification of classical spin liquids (CSLs) is based on the structure of the bottom bands and higher dispersive bands in the spectra of their Hamiltonians.
 (b)  The category of algebraic CSLs features gap closings between the bottom flat bands and higher dispersive bands. The eigenvector configuration at the band touching point determines the emergent Gauss's law.
 (c)  The category of fragile topological CSLs (FT-CSLs) have no   gap-closing points between the bottom flat bands and higher dispersive bands. They are classified by their eigenvector homotopy.
 (d) The landscape of CSLs consists of regions of FT-CSLs whose boundaries are algebraic  CSLs.}
 \label{fig:Fig_overview}
\end{figure*}

\section{Introduction}
The discovery of topological states of matter is one of the central advances in condensed matter physics (and beyond) in the last half century \cite{Klitzing80IQHE, Tsui1982PRL, Laughlin1983PRL, FuPRB2007, Kitaev2009, Ryu2010NJP, Wen2002PhysRevB.65.165113, Castelnovo2012ARCMP, Bradlyn2017Nature, BernevigHughesBook, MoessnerMooreBook}. In recent years, two major efforts have been to broaden and deepen the scope of this concept. This is evidenced by the appearance of symmetry-protected \cite{Pollmann2012PhysRevB.85.075125,Chen2013PRB, Chan2016PRB, SongPRX2017}, Floquet \cite{Kitagawa2010Floquet, Lindner2011Floquet, Mikami2016Floquet}, and other forms of topological phases \cite{MesarosPRB2013, Romhanyi2015, Schindler2018, Ezawa2018, Barkeshli2019PRB, thorngren-else2018, Corticelli2022}, as well as copious experimental advances actually realising topological phases in the laboratory and investigating their physical properties
\cite{Zhang2009NPhys, Brune2011HgTe, Mourik2012, Rechtsman2013, Jotzu2014, Chisnell2015PRL, Xu2015Weyl, Bian2016, McClarty2017NPhys, Groning2018, Imhof2018, ElHassan2019, Xue2019}. 

The wide variety of observations and discovery has naturally motivated an attempt to classify as comprehensively as possible the various phases that are imaginable. This programme has made tremendous progress, e.g.\ for electronic band structures alone, we
can now distinguish various kinds of stable \cite{Schnyder2008PRB, Kitaev2009, Chiu2016RMP} and fragile \cite{Po2018} topological insulators and multiple
forms of topological semi-metal \cite{Burkov2011, Matsuura2013, Soluyanov2015}.
Indeed, such classification schemes have themselves taken on a wide variety of guises. They range from the rather compact description of the ten-fold way \cite{Kitaev2009} to the 
classification of $Z_2$ quantum spin liquids, which has found a bewildering variety of possibilities \cite{essin-hermele2013}, to a more general classification of two-dimensional~\cite{Barkeshli2019PRB} and three-dimensional~\cite{thorngren-else2018} topological phases with internal or crystalline symmetries. 

Spin liquids have been at the forefront of topological physics for quite some time, with the resonating valence bond liquid \cite{ANDERSON1973RVB} proposed already in the early 1970s, but not discovered and identified as a topological phase until much later \cite{Kivelson87PRB, Moessner01RVB}. 

In a separate development, the search for disordered magnetic ground states was pursued in the context of spin glasses, where the role of magnetic frustration was identified as a crucial ingredient for destabilising conventional  ordered states \cite{Villain1979}. Since the foundational works of Anderson and Villain, the field of frustrated magnets has grown into a huge field of its own, and proposals of spin liquids as well as candidate materials have become increasingly plentiful \cite{Balents2010Nature,Gingras2014RPPh,Norman2016, Winter2017JPCM, Takagi2019, Knolle2019AnnRev, Broholm2020, UdagawaJaubertBook}. 

Spin liquids can appear in both classical and quantum models.
Classical spin liquids (CSLs), tend to go along with an
extensively degenerate manifold of ground states, amongst which the system fluctuates \cite{Moessner1998PhysRevB.58.12049, MC_pyro_PRL, Isakov2004PhysRevLett.93.167204, Henley05PRB, Henley2010ARCMP, Garanin1999PhysRevB.59.443, Rehn16PRL, Rehn17PRL, Benton16NComms, Taillefumier2017PhysRevX.7.041057, Yan20PRL, Benton21PRL, davier2023combined, Gembe2023arxiv}.
Quantum spin liquids (QSLs), on the other hand, generally have only a small degeneracy of ground states, which can often be thought of as  entangled superpositions of classical basis states \cite{RVB_Fazekas, RK_1988, MS_PRB2001,Moessner01RVB, Hermele2004PhysRevB.69.064404, Gingras2014RPPh, Sibille2018NatPh,Gaudet2019PhysRevLett.122.187201, Gao2019, Sibille2020, poree2023fractional}.
Classification schemes for topological QSLs exist via 
projective symmetry group (PSG) analysis \cite{Wen2002PhysRevB.65.165113} and, more recently, via the braided tensor category classification in 2+1 dimensions \cite{Barkeshli2019PRB}. 

Despite their apparently simpler character, there exists, as yet, no similarly comprehensive formalism for CSLs.
However, CSLs are of considerable interest in their own right. With their extensive degeneracies, they represent the extreme limit of the consequences
of frustration. Even though CSLs are always found at fine-tuned points in parameter space at $T=0$, their large entropy allows them to spread out in the surrounding phase diagram at finite $T$. They are thus relevant to the finite temperature behavior of real frustrated magnets. They may also serve as a starting point in discovering QSLs, with 
several of the most prominent QSL models having a classical counterpart with a CSL ground state \cite{Hermele2004PhysRevB.69.064404, Harris1997PhysRevLett.79.2554, Kitaev2006AnnPhys, Samarakoon17, Balents02PRB, Rehn17PRL}.

Amongst classical spin models with continuous spins -- of which the Heisenberg model is the most familiar member -- there exist a number of well-established CSLs (see Table~\ref{table:summary.of.models} for a survey). The first was the Heisenberg antiferromagnet on the pyrochlore lattice, which exhibits an emergent U(1) gauge field in the low-energy description of its so-called Coulomb phase \cite{MC_pyro_PRL, Moessner1998PhysRevB.58.12049,Henley05PRB,Henley2010ARCMP}. 

For a long time, it has seemed that the number of distinct CSLs, in the sense of a classification, is quite limited.  However, recent work has begun to uncover a landscape of classical spin liquids beyond the ``common'' U(1) Coulomb liquids, both at the level of effective field theories and microscopic models. In this vein, there have been proposals of 
short-range correlated spin liquids \cite{Rehn16PRL}, higher-rank Coulomb phases \cite{Yan20PRL, Benton21PRL, niggemann-arxiv}, and pinch-line
liquids \cite{Benton16NComms}. This has brought the tantalising promise that there may be quite a large uncharted landscape of possibilities waiting to be discovered. 

The present work is devoted to realising this promise. We provide a classification scheme for spin liquids occurring as ground state ensembles in classical continuous-spin Hamiltonians (ref. Fig.~\ref{fig:Fig_overview}) and apply it to a number of existing and new models (ref.  Table~\ref{TABLE_algebraic_class} and Table~\ref{table:summary.of.models}).
This enables us to understand and distinguish different kinds of
CSL in a way that goes beyond simply distinguishing algebraic from short-range correlations.
We identify distinct kinds of algebraic and short-range
correlated CSL and zero-temperature  transitions between them, and uncover simple models exhibiting previously
unseen forms of spin liquid.

In this article, we develop the classification
theory in some detail, with numerous examples.
A shorter  companion paper \cite{yan2023ArxiVclassification}, which illustrates
the main ideas in the context of a single
model on the kagome lattice, accompanies this article, and may be of use to any readers who do not require such a comprehensive exposition.

The example models we construct are themselves significant as
they provide simple settings in which to realize novel
physics. This includes a model realizing anisotropic Gauss's laws, in which derivatives with respect to different directions enter the Gauss's law with different powers \cite{Bulmash2018, Hart2022PhysRevB} and concomitant subdimensional excitations; a
spin-liquid with a network of line-like singularities (pinch lines) in the structure factor; and a series of topological CSLs 
connected by zero-temperature  transitions.
We therefore establish the utility of our classification scheme in the construction of new models realizing interesting phenomena. 

Previous works have classified 
highly frustrated classical spin systems via constraint counting \cite{MC_pyro_PRL}, via linearization around particular ground state configurations \cite{Roychowdury2018PRB} and
via supersymmetric connections between models \cite{Roychowdhury-arXiv}.
In recent work by two of the present authors, the possibility of distinct types of algebraic spin liquids distinguished by topological properties was explicitly demonstrated \cite{Benton21PRL}.
Here we present a scheme 
which generalizes across different kinds of spin liquid and assists in the construction of new ones.
It is based on the physics of the spin liquid as a whole, rather than individual spin configurations 
within it and, in the case of algebraic spin liquids,
unveils the connection between the microscopic model and the Gauss's law which governs the
long distance physics.

The classification scheme is based on a soft spin description
of the CSL state.
In such a description one neglects the spin-length constraints $|{\bf S}|_i=1$, replacing it instead with an averaged constraint $\langle {\bf S}_i \cdot {\bf S}_i \rangle=1$.

The soft spin approximation is known to provide a 
good description of CSLs for many known examples \cite{Isakov2004PhysRevLett.93.167204, Conlon2010PRB, Conlon2009PRL, Rehn16PRL, Rehn17PRL, Benton21PRL}.
Nevertheless, classifying CSLs according to their properties within an approximate treatment such as the soft spin approximation,
may seem unsatisfactory.
It is, however, in keeping with the spirit
of other classification schemes such as the use of PSGs to classify QSLs. The PSG analysis is based around the properties
of a mean field description of a given QSL, but remains useful
because the qualitative nature of the phase is more robust
than the quantitative accuracy of the mean field theory.
Here, similarly, we expect our classification to correctly distinguish between CSLs, the limitations of the soft-spin
description notwithstanding.

If the Hamiltonian is bilinear in spins, then one may diagonalize it in momentum space, leading to a spectrum with a band structure that carries information about the low energy spin liquid state.
Our classification is based on the algebraic and topological properties of this band structure, and is schematically
illustrated in Fig.~\ref{fig:Fig_overview}.

The common feature of the soft spin description of CSLs is the presence
of one or more flat bands at the bottom of the spectrum (Fig.~\ref{fig:Fig_overview}(a)).
These flat bands correspond to the extensive number of degrees of freedom which remain free in the CSL ground state.
The most basic distinction we can make between CSL soft spin band structures is whether or not the flat bands at the bottom of the spectrum are separated by a gap from the higher energy bands.
Spectra with (without) a gap correspond to CSLs with short-ranged (algebraically-decaying) spin correlations.

We   classify   CSLs without a gap for the bottom band via the algebraic properties of their band structures around the gapless points in the Brillouin Zone (BZ). In particular, a Taylor expansion of the eigenvector(s) of the 
dispersive band(s) which come down to meet the low energy flat band(s) at the gapless point defines an effective Gauss's law which constrains the long wavelength fluctuations of the CSL (e.g. $\nabla \cdot {\bf E}=0$ in the ordinary Coulomb phase). The form of this Gauss's law 
distinguishes different kinds of such CSLs.
Table~\ref{TABLE_algebraic_class} lists representatives with different generalized Gauss's laws.
We name such CSLs ``algebraic CSLs'', due to the fact that their correlation decays algebraically, and the emergent generalized Gauss's law depends on the algebraic structure of the gapless points (Fig.~\ref{fig:Fig_overview}(b)).

For the  short-range correlated CSLs , the classification is based on the topology 
of the soft spin band structure. 
Depending on the symmetries present, and the number of sites in
the unit cell, the bands may possess topological invariants which
are insensitive to small changes to the CSL ground state constraint.
These topological invariants can be used to distinguish different classes of
such CSLs.
We find that the nontrivial topology of these CSLs   is generically
fragile, in the sense that it can be rendered trivial by adding additional spins in the unit cell. 
This motivates us to introduce the term ``fragile 
topological CSL" (FT-CSL) as a descriptor of
short-range correlated CSLs (Fig.~\ref{fig:Fig_overview}(c)).

By tuning the Hamiltonian, it is possible to drive zero-temperature
transitions between FT-CSLs. At these transitions, algebraic CSLs emerge.
We hence arrive at a landscape of CSLs where the phases are occupied by the FT-CSLs, and the phase boundaries are algebraic CSLs, as shown in Fig.~\ref{fig:Fig_overview}(d).

To illustrate all these ideas we
introduce a number of new models which are of autonomous interest beyond the classification scheme, in that they represent hitherto unknown types of classical spin liquids, worthy of study in their own right.
A summary of different algebraic and fragile topological CSLs known in literature is presented in Table~\ref{table:summary.of.models}.

Our approach to the analysis of CSLs presents a comprehensive advancement in our understanding of these frustrated systems. 
It reveals a landscape of classical spin liquids as fragile topological CSLs separated by the algebraic CSLs, and encompasses all CSL models to the best of our knowledge (in the soft spin setting at least).
While building the classification scheme, 
we have established close connections between CSLs and other fields of physics and mathematics  including flat bands in electronic band theory, symmetry protection and fragile topology, and homotopy theory.
Our classification scheme can also be  easily reversely engineered to design  new CSL models with desired properties.

The article is organised as follows.
The next section (Sec.~\ref{sec.summary.main.result}) provides a non-technical overview of our central results.  
The main content   starts from Sec.~\ref{sec.motivation}, which reviews a few recently discovered new models of classical spin liquids, and motivates us to pose the question of classification.
In Sec.~\ref{sec.set.up.classification},
we formulate the problem on a more  mathematical footing, to make it amenable to the algebraic and topological treatments later.
Sec.~\ref{sec:algebraic} discusses the abstract  aspect of one of the two main categories of CSLs: the algebraic CSLs,
followed by Sec.~\ref{sec:algebraic.csl.models} which provides a handful of examples for concrete demonstration of the physics.
Following a similar structure,
Sec.~\ref{sec.topological.csl} discusses the abstract  aspect of the  fragile topological CSLs,
followed by Sec.~\ref{sec.kagoem.star} which provides a concrete example. 
We then briefly discuss wider applications of our classification scheme  in Sec.~\ref{sec.app.to.exp} and show how 
previously established examples of CSLs fit into our scheme.
Finally, we conclude with a summary and outlook of future directions and open issues in Sec.~\ref{sec.summary.outlook}.\\

\section{Sketch of the main results }
\label{sec.summary.main.result}

Here, we telegraphically list our main results to guide readers through the  technical details later. 
A self-contained, non-technical narrative  can be found in the short sister paper Ref.~\onlinecite{yan2023ArxiVclassification}.

We study  spin models in the limit  of a large number of spin components $\mathcal{N}$. This is effectively a ``soft spin'' approach, where the spin length constraint is enforced `on average' by the central limit theorem for $\mathcal{N}\rightarrow\infty$. This amounts to treating each spin components as a scalar, and this has been shown to be a good approximation for many, but not all, Heisenberg candidate CSLs.
CSLs in such a description tend to have an extensive degeneracy of exact ground states.

Such CSL Hamiltonians can be generally be written in what we call the \textit{constrainer form}:
\[ 
\label{eq:H_constrainer}
\mathcal{H} =     \sum_{\mathbf{R} \in \text{u.c.}} \sum_{I=1}^M \, [\mathcal{C}_I (\mathbf{R} )]^2\  ,
\] 
where for a given constrainer index $I$, $\mathcal{C}_I(\mathbf{R})$ is the sum over a local cluster of spins around the unit cell located at $\mathbf{R}$. The Hamiltonians we consider are the translationally invariant sums of such squared constrainers. 
 
For simplicitly,
we will mostly work with models with one constrainer ($M=1$) and $N$ sublattice sites per unit cell, and will note where deviations from this setup affect the classification.
In this case, there are $N-1$ bottom flat bands at zero energy that satisfy the constraint, and one higher dispersive band that violates it.
The dispersive band's eigenvector, denoted $\mathbf{T}(\mathbf{q})$, can be algebraically determined by Fourier transforming the constrainer $\mathcal{C}(\mathbf{R})$.
The dispersion of the higher band is exactly $\omega_T(\mathbf{q}) = |\mathbf{T}(\mathbf{q})|^2$.

The  overall spectra encode the information of the CSLs. 
They can be divided into two broad categories.

\vspace{2mm}
\noindent\textbf{1. Algebraic CSL:} There is one or more gap-closing point between the bottom flat band and higher dispersive bands.
In this case, the CSL is an algebraic CSL, i.e.
the spin correlations decay algebraically.
Furthermore, 
the ground states can be described by a charge-free Gauss's law, determined by the Taylor expansion of $\mathbf{T}(\mathbf{q})$, where $\mathbf{q}$ denotes the distance in momentum space from the band touching. 
More specifically, if the lowest order term in the Taylor expansion is 
\[ 
\label{EQN_T_expansion_general}
T_a (\mathbf{q} )  
 = \sum_{j=0}^{m_a} c_{aj} (-iq_x)^{j} (-iq_y)^{m_a-j}, \quad a = 1,\dots, N\ ,
\]
the charge-free Gauss's law is then given by 
\[
\label{EQN_Gauss_general_real_3}
\begin{split}
&\mathbf{T^\ast}(\mathbf{q}) \cdot \tilde{\mathbf{S}}  = 0 \quad  \rightarrow  \\
& \sum_{a=1}^N\left(\sum_{j=0}^{m_a} c^\ast_{aj} (\partial_x)^{j} (\partial_y)^{m_a-j} S_a \right) \equiv \sum_{a=1}^ND_a^{(m_a)} S_a = 0\ , 
\end{split}
\]
where we have defined a generalized differential operator $D_a^{(m_a)}$ of order $m_a\geq 1$ on sublattice $a$.
A similar picture applies
for models with multiple constraints per unit cell,
and hence more than one ${\bf T(\bf q)}$, with the subtlety that one must take care of the orthogonality between different ${\bf T(\bf q)}$ around the band touching.

\vspace{2mm}
\noindent\textbf{2. Fragile Topological CSL:} 
The bottom flat band is fully gapped from the higher dispersive band.
In this case, $\mathbf{T}(\mathbf{q})$ is a non-zero and smoothly defined vector field in the target manifold $\CP^{N-1}$ (if it is complex) or $\RP^{N-1}$ (if it is real)  over the entire BZ. 
It can then wind around the BZ (a $d$-torus, ${T}^d$ in $d$ dimensions) in a non-trivial manner, captured by the homotopy class 
\[
\label{eq:Tq-map}
\hat{\mathbf{T}}(\mathbf{q}): {T}^d \rightarrow \CP^{N-1} (\text{or }\RP^{N-1});\  \mathbf{q} \mapsto \hat{\mathbf{T}}(\mathbf{q})\ .
\]
In the case where there is more than one constraint per unit cell,
 the target manifold
may be something other than $\CP^{N-1}$ or  $\RP^{N-1}$.
Adiabatic changes to the Hamiltonian which retain the constrainer form and do not close the gap between the bottom flat band and the upper bands cannot change the homotopy class.
The homotopy class only changes when the gap closes.
That is, \textbf{at the  boundaries of fragile topological CSLs are algebraic CSLs.} 

FT-CSLs can be rendered trivial by the addition of extra degrees of freedom to the unit cell, hence our use of the term `fragile', in keeping with the notion of fragile topology in electronic band theory \cite{Po2018}.

In the main text we provide numerous examples to show how the abstract theory above can be applied to concrete models.

\section{Motivating the classification problem}
\label{sec.motivation}
In this section we motivate the question of classification by reviewing two known examples of CSLs.
The two models we call  the honeycomb-snowflake model proposed in Ref.~\onlinecite{Benton21PRL} and the  Kagome-Hexagon model, Ref.~\onlinecite{Rehn17PRL}. 
These are representatives for the two different categories of CSLs. 
The honeycomb-snowflake model with a varying parameter hosts several algebraic CSLs that realize different generalized $U(1)$ Gauss's laws.
Correspondingly, the spin correlations decay algebraically.
The Kagome-Hexagon model hosts a qualitatively different CSL -- the fragile topological CSL -- that does not exhibit any $U(1)$ Gauss's laws,
and has exponentially decaying spin correlations.

After reviewing the two models,
we summarize their common features to extract the most general set-up for the CSL models.
At the end of this section, we will be ready to establish a classification scheme that, once a specific Hamiltonian is given, can mechanically analyze the CSL physics from that Hamiltonian. 

\subsection{Honeycomb-snowflake model}
\label{Sec_BM_Honeycomb_model}

\begin{figure}[ht]
 \centering
  \includegraphics[width=\columnwidth]{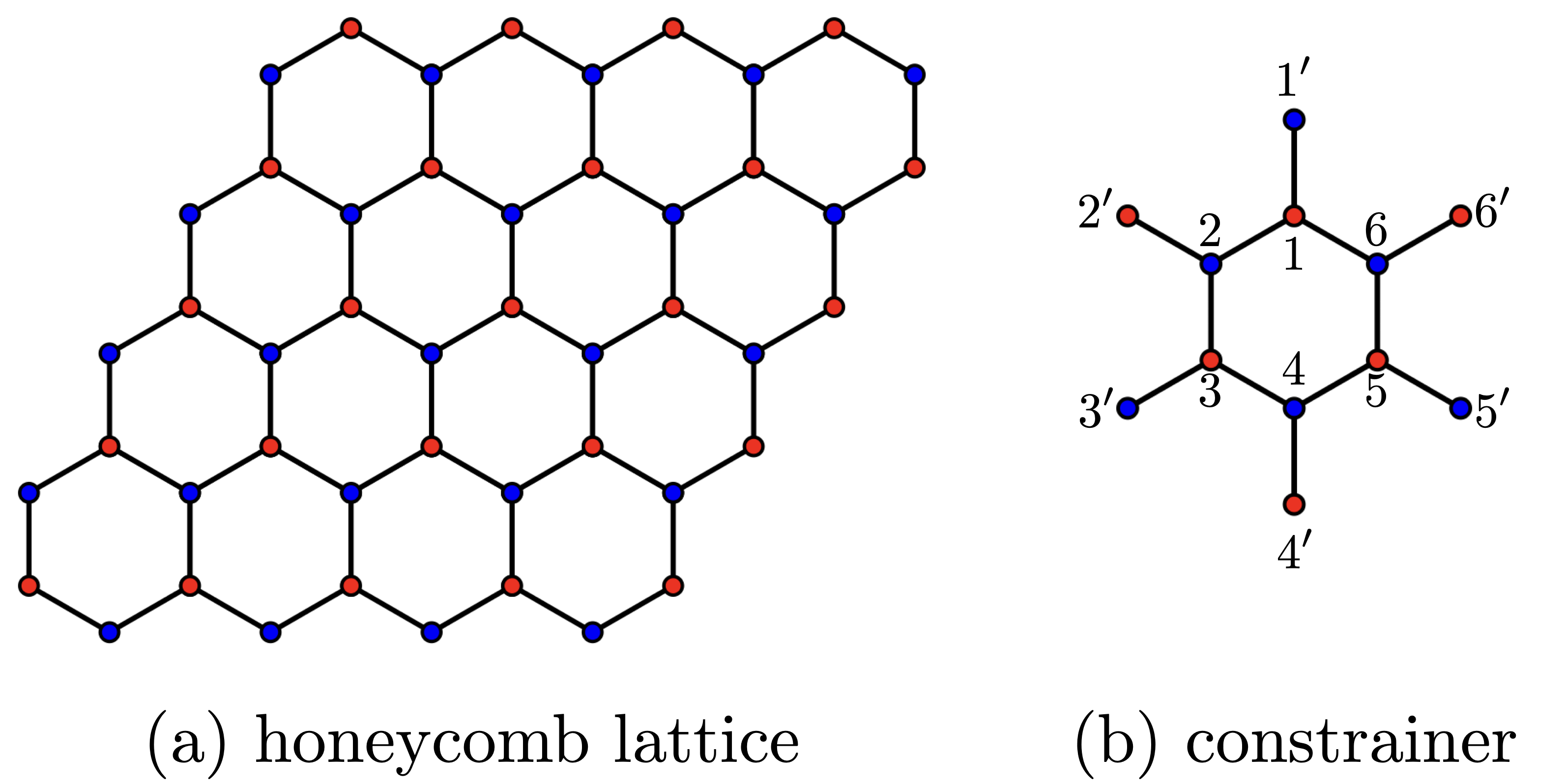} 
 \caption{(The honeycomb-snowflake  model \cite{Benton21PRL}, Eq.~\eqref{eq:Hhex}, exhibiting a series of distinct spin liquids as the model parameter $\gamma$ is varied. (a) The honeycomb lattice, composed of two sublattices, colored red and blue respectively. 
 (b) The constrainer defining the ground states of the model, applied to each hexagonal plaquette in the lattice. 
 The sum of spins on each hexagon ($1$ to $6$) plus the coefficient $\gamma$ multiplied by the sum of spins linked to the exterior of the hexagon ($1'$ to $6'$) must vanish on every hexagon (Eqs. (\ref{eq:Mhex})-(\ref{eq:honeycomb_constraint})). }
 \label{fig:BMmodel}
\end{figure}
 
\label{subsubsec:BM_intro}

The honeycomb-snowflake model 
proposed in Ref.~\onlinecite{Benton21PRL}
serves to demonstrate how
a series of distinct algebraic CSLs
can be accessed
by varying local constraints
on a classical spin system.
Its Hamiltonian is defined as a squared sum of Heisenberg spins around  the hexagonal plaquettes
on the Honeycomb lattice [Fig. \ref{fig:BMmodel}]:
\[
\label{eq:Hhex}
\mathcal{H}_\mathsf{HS}=
\frac{J}{2} \sum_{\mathbf{R} \in \text{all hexagons}} \ \sum_{\alpha = x, y ,z} [\mathcal{C}_{\mathsf{HS}, \alpha}^{  \gamma}(\mathbf{R} )]^2 \ .
\]
The sum  of $\mathbf{R}$ is  taken over all hexagonal plaquettes of the lattice or, equivalently, over all unit cells.
The sum of ${\alpha = x, y ,z}$ is taken over all three spin components.
The terms $\mathcal{C}_{\mathsf{HS}, \alpha}^{  \gamma}(\mathbf{R} )$ defined on the   hexagons are  weighted sums 
of spins around each  ``snowflake''  shown in  Fig.~\ref{fig:BMmodel}(b):
\[
\mathcal{C}_{\mathsf{HS}, \alpha}^{  \gamma}(\mathbf{R} ) 
= \sum_{j  = 1  }^6 {S}^\alpha_j + \gamma \sum_{j = 1'  }^{6'}{S}^\alpha_j \ 
\label{eq:Mhex} 
\]
The first sum in Eq. (\ref{eq:Mhex}) is over spins on the hexagon at $\bfR$ (sites in Fig. \ref{fig:BMmodel}(b)  labelled $1$ to $6$) and the second is over neighboring spins connected to exterior of the hexagon (sites in Fig. \ref{fig:BMmodel}(b)  labelled $1'$ to $6'$). 
$\gamma$ is a dimensionless parameter which we use to tune the model.
Ground states of Eq. (\ref{eq:Hhex}) satisfy the constraint: 
\begin{eqnarray}
\mathcal{C}_{\mathsf{HS}, \alpha}^{  \gamma}(\mathbf{R} )  =    0 \quad  \forall \ \mathbf{R},\ \alpha \ .
\label{eq:honeycomb_constraint}
\end{eqnarray}
The case $\gamma=0$ corresponds to the model of Ref. \cite{Rehn16PRL}.

Let us now outline a description of the honeycomb-snowflake model, equivalent to that in \cite{Benton21PRL}, based on the gap closings in the spectrum of the Fourier transformed Hamiltonian.
First, we observe that the Hamiltonian is identical for the three components $\alpha = x, y, z$.
If we relax the spin norm constraint, $|{\bf S}_i|=1$,
and treat it only on average ($\langle {\bf S}_i\cdot {\bf S}_i \rangle=1 $), the spin components can be thought of as
essentially indenpendent scalar variables.
This step can be justified more formally taking the limit of a large number of spin components, $\mathcal{N}$.
The theory in which the spin norm is fixed only on average then corresponds to the leading order of a $1/\mathcal{N}$ expansion.
This approach has been for example successful, even quantitatively, in describing pyrochlore spin liquids with $O(3)$ Heisenberg  ($\mathcal{N}=3$), and even  Ising spins ($\mathcal{N}=1$) \cite{Isakov2004PhysRevLett.93.167204}, and has been widely used in the treatment of spin liquids since its introduction to the field in Ref.~\onlinecite{Garanin1999PhysRevB.59.443}.
In the remainder of the paper, we work within this large-$\mathcal{N}$ picture.
This allows us to build our classification scheme, and in this sense we are working in the same spirit as other classification schemes in Condensed Matter Physics which are also derived from mean-field or non-interacting theories; with the expectation that the classification labels are  robust even when the underlying approximate theory is not 
quantitatively accurate. 
Exceptions to this can, and do, occur however -- such as in the case of the $O(3)$ kagome Heisenberg model.
While a large-$\mathcal{N}$ picture predicts a spin liquid, the order-by-disorder effects drive the 
$O(3)$ system into an ordered phase at very low temperature \cite{Chalker1992PhysRevLett.68.855,Harris1992PhysRevB.45.2899,Ritchey1993PhysRevB.47.15342,Zhitomirsky2008PhysRevB.78.094423,Chern2013PhysRevLett.110.077201}.
The approach we present here thus provides a tool for classifying CSLs but does not prove the stability of a CSL in any given hard-spin model, which is a task that generally requires simulations.

Working within the large-$\mathcal{N}$ theory, we can drop the component index label
$\alpha$ and regard each spin $S_i$ as now a scalar instead of a vector.
 
Taking the Fourier transform of Eq. (\ref{eq:Hhex}) results in a Hamiltonian written as  a $2 \times 2$ interaction matrix $\mathbf{J}({\bf q})$,  
\[ 
\mathcal{H}_{\sf HS}=\frac{1}{2} \sum_{{\bf q}} \sum_{a,b = 1}^{2}  \tilde{S}_a(-{\bf q})   {J}^\gamma_{ab} ({\bf q})  \tilde{S}_b({\bf q})\ ,
\]
where $a$, $b$ index the two translationally inequivalent sublattices of the honeycomb lattice, 
and $\tilde{\mathbf{S}}(\mathbf{q}) = (\tilde{S}_1(\mathbf{q}),\tilde{S}_2(\mathbf{q}))$ is the lattice Fourier transform of the spin fields on the  sublattice sites $1,2$.
The interaction  matrix $\mathbf{J}^\gamma({\bf q})$
depends on the parameter $\gamma$,
and can be computed straightforwardly.
Its explicit form is lengthy and not of importance for now, but can be found in Eq.~\eqref{EQN_BM_Hamiltonian_J} in Sec.~\ref{Sec_BM_classification_application} when we revisit this model.

\begin{figure*}
 \centering
 \includegraphics[width=0.95\textwidth]{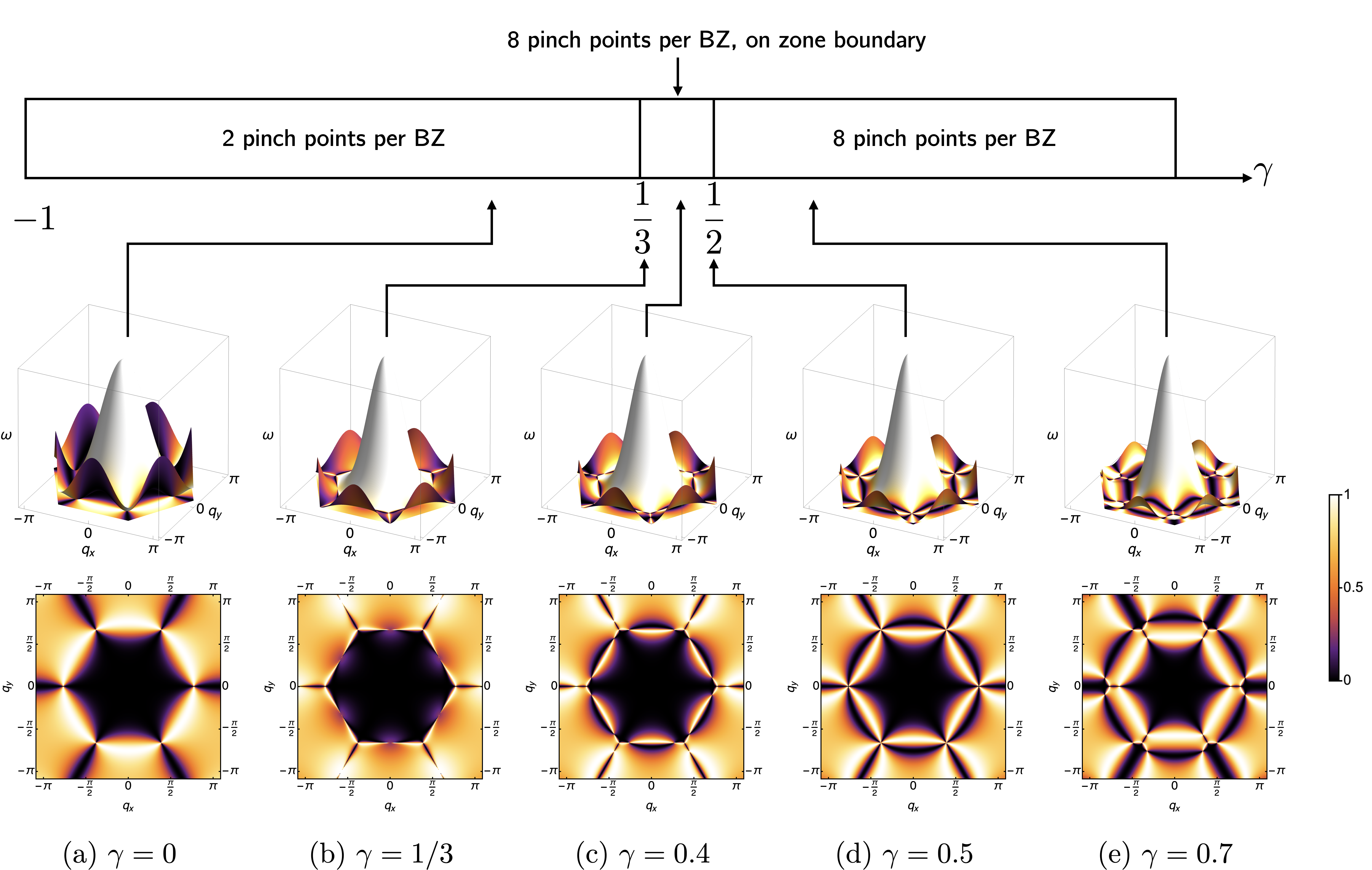}\\ 

 \caption{ 
  Phase diagram of the honeycomb-snowflake  model (Eq.~\eqref{eq:Hhex}) as a function of $\gamma$, showing a series of algebraically correlated CSLs. Transitions between spin liquids occur either by creation/annihilation of pinch points in the spectrum ($\gamma=1/3$) or by merging of them ($\gamma=1/2$)  leading to a higher-rank Coulomb liquid with multi-fold pinch points.
 The phase diagram is based on Ref.~\onlinecite{Benton21PRL}.
 (a-e):
 band dispersion $\omega({\bf q})$ (upper panels) and structure factor $S({\bf q})$ (lower panels) of the honeycomb-snowflake model with varying $\gamma$.  
 There is always a gap closing at the K point of the Brillouin zone and additional ones are created or annihilated in pairs as $\gamma$ is varied, giving rise to topological transitions between distinct CSLs.
 \label{fig:honeycomb_bands}}
\end{figure*}

Diagonalizing $\mathbf{J}^\gamma({\bf q})$ yields a 2-band spectrum, in which the lower band is flat at energy $\omega = 0$, and the upper band is dispersive, its dispersion  denoted as $ \omega({\bf q})$.
The gap between the flat and dispersive bands closes at multiple points in the
Brillouin zone [Fig. \ref{fig:BMmodel2}].
The Hamiltonian can then be represented as
\begin{eqnarray}
\mathcal{H}_{\sf HS}=\frac{1}{2} \sum_{{\bf q}} \sum_{ab=1}^2 \omega({\bf q})\,  \tilde{ S}_a(-{\bf q}) \,   \hat{  T}_a({\bf q}) \hat{  T}_b^{\ast} ({\bf q}) \,\tilde{ S}_b({\bf q})\ ,
\label{eq:Hhex_q}
\end{eqnarray}
where $\omega({\bf q})$ is the eigenvalue of the dispersive (upper) band
and $ \hat{\bf T} ({\bf q})$ is the corresponding normalized eigenvector of the top band.

The upper eigenvector can be used to give a momentum space description of the ground state constraints, Eq. (\ref{eq:honeycomb_constraint}).
Any Fourier-transformed spin configuration   on the two sublattices $a=1,2$ obeying the condition
\[ 
\sum_{a=1,2} \sqrt{\omega(\mathbf{q})}\hat{T}^{\ast}_a ({\bf q})\tilde{S}_a ({\bf q})   \equiv  \sum_{a=1,2}{T}^{\ast}_a  ({\bf q})  \tilde{S}_a ({\bf q})  =  0  \quad
\forall \ {\bf q}\ ,
\label{eq:momentum_space_constraint}
\]
is a ground state.
This is the momentum space representation of Eq. (\ref{eq:honeycomb_constraint}).

Eq. (\ref{eq:momentum_space_constraint}) can be seen as an orthogonality condition between the vector of sublattice Fourier transforms of the spin
configuration $\tilde{\mathbf{S} } ({\bf q}) $ and the upper band eigenvector $\mathbf{T} ({\bf q}) $.
(The upper band eigenvector is thus equivalent to the  constraint vector ${\bf  L} ({\bf q})$ introduced in Ref.~\onlinecite{Benton21PRL}).

The ground state phase diagram of the honeycomb-snowflake model is shown in Fig.~\ref{fig:honeycomb_bands}. 
Three distinct  algebraic  CSLs emerge as $\gamma$ is varied (the CSLs at large negative $\gamma$ and large positive $\gamma$ are equivalent, as may be inferred from Eqs. (\ref{eq:Hhex})-(\ref{eq:Mhex})).
In Ref.~\onlinecite{Benton21PRL}, the distinction between these CSLs was understood in terms of topological defects in ${\bf T}({\bf q})$.

It was found that the CSLs with a pinch point (singular pattern of the structure factor at the $K$ point) \cite{Harris1997PhysRevLett.79.2554,yan2023identifyingArxiv,yan2023pinchArxiv} host a spin liquid described by the Gauss' law of a Maxwell U(1) gauge theory:
\[
\partial_\alpha E_\alpha = 0\ .
\]
Here, $\mathbf{E} = (E_x, E_y)$ is an emergent vector electric field degree of freedom (DOF).

At $\gamma = 1/2$, four of the pinch points merge at the $K$ point, forming a 4-fold pinch point (4FPP) \cite{Prem18PRB,Niggemann2023PhysRevLett}, and  a more exotic  Gauss's law describing the system in terms of a rank-2 electric field with a scalar charge \cite{PretkoPRB16} was found:
\[
\partial_\alpha \partial_\beta E_{\alpha\beta} = 0,
\quad \text{where }
\mathbf{E}
=
\begin{pmatrix}
  E_{xx} & E_{xy}  \\
    E_{xy} & - E_{xx}\\
\end{pmatrix}\ .
\]
We will come back to the emergence of different Gauss's laws in Sec.~\ref{Sec_BM_classification_application}.

Finally, let us explain the plots in Fig.~\ref{fig:honeycomb_bands}, and the similar plots which appear for other models throughout the paper.
Fig.~\ref{fig:honeycomb_bands} shows the spectrum of ${\bf J}^{\gamma}({\bfq})$. Additionally, on each band $i$, we have also plotted the spin correlations defined as 
\[
\label{eq:S(q,w)}
S(\omega, \bfq)_i = \delta(\omega - \omega_i(\bfq)) \left| \sum_{a=1,2} v_i^a(\bfq)\right |^2\ ,
\]
where $\bm{v}_i (\bfq)= (v_i^1(\bfq), v_i^2(\bfq))$ is the eigenvector of the band $i$.

In the $T=0$ limit of a large-$\mathcal{N}$ approximation the  equal time structure factor is the sum of the structure factors $S(\omega, \bfq)_i$ over the flat bands only \cite{Henley05PRB}:
\[
\label{eq:S(q)}
S(\bfq)_\text{$T=0$} =\sum_{i\text{ s.t }\omega_i = 0}  \left| \sum_{a=1,2} v_i^a(\bfq)\right |^2\ .
\]
In particular, the spin structure factor  measured in inelastic neutron scattering  contains valuable information about these pinch points and can be used to experimentally determine the nature of the CSL.

The dispersion $\omega({\bf q})$ vanishes at multiple points in the Brillouin zone (BZ). At these points, the
upper band eigenvector $ {\bf T} ({\bf q})$ (and hence Eq. (\ref{eq:momentum_space_constraint})) is not uniquely defined and
there are corresponding singularities in $S({\bf q})$.
These singularities in the structure factor give
rise to an algebraic form of the spin correlations when Fourier transformed back into real space,
and also dictate  the Gauss's law constraining the spin fluctuation of the ground states.
The presence of non-trivial gap closings is, therefore, an essential part of the physics of these CSLs.

\subsection{Kagome-Hexagon model}
\label{subsec:RSMintro}
 
\begin{figure*}[ht]
 \centering
 \includegraphics[width=0.95\textwidth]{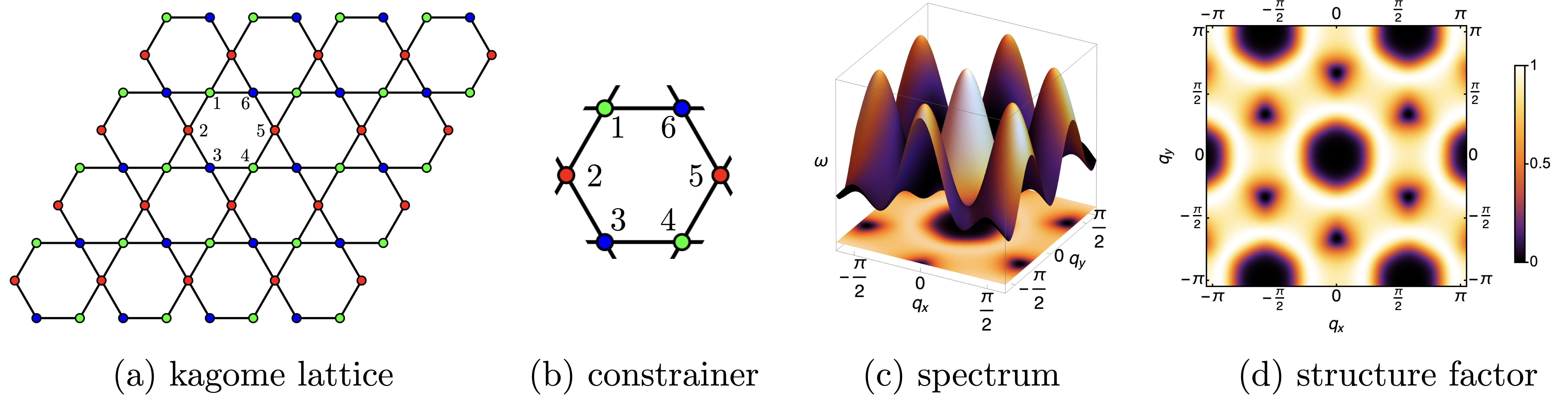}
 \caption{
 (a)  The kagome lattice. It has three sites in the unit cell, forming three sublattices indicated here in red, blue and green. 
 (b) Constrainer of the Kagome-Hexagon model. Classical spins are arranged on a kagome lattice, with ground states defined by the constraint that the sum of spins on each heaxagonal plaquette   must vanish (Eqs. (\ref{eq:HRSM})-(\ref{eq:kag_constraint})).
 (c) Spectrum $\omega({\bf q})$ that arises from diagonalizing the Hamiltonian (Eq.~\eqref{eq:HRSM}) in momentum space. There are two degenerate flat bands at the bottom of the spectrum and a dispersive upper band with no band touchings between the upper and lower bands.
 (d) Spin structure factor showing an absence of singularities.
 }
 \label{Fig_RSM_all}
\end{figure*}

We now discuss the kagome-hexagon model \cite{Rehn17PRL} as an example of fragile topological CSL with short-ranged correlations at low temperature. Its Hamiltonian is defined on the kagome lattice:
\begin{equation}
 \mathcal{H}_{\sf KH} = \frac{J}{2}
\sum_{\mathbf{R} \in \text{all hexagons}}\  \sum_{\alpha= x, y ,z} [\mathcal{C}_\alpha(\mathbf{R})]^2\ ,
\label{eq:HRSM}
\end{equation}
where the sum of $\mathbf{R}$ runs over hexagonal plaquettes on the kagome lattice
(indicated in Fig.~\ref{Fig_RSM_all}(a,b)), or equivalently the centers of all unit cells.
$\mathcal{C}_\alpha(\mathbf{R})$ is the sum of the six spins around each hexagon  as labeled in Fig.~\ref{Fig_RSM_all}(a,b):
\begin{eqnarray}
\mathcal{C}_\alpha(\mathbf{R}) =\sum_{i \in \text{hex at }\mathbf{R}}   S^\alpha_i\ .
\label{eq:RSM_M}
\end{eqnarray}
Ground states are hence defined by the constraint:
\begin{eqnarray}
\mathcal{C}_\alpha(\mathbf{R}) = 0  \quad  \forall \ \mathbf{R},\ \alpha \ .
\label{eq:kag_constraint}
\end{eqnarray}
on every hexagonal plaquette.
Again, the Hamiltonian is the same for the three components $\alpha = x, y, z$, 
and within the large-$\mathcal{N}$
approximation we treat this as three copies of a theory
in which the spins are independent scalars.
Multiplying out Eq. (\ref{eq:HRSM}) we can rewrite the Hamiltonian in 
terms of bilinear exchange interactions. These interactions couple first, second and third neighbor spins across the hexagon with equal strength.

Taking the lattice Fourier transform of the interactions results in a 
$3\times3$ interaction matrix $\mathbf{J}({\bf q})$, since there are three sites per unit cell. Diagonalizing $\mathbf{J}({\bf q})$ yields   a spectrum with three bands, of which the lowest two are flat and degenerate, with the upper band being dispersive (Fig. \ref{Fig_RSM_all}(c)). 

There are no band touchings between the upper band and the two flat bands at any point in the Brillouin zone.
Accordingly, the real space correlations remain short ranged with a
correlation length on the order of the nearest-neighbor distance at $T=0$.
Also, the ground state fluctuations are not described by any effective Gauss's law.

The CSL state of this model seems to be qualitatively
distinct from a trivial paramagnet, as evidenced by the fractionalization of ``orphan'' spins around a cluster of introduced vacancies \cite{Rehn17PRL}. 
But this raises questions: are there different types of 
non-trivial, short range correlated CSLs?
If so, how do we distinguish these and are they separated
by sharp transitions?
These are some of the core questions this work will address.

\subsection{The question of classification}
Having examined some sample models, 
we can now sharpen the question of classification.
The common feature of the CSLs is that they are described by the type of Hamiltonian $\mathcal{H} =\sum_{\mathbf{R}}[ \mathcal{C}(\mathbf{R})]^2$, where $\mathcal{C}$ is a sum over a local cluster of spins. 
The Hamiltonian can be diagonlized in momentum space, characterized by a matrix $\mathbf{J}(\mathbf{q})$. 
Its spectrum has one or several flat band(s) at the bottom, below one or several bands that are generally dispersive.

However, 
depending on the structure of the spectrum,
different CSLs can have very distinct properties.
It is thus important to understand the mechanism that leads to such distinction, and provide a classification scheme that puts all CSLs in their place. 

The first fundamental difference between CSL models can be seen by comparing the honeycomb snowflake model with the Kagome-Hexagon model.
Some CSLs, such as in the former model, have gap closing points between the bottom flat band(s) and the higher band(s), while others do not. 
This leads to the most crucial differences between the two categories of CSLs.
The former has an emergent generalized Gauss's law describing the ground state fluctuations, and 
algebraically-decaying spin correlations.
The latter exhibits no emergent Gauss's law,
and its spin correlations decay exponentially.

Within each category, 
we still need to make finer distinctions between CSLs.
For the first category, which we call algebraic CSLs, the main question is how many and what kind of generalized Gauss's laws describe the ground state fluctuations, and how a particular type of Gauss's law appears.
We will show that the number and structure  at the gap-closing points determines this, and also explains the transitions between different algebraic CSLs.

For the second category, which we call fragile topological CSLs, there is no generalized Gauss's law. 
It is then natural to ask what can distinguish the different members of this group.
We will show that the homotopy class of the eigenvector, defined as a map over the BZ, see Eq.~\eqref{eq:Tq-map}, is a good topological quantity for the classification of the fragile topological CSLs.

\section{Setting up the classification formalism}
\label{sec.set.up.classification}
\subsection{Constrainer Hamiltonian and its spectrum}
\label{Sec_Constrainer_Hamiltonian}

Let us now define the CSLs in sufficiently general and accurate terms for the development of a robust classification scheme.

First, we work with spins in the large $\mathcal{N}$ limit, or equivalently with \textit{soft spins}.
That is, we treat every spin component  $S_i^\alpha$ as a real number,  $S_i^\alpha \in \mathbb{R}$.
We ignore non-linear constraints that may apply to a real system.
The   common types of non-linear constraints   include those on Ising and Potts variables that take a  finite range of discrete values; or on classical Heisenberg spins that are three-dimensional vectors of unit length ($\mathbf{S}^2 = 1$).
Very often, the soft spin treatment provides a good  approximation to real spins, and can correctly capture the physics of the actual CSLs. Exceptions do exist, and 
we will discuss this point in later sections of this Article.

For Heisenberg models,
each vector spin has three DOFs $S^x, S^y, S^z$. 
However, because they decouple  from each other in   the soft-spin limit, and are described by the same Hamiltonian individually,
we can just analyze one copy of them.
From now on, we therefore treat spin components as independent scalars, and collectively denote them as $S_a(\mathbf{R})$ where $a = 1, \dots, N$ are the number of DOFs in a unit cell labelled by its position $\mathbf{R}$.
 
Second, we work with bilinear Hamiltonians with finite range of interactions,
which is natural for most physical systems.
We specifically investigate the CSLs where the dimension of the ground state manifold grows linearly with the system volume.
We thereby exclude spiral spin liquids \cite{BergmanNatPhys, Mulder2021Spiral, yao21spiral, Yan2022Spiral, Gao2022Spiral}, which have subextensive
degeneracies.
An equivalent statement is that  we study the systems where the spectrum of the Hamiltonian has one or more flat bands at its bottom.

Such CSLs can be written in forms of what we call \textit{constrainer Hamiltonians}.
Such Hamiltonians are written in real space as 
\[
\label{eq:constr}
\mathcal{H} =     \sum_{\mathbf{R} \in \text{u.c.}}\sum_{I=1}^M   [\mathcal{C}_I(\mathbf{R} )]^2 \ ,
\]
where $\mathcal{C}_I(\mathbf{R})$ is a linear combination of spins around the unit cell centered at $\mathbf{R}$ (not necessarily restricted to nearest neighbors).   
Different spins can have different real-valued coefficients (weights) in this sum.
 The index $I$ in the summation runs over all constrainers (there could be more than one) in a unit cell.
Here we denote the number of constraints per unit cell by
$M$, and the number of spin sites in the unit cell by $N$.

In real space, 
the ground states of a classical spin liquid are the spin configurations s.t. $C_I(\mathbf{R} ) = 0$  at  all unit cells and for all $I$'s, hence the name \textit{constrainer}. 
Given $N$ DOFs in a unit cell and $N > M$,
then generally such ground states exist, at least within the
large-$\mathcal{N}$ approximation, because there are more DOFs than constraints.

The constrainer Hamiltonian formalism includes all the canonical classical Heisenberg spin liquids.
For such models, one can always add a term $\sum_i (\mathbf{S}_i)^2$ or $\sum_i ({S}^z_i)^2$  with the correct coefficient to turn the Hamiltonians into the constrainer formalism. 
This added term does not affect the physics because the spin length is fixed in the hard-spin Heisenberg model.

Let us now write down the constrainer Hamiltonian in a more explicit form by specifying $\mathcal{C}_I(\mathbf{R})$. First, it is convenient to encode a given constrainer $\mathcal{C}_I(\mathbf{0})$ for the unit cell at the origin ($\mathbf{R}=\mathbf{0}$) in a $N-$component vector 
encoding the information of how different sublattice sites are summed in the constrainer.
It is  written as
\[
\label{eq:C(0)}
\mathbf{C}_I(\mathbf{0}, \mathbf{r}) = \left(\begin{array}{c}
\sum_{j \in \text{1st sub-lat. sites   } } c_{1,j} \delta_{\mathbf{r}, \mathbf{a}_{1,j}} \\
\sum_{j \in \text{2nd sub-lat. sites   } } c_{2,j} \delta_{\mathbf{r}, \mathbf{a}_{2,j}} \\
\vdots\\
\sum_{j \in \text{$N-$th sub-lat. sites   } } c_{N,j} \delta_{\mathbf{r}, \mathbf{a}_{N,j}} 
\end{array}\right)\  .
\]
Here, $\mathbf{r}$ is a variable that we use to visit all sites on the lattice to see if a spin  at the location  $\mathbf{r}$ is  involved in $\mathcal{C}_I(\mathbf{0})$ (it is if and only if $\mathbf{a}_{b,j}$ appears in $\mathbf{C}_I(\mathbf{0}, \mathbf{r})$). 
The first component $[{C}_I(\mathbf{0}, \mathbf{r})]_{1}$ records information 
of all the first sub-lattice sites in different unit cells that are involved in $\mathcal{C}_I(\mathbf{0})$. 
Their locations are at $\mathbf{a}_{1,j}$'s relative to the center at $\mathbf{0}$, 
where $\mathbf{a}_{1,j}$, pointing to nearby unit cells, is always an integer multiples of lattice vectors plus a constant shift to the center at $\mathbf{0}$.
The coefficients for different  spins summed in $\mathcal{C}_I(\mathbf{0})$ are  $c_{1,j}$'s.
Similarly, the $b^\text{th}$ component $[{C}_I(\mathbf{0}, \mathbf{r})]_{b}$ records the information of how the $b^\text{th}$ sublattice sites are summed in $\mathcal{C}_I(\mathbf{0})$.
Hence, given $\mathbf{C}_I(\mathbf{0}, \mathbf{r}) $, we have the complete information of how the constrainer $\mathcal{C}_I(\mathbf{0})$ is defined.

For the constrainer in a unit cell at a general location $\mathbf{R}$, we need to perform a translation  on $\mathbf{C}_I(\mathbf{0}, \mathbf{r}) $    to get
\[
\mathbf{C}_I(\mathbf{R},\mathbf{r}) = \mathbf{C}_I(\mathbf{0} ,\mathbf{r} -\mathbf{R})\ .
\]
The real space Hamiltonian is written explicitly as
\[\label{EQN_abstract_Ham_general}
\begin{split}
 \mathcal{H}   &= \sum_{\mathbf{R} \in \text{u.c.}}\sum_{I=1}^M   [\mathcal{C}_I(\mathbf{R} )]^2 
\\
&= \sum_{\mathbf{R} \in \text{u.c.}}\sum_{I=1}^M\left[
\sum_{\mathbf{r}} \mathbf{S}(\mathbf{r})\cdot \mathbf{C}_I(\mathbf{R},\mathbf{r})
\right]^2 \ .  
\end{split}
\]
Here, $\mathbf{S}(\mathbf{r}) = (S_1, \dots, S_N) (\mathbf{r})$ is the vector array formed of the $N$ sublattice sites.
For example, $S_b(\mathbf{r})$ is the $b$-th sublattice site at location $\mathbf{r}$.
The term $\sum_{\mathbf{r}} \mathbf{S}(\mathbf{r})\cdot \mathbf{C}_I(\mathbf{r},\mathbf{R})$ is the explicit form of the constrainer $\mathcal{C}_I(\mathbf{R})$ as shown in Eqs.~(\ref{eq:Mhex},\ref{eq:RSM_M}).
With $\mathbf{C}_I(\mathbf{R},\mathbf{r})$ given, we now do not need to rely on pictorial description of the constrainers. 
Instead, we now have their algebraic description ready for mathematical treatment in what follows.

Now let us diagonalize the Hamiltonian in momentum space. A billinear Hamiltonian can be diagonalized in momentum space as 
\[ 
\mathcal{H} =\frac{1}{2} \sum_{{\bf q}} \sum_{a,b = 1}^{N} \tilde{S}_a(-{\bf q}) {J}_{ab} ({\bf q}) \tilde{S}_b({\bf q})
\]
Here, $\tilde{S}_a$ is the Fourier-transformed spin field $S_a$, and $a = 1, \dots , N$ labels the sub-lattice sites.
$\mathbf{J}$ is the $N$-by-$N$ matrix of the interactions.

For constrainer Hamiltonians, there is a simple expression for $\mathbf{J}$ based on $\mathcal{C}_I$. Each constrainer $\mathcal{C}_I$ can be Fourier transformed into momentum space as the \textit{FT-constrainer} $\mathbf{T}_I(\mathbf{q})$,
\[
\label{eqn:FT_constrainer_def}
\begin{split}
 \mathbf{T}_I(\mathbf{q})& =
\sum_{{\bf r}} \ e^{-i\mathbf{r} \cdot \mathbf{q}} \mathbf{C}_I(\mathbf{0},\mathbf{r})\ .
\\
&=
\left(\begin{array}{c}
\sum_{j \in \text{1st sub-lat. sites in }C_I} c_{1,j} e^{-i \mathbf{q}\cdot \mathbf{a}_{1,j} } \\
\sum_{j \in \text{2nd sub-lat. sites in }C_I} c_{2,j} e^{-i \mathbf{q}\cdot \mathbf{a}_{2,j} } \\
\vdots\\
\sum_{j \in \text{$N-$th sub-lat. sites in }C_I} c_{N,j} e^{-i \mathbf{q}\cdot \mathbf{a}_{N,j} } 
\end{array}\right)\  .
\end{split}
\]
$J_{ab} ({\bf q})$ explicitly reads
\[
J_{ab} ({\bf q}) = \sum_{I=1}^M T_{I,a}(\mathbf{q}) T_{I,b}^*(\mathbf{q})\ .
\]

Note that using the constrainer at either  $\mathbf{0}$ or a general unit cell position  $\mathbf{R}$  to define FT-constrainer $\mathbf{T}_I(\mathbf{q})$ does not affect $\mathbf{J}(\mathbf{q})$, since it only adds an overall phase to $\mathbf{T}_I(\mathbf{q})$ that is cancelled in $\mathbf{J}(\mathbf{q})$.

In momentum space, we can examine the spectrum of $\mathbf{J}(\mathbf{q})$ (we will now slightly abuse the notation and refer to both $\mathcal{H}$ and $\mathbf{J}(\mathbf{q})$ as the Hamiltonian).
Given $M$ constrainers in the Hamiltonian, there will generally be $M$ upper bands and $N- M$ bottom flat bands. The upper bands may touch the bottom flat ones at some special points (or in some cases along special lines or planes). 
The higher bands' eigenvectors are those in the space spanned by all $\mathbf{T}_I(\mathbf{q})$'s,
but not necessarily   $\mathbf{T}_I(\mathbf{q})$'s themselves:
note that two different constrainers $\mathbf{T}_I(\mathbf{q})$ and $\mathbf{T}_J(\mathbf{q})$ are not required to be orthogonal to each other.
The bottom degenerate flat bands' eigenvectors are those orthogonal to all $\mathbf{T}_I(\mathbf{q})$'s.

The information of the ground states of the CSL is encoded  in the bottom bands and their eigenvectors.
Equivalently, one can also access such information from the higher bands and their eigenvectors $\mathbf{T}_I(\mathbf{q})$, since the two sets of eigenvectors are orthogonal to each other
and span the full $N$-dimensional vector space. 
It is often easier to look at the higher bands since all $\mathbf{T}_I(\mathbf{q})$'s are known explicitly from the definition of the constrainer $\mathcal{C}_I$'s.

Let us now analyze the structure of the ground states.
First, we note that they span a linear subspace in the space of all spin configurations, and the ground state fluctuations span an isomorphic linear space.
Starting from a ground state that satisfies $\mathcal{C}_I(\mathbf{R}) = 0$  for all $\mathbf{R}$'s and $I$'s,
we can then consider a fluctuation that  keeps  $\mathcal{C}_I(\mathbf{R}) = 0$.
Note that the $\mathcal{C}_I$'s are  linear in the spin variables,
thus, at the level of the soft spin approximation,
any ground state and any such fluctuation can be added linearly with the system remaining in a ground state.
Mathematically speaking, all the ground states span a linear (vector) space,
so the ground states manifold and the manifold of fluctuations between ground states are isomorphic.
In more physical terms, we can start with any initial  ground state, and then every other ground state is bijectively mapped to a fluctuation from the initial ground state to it.

Just like the constrainers describe energetically costly spin 
configurations in real space, and their Fourier transforms describe the higher bands in the spectrum,
their counterparts describe the ground state fluctuations.
Let us consider the \textit{local} fluctuations that satisfies the $\mathcal{C}_I(\mathbf{R}) = 0$ condition for all $\mathbf{R}$'s and $I$'s.
Since the bottom band is $(N-M)$-fold degenerate, we know that there should be $(N-M)$ such linearly-independent local fluctuations. We name these \textit{fluctuators} and abstractly denote them as $\mathcal{F}_I(\mathbf{R})$, where $I=1,\dots, N-M$.
We express each fluctuator as an $N-$component  operator acting linearly in the spin vector space, just as we did with the constrainers $\mathcal{C}_I$'s in Eq.~\eqref{eq:C(0)}, and denote them as $\mathbf{F}_I(\mathbf{R},\mathbf{r})$:
\[
\mathbf{F}_I(\mathbf{R},\mathbf{r}) 
= \left(\begin{array}{c}
\sum_{j \in \text{1st sub-lat. sites   } } d_{1,j} \delta_{\mathbf{r}-\mathbf{R} , \mathbf{a}_{1,j}} \\
\sum_{j \in \text{2nd sub-lat. sites  } } d_{2,j} \delta_{\mathbf{r}-\mathbf{R}, \mathbf{a}_{2,j}} \\
\vdots\\
\sum_{j \in \text{$N-$th sub-lat. sites  } } d_{N,j} \delta_{\mathbf{r}-\mathbf{R}, \mathbf{a}_{N,j}} 
\end{array}\right) \ .
\]
The components in the fluctuator $\mathbf{F}_I(\mathbf{R},\mathbf{r})$ describe quantitatively the local spin fluctuations  that keep  all constrainers   zero, i.e., the flucturator is a zero eigenmode  of the Hamiltonian.

Fluctuators and constrainers are  orthogonal:
\[
\sum_{\mathbf{r}}\mathbf{F}_I(\mathbf{R}_1,\mathbf{r}) \cdot \mathbf{C}_J(\mathbf{R}_2,\mathbf{r}) = 0\quad 
\forall \ \mathbf{R}_1,\ \mathbf{R}_2,\ I,\ J\ .
\]

The \textit{FT-fluctuator}, defined as the Fourier transform to the momentum space
\[
\begin{split}
 \mathbf{B}_I(\mathbf{q})& = 
\sum_{\mathbf{r}} \ e^{-i\mathbf{r} \cdot \mathbf{q}} \mathbf{F}_I(\mathbf{0},\mathbf{r})    \\
&=
\left(\begin{array}{c}
\sum_{j \in \text{1st sub-lat. sites   } } d_{1,j} e^{-i \mathbf{q}\cdot \mathbf{a}_{1,j} } \\
\sum_{j \in \text{2nd sub-lat. sites  } } d_{2,j} e^{-i \mathbf{q}\cdot \mathbf{a}_{2,j} } \\
\vdots\\
\sum_{j \in \text{$N-$th sub-lat. sites   } } d_{N,j} e^{-i \mathbf{q}\cdot \mathbf{a}_{N,j} } 
\end{array}\right) \ ,
\end{split}
\]
is then orthogonal to all the FT-constrainers $\mathbf{T}_J(\mathbf{q})$.
The FT-fluctuators are exactly the eigenvectors spanning the $(N-M)$ degenerate bottom flat bands.

The sample models studied in this \paper have only one constrainer ($M=1$), so we can drop the index $I$:
\[
\mathcal{H} = \sum_{\mathbf{R}  \in \text{u.c.}}   \mathcal{C}(\mathbf{R}  )^2  \ ,
\] 
In this way, the physics can be clearly demonstrated without too much notational complication.
Correspondingly, the Hamiltonian matrix
\[
\label{eq:Jab(q)}
J_{ab} ({\bf q}) = T_{a}(\mathbf{q}) T_{b}^*(\mathbf{q})\ .
\]
has $N-1$ flat bands with eigenvalue zero and $1$ dispersive higher band. 
$\mathbf{T} (\mathbf{q})$ as the only FT-constrainer 
is   also the unnormalized eigenvector of the higher band.
The higher band dispersion is  
\[
\omega_\text{top} (\mathbf{q}) = |\mathbf{T} (\mathbf{q})|^2\ .
\]
The top band may or may not touch the bottom bands, depending of the specific form of the constrainer $\mathbf{C}(\mathbf{R},\mathbf{r})$ and its Fourier transform $\mathbf{T} (\mathbf{q})$.
Since there is only one top band but several bottom bands, 
it is easier to analyze the top band rather than the bottom ones. The physics is easily generalizable to the cases with multiple higher bands. 

Depending on whether the top band touches the bottom bands, 
the CSL falls into one of two broad categories.

The algebraic CSLs have band-touching points, and are controlled by the physics around those points;
they have algebraically-decaying correlations described by 
emergent, generalized U(1) Gauss's laws.

The fragile topological CSLs have no band-touching points, 
and the correlations in the bulk are short-ranged.
However, as we shall demonstrate below, they have quantized topological properties that cannot be changed without closing the gap between the top and bottom flat bands.
All the topological information is encoded in the FT-constrainer $\mathbf{T} ({\bf q})$. 
After introducing several mathematical tools, we will demonstrate in detail how to extract the information about the algebraic and fragile topological CSLs from the Hamiltonian.

\subsection{Tools from flat band theory} 
\label{SubSec.flat.band.theory}

\begin{table*}[th]
\caption{Connection between the physics of flat band theory and classical spin liquid,  using the language of compact local states (CLS) and non-local loop states (NLS) \cite{RhimPhysRevB.99.045107}.
\label{TABLE_flat_band_theory}
} 
\begin{tabular}{c @{\hskip 15pt} c @{\hskip 15pt} c} 
	\toprule
	 Flat band theory & Classical spin liquid \\[3pt] \midrule 
	CLS 	: local eigenstate of the flat band & local spin fluctuation within ground states \\[3pt]
	NLSs 	: non-local eigenstate of flat band & non-local spin fluctuation within ground states \\[3pt]
	a singular band touching point &
	effective Hamiltonian indicates the Gauss's law\\[3pt]
	multiple singular band touching points &
	coexistence of different Gauss's laws \\[3pt]
	merging/splitting of singular band touching points &
	transition between different algebraic CSLs\\[3pt]
 no band touching on the flat bands & fragile topological CSLs\\[3pt] 
	\bottomrule
	\end{tabular}
\end{table*}

Since our analysis focuses on flat bands, known  results from flat band theory (for fermionic/bosonic hopping models) can be applied here.
In this subsection we review these results, with a view to applying them later.

The key properties for the CSLs  are encoded in the flat bands at the bottom of the spectrum of the Hamiltonian matrix $\mathbf{J}(\mathbf{q})$.
In the context of classical spin systems, the bottom bands are related to the fluctuations between ground states, as  discussed in Sec.~\ref{Sec_Constrainer_Hamiltonian}  .
The real space local fluctuators $\mathbf{F}_I(\mathbf{R},\mathbf{r})$, or equivalently the momentum space FT-fluctuators $\mathbf{B}_I(\mathbf{q})$, describe  these fluctuations.

One can  write down a hopping model described by the same Hamiltonian $\mathbf{J}(\mathbf{q})$.
Here, flat bands are also of great interest and have been studied intensively \cite{Bergman2008PRB, ZhangPRB2010, Chalker2010PRB, Tang2011PRL, Neupert2011PRL, Zhi2018SciAdv, Chebrolu2019PRB, ChiuPRR2020, Kang2020NatComm, RhimPhysRevB.99.045107}.
In reviewing these results we largely follow  Ref.~\onlinecite{RhimPhysRevB.99.045107}. The key concepts are summarized in Table.~\ref{TABLE_flat_band_theory}.

The key to the physics of a flat band in a free hopping model  is that a flat band in momentum space corresponds to a \textit{compact local state} (CLS, not to be confused with classical spin liquid, CSL) in real space. 
The compact local state is an exact eigenstate of the Hamiltonian, 
and is only supported on a finite, local region of the lattice.  
Their existence is proven in Appendix A of Ref.~\onlinecite{RhimPhysRevB.99.045107}.
Such a locally supported state usually does not exist for a dispersive band. 
Compact local states in real space, and the flat band in momentum space, are two facets of the same physics.
For a rigorous proof of this statement, see Sec.~II.A of Ref.~\onlinecite{RhimPhysRevB.99.045107}.

The connection to CSLs is the following:
the compact local state in the hopping model corresponds to the fluctuator in CSL.
Let us use nearest-neighbor-hopping kagome model,
\[\label{eqn:kagome.hopping}
 \mathcal{H}_{\sf{kagome,hopping}} = \sum_{\braket{i,j}}c_i c_j^\dagger + \text{h.c.}
\]
as an example. 
Its CSL version is the nearest neighbor AFM kagome model, which is a classical spin liquid in the large-$\mathcal{N}$ description
(although order-by-disorder at very low temperatures 
cuts off the spin liquid behaviour for $O(3)$ Heisenberg spins \cite{Chalker1992PhysRevLett.68.855,Harris1992PhysRevB.45.2899,Ritchey1993PhysRevB.47.15342,Zhitomirsky2008PhysRevB.78.094423,Chern2013PhysRevLett.110.077201}). A more detailed analysis of the kagome model will be presented in Sec.~\ref{subsubsec_kagome}.
Here we state a few basic facts of it. Its Hamiltonian is 
\[
\label{eqn:kagomeAFM}
\begin{split}
  \mathcal{H}_{\sf{kagome}}& = \sum_{\langle i,j\rangle} S_i S_j + 2 \sum_i S_i^2 \\
  &= \sum_\bigtriangleup\left( \sum_{i \in \bigtriangleup} S_i\right)^2 + \sum_\bigtriangledown\left( \sum_{i \in \bigtriangledown} S_i\right)^2  \ .
\end{split}
\]

\begin{figure}
 \centering
 \includegraphics[width=\columnwidth]{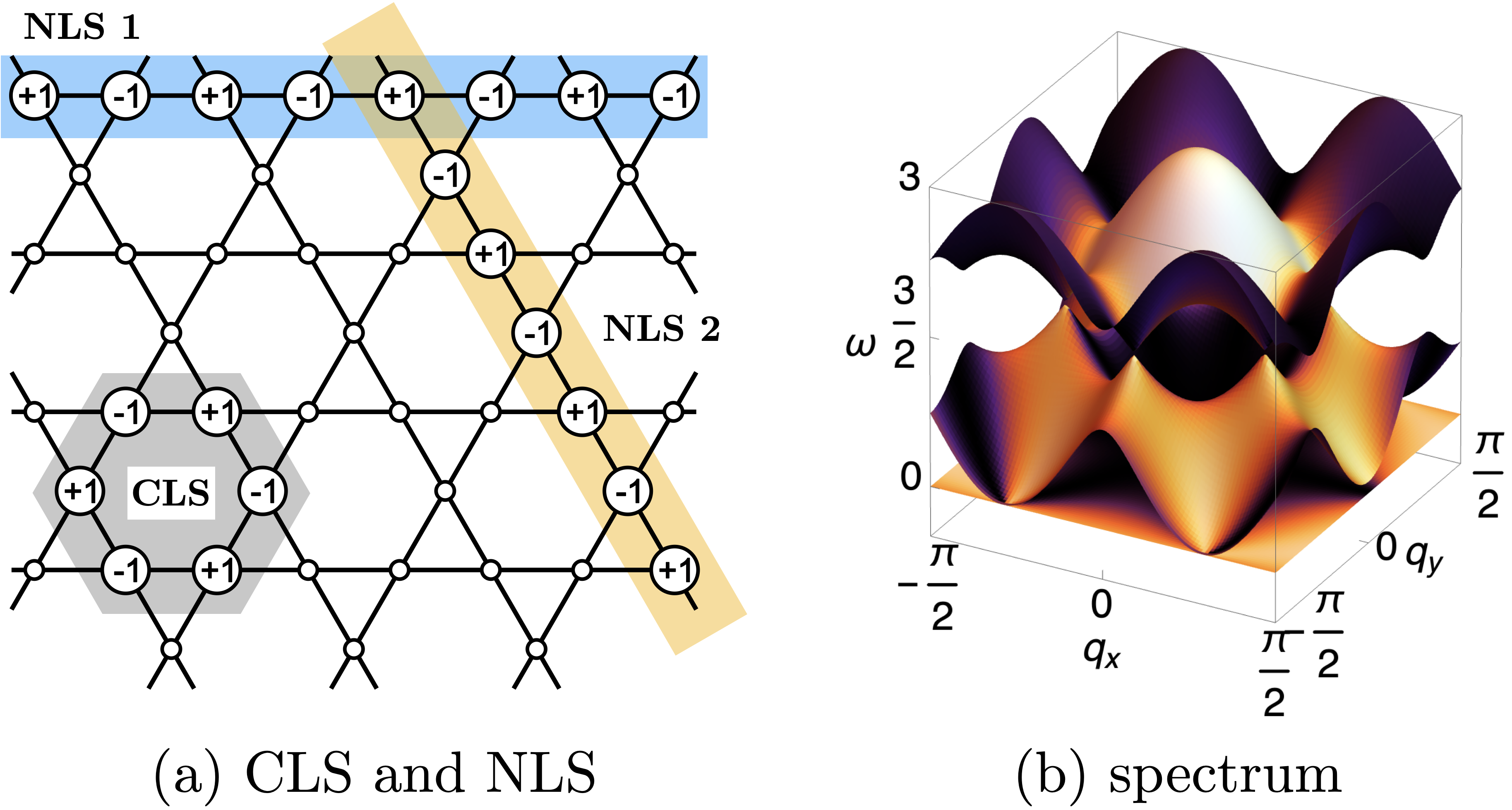}
 \caption{ (a) Compact local states (CLS) and non-local loop states (NLS) of the kagome model (Eq.~\eqref{eqn:kagome.hopping}), which can also be interpreted as the local and loop fluctuators in the classical spin liquid model (Eq.~\eqref{eqn:kagomeAFM}). 
 One can check that the hopping amplitude from these states to any other site is zero.
The CSLs are not linearly independant: adding all of them on the entire lattice yields zero.
 (b) Spectrum of the Hamiltonian in Eq.~\eqref{eq:kagome.Hamiltonian}.
 }
 \label{Fig_Kagome_ice}
\end{figure}

Given the hopping Hamiltonian Eq.~\eqref{eqn:kagome.hopping}, one can find by inspection the compact local state of the model. 
The wavefunction of the compact local state at location $\mathbf{R}$ can be generically encoded in an $N$-component fluctuator vector $\mathbf{F}_{\mathbf{r}, \mathbf{R} }$  via the relation
\[
\label{eq:compact-local-wavefunction}
\left|\chi_{\mathbf{R}}\right\rangle=\sum_{\mathbf{r} } \sum_{a=1}^{N} [{F}_{\mathbf{R}, \mathbf{r}}]_a \cdot\left|\mathbf{r}, a\right\rangle\  .
\]
Here, the $a^\text{th}$ component of $\mathbf{F}_{\mathbf{R}, \mathbf{r}}$ encodes the information of the $a^\text{th}$ sub-lattice site's contribution to the compact local state.
And $|\mathbf{r}, a\rangle$ denotes an electron occupying the sublattice site $a$ at unit cell $\mathbf{r}$. 
In the case of kagome model (Eq. (\ref{eq:kagome.Hamiltonian})), $N=3$, and $\mathbf{F}_{\mathbf{0}, \mathbf{r}}$ is 
\[
\mathbf{F}_{\mathbf{0}, \mathbf{r}}=\frac{1}{\sqrt{6}}\left(\begin{array}{c}
\delta_{\mathbf{r},-\mathbf{a}_{1}}-\delta_{\mathbf{r}, 0} \\
\delta_{\mathbf{r}, 0}-\delta_{\mathbf{r}, \mathbf{a}_{2}} \\
\delta_{\mathbf{r}, \mathbf{a}_{2}}-\delta_{\mathbf{r},-\mathbf{a}_{1}}
\end{array}\right)\  .
\label{EQN.A.kagome.ice}
\]
The corresponding compact local state wavefunction
is illustrated in Fig.~\ref{Fig_Kagome_ice}.
One can apply the hopping Hamiltonian to it and find that
the hopping amplitude of the compact local state to any other site is exactly zero.

We can now  illustrate the connection between the compact local state in Eq.~\eqref{eq:compact-local-wavefunction} and the bottom band eigenvector of the CSL model in Eq.~\eqref{eqn:kagomeAFM}.  Indeed, the compact local state in Fig.~\ref{Fig_Kagome_ice} has the property, when reformulated in the language of spin components, that the sum of spins on each triangle remains $0$, as expected from Eq.~\eqref{eqn:kagomeAFM}.
More formally,
Fourier transforming the fluctuator \eqref{EQN.A.kagome.ice} into momentum space yields 
\[\label{eqn.ft.fluctuator.kagome}
\begin{split}
  {\mathbf{B}} (\mathbf{q})= &
\int \text{d}\mathbf{r} e^{-i\mathbf{r} \cdot \mathbf{q}} \mathbf{F}_{\mathbf{0}, \mathbf{r}}  \\
=& \frac{1}{\sqrt{6}}\left(\begin{array}{c}
e^{i \mathbf{a}_{1} \cdot \mathbf{q}}-1 \\
1-e^{-i \mathbf{a}_{2} \cdot \mathbf{q}} \\
e^{-i \mathbf{a}_{2} \cdot \mathbf{q}}-e^{i \mathbf{a}_{1} \cdot \mathbf{q}}
\end{array}\right)\  .   
\end{split}
\]

On the other hand, diagonalizing both Hamiltonians in momentum space, we obtain   (up to adding an additional $c\mathbf{1}_{3\times 3}$ to shift all bands by a constant) 
\[
\label{eq:kagome.Hamiltonian}
\mathbf{J}_{\sf{kagome}}(\mathbf{q})=  \left(\begin{array}{ccc}
 2 & e^{-i \mathbf{a}_{3} \cdot \mathbf{q}}+1 & e^{i \mathbf{a}_{2} \cdot \mathbf{q}}+1 \\
e^{i \mathbf{a}_{3} \cdot \mathbf{ q}}+1 &  2 & e^{-i \mathbf{a}_{1} \cdot \mathbf{q}}+1 \\
e^{-i \mathbf{a}_{2} \cdot \mathbf{q}}+1 & e^{i \mathbf{a}_{1} \cdot \mathbf{q}}+1 &  2
\end{array}\right) \ , 
\]
where $\mathbf{a}_{1} = (1,0),~ \mathbf{a}_{2}=(-1 / 2, \sqrt{3} / 2),$ and $\mathbf{a}_{3}=-\mathbf{a}_{1}-\mathbf{a}_{2}$ encode the lattice geometry.
We can directly confirm  the flat band is at $\omega = 0$ with eigenvector
\[
\hat{\mathbf{B}}{(\mathbf{q})}=c_{\mathbf{q}}\left(\begin{array}{c}
e^{i \mathbf{a}_{1} \cdot \mathbf{q}}-1 \\
1-e^{-i \mathbf{a}_{2} \cdot \mathbf{q}} \\
e^{-i \mathbf{a}_{2} \cdot \mathbf{q}}-e^{i \mathbf{a}_{1} \cdot \mathbf{q}}
\end{array}\right) \ ,
\label{EQN.v.kagome.ice}
\]
where 
\[
c_{\mathbf{q}}=\left(6-2\cos q_{x}-4 \cos (q_{x} / 2) \cos (\sqrt{3} q_{y} / 2)\right)^{-1 / 2}
\]
is the nomalization factor.
This is exactly the normalized FT- fluctuator in Eq.~\eqref{eqn.ft.fluctuator.kagome}. 

We have thus established that the compact local state formulation of the flat band hopping Hamiltonian is related, via the Fourier tranform, to the momentum-space eigenvector (a.k.a. the fluctuator) of the ground state spin configuration in a CSL.
Having established this connection, we can now translate known properties of compact local states into the language of spin liquids.

As mentioned before, 
the leading-order criterion for classification is whether there is  band touching between the flat bands and upper  bands or not.
This is also one of the main topics in the study of compact local states.
Depending on the hopping model,
there can be three scenarios: no band touching, non-singular band touching,
and singular band touching.
Each of these cases is reflected in the structure of the corresponding compact local states.

\subsubsection*{Non-singular band touching}
Roughly speaking, a non-singular band touching is an ``accidental'' band touching that does not 
qualitatively affect the physics of the flat band. 
More precisely, it can be defined in terms of the CLSs 
being linearly independent of the eigenvectors of the
non-flat band. 
The simplest example of this is two completely decoupled systems I and II, each with its own bands.
Obviously, if a band in system I touches the flat band in system II,
there is nothing special happening at the band-touching point, 
and the band touching can be lifted trivially.
Another example is  that the vicinity of the band touching 
in a two band system
can be written as $p(k_x, k_y)(\sigma_z + \sigma_0)$, where 
$p(k_x, k_y)$ vanishes at the band touching. 
Since the matrices $\sigma_z$ and $\sigma_0$ commute, the two modes can be trivially separated by shifting the dispersive band upwards via addition of a term $E_0 \sigma_z$. Such non-singular band touching can thus be smoothly deformed to a gapped spectrum.

Let us first discuss the physics of flat band  with no band touching or with non-singular band touchings only. 
In this case,  the eigenvector $\mathbf{B}(\bfq)$ of the bottom band is well-defined globally so that a \textit{vector bundle} associated with the flat band exists \textit{globally} -- this is known as a \textit{trivial} vector bundle.  
(The reader may be familiar with nontrivial (complex) vector bundles, which can possess nontrivial Chern number. The perfectly flat bands resulting from the constrainer formulation of the CSL can be shown to have zero Chern number if they are separated by the gap, see Section~\ref{sec.topological.csl} below and Ref.~\onlinecite{Chen2014}.) 
To make $\mathbf{B}(\bfq)$ well-defined without any singularity requires $|\mathbf{B}(\bfq)|> 0 $ in the entire BZ. 
This is exactly the condition that the bottom band is separated by a gap from the dispersive higher bands.

In real space, that means the $L_x L_y$ compact local states
generated by applying lattice translations to a single CLS (assuming the lattice has $L_x L_y$ unit cells) are all linearly independent, 
so they span the entire flat band~\cite{RhimPhysRevB.99.045107}, encoding the $L_x L_y$ states on this band exactly.
The same applies to the CSL models:
if the total $L_x L_y$ fluctuators $\mathbf{F}_{\mathbf{R}, \mathbf{r}}$ on different unit cells are linearly independent, 
then the corresponding FT-fluctuator $\mathbf{B}(\bfq)$ is non-vanishing everywhere in the BZ,
and there is no band-touching between the bottom bands and the top ones.
The Kagome-Hexagon model (Eq.~\eqref{eq:HRSM}) is an example of such a system
(with a slight complication that the flat bands are twofold degenerate).

\subsubsection{Singular band touching}

Let us move on to the case of singular band-touching between the bottom flat bands and higher bands.
In this  case, 
the bottom band eigenvector $\mathbf{B}(\bfq)$ vanishes  at certain $\bf{q}$'s,
which are the band touching points. 
A single band accounts for $L_x L_y$ states of the Hamiltonian,
so a flat band with band touching points should account for $L_x L_y + n $ states, where the additional $n$  states come from the degeneracy at the band touching points.
The exact value of $n$ depends on the type of band touching ponts. 
Therefore, the $L_x L_y$ compact local states (related by spatial translations)  are not enough to account for all the $L_x L_y + n $  states on the flat band.
Moreover, in the presence of singular band touchings, it can be shown (see e.g. Ref.~\onlinecite{RhimPhysRevB.99.045107}) that the $L_x L_y $ compact local states 
are \textit{not} linearly independent.

Where are the missing states?
It turns out that there are new, non-local loop (or other topological) states (NLS), which are eigenstates of the Hamiltonian.
They are new in the sense of being linearly independent from the compact local states. 
They account for the  states on the flat band  which are missing due to the linear dependence of the compact local states \cite{Bergman2008PRB},
as well as additional $n$ states from the band touching points.
Singular band touching, linear-dependence of compact local states, and the existence of nontrivial loop states are different facets of the same physics.
 
The physics can be translated to CSLs too.
In this context,
the local fluctuators $\mathbf{F}({\mathbf{R}, \mathbf{r}})$ are not linearly independent,
and $\mathbf{B}(\bfq)$ becomes zero at the singular band touching points.
There are loop fluctuators $\mathcal{F}_\text{loop}$ accounting for the degeneracy at the band touching points.  
The consequence of them -- emergence of the generalized U(1) structure -- will be analyzed in detail in the next section.

The kagome model (Eq.~\eqref{eq:kagome.Hamiltonian})
and honeycomb-snowflake model (Eq.~\eqref{eq:Hhex}) exhibit singular band touchings. 
Let us consider the kagome model as an
example.
There is one band touching per BZ, and  the total number of zero energy states is $L_xL_y+1$. 
We can see that the eigenvector $\mathbf{B}(\bfq)$ (Eq.~\eqref{eqn.ft.fluctuator.kagome}) becomes zero at $\bf{q} = 0$, where a singularity exists,
i.e., $\mathbf{B}(\bfq)$ is not smooth there. 
This is in contrast to the non-singular band touching point, where $\mathbf{B}(\bfq)$ can be written down smoothly.
In real space, 
that means an equal weighted sum  (i.e. phase distribution of  $\bf{q} = 0$) of all the $L_x L_y$ compact local states (Fig.~\ref{Fig_Kagome_ice})
vanishes, meaning they are linearly dependent \cite{Schulenberg2002PRL, Zhitomirsky2004PRB}.
Removing any one of them results in $L_x L_y-1$ linearly independent states.
In addition to the compact local states, there are two non-local loop states supported on winding loops on the lattice (Fig.~\ref{Fig_Kagome_ice})  that are also eigenstates of the Hamiltonian.
They cannot be constructed from the compact local states,
so we have in total $L_x L_y+1$ states at the energy of the flat band. 
They account for all the states on the flat band and at the point of band touching.

Finally, we comment that a complete set of compact local states
accounting for all states on the flat band can always be found in 1D systems. 
Therefore all flat bands in 1D system have no band touching, or at most non-singular ones \cite{Maimaiti2017PhysRevB,Maimaiti2019PhysRevB}.
We will therefore concentrate on 2D and 3D examples in this article.

\section{Algebraic CSL classification: emergent Gauss's laws} 
\label{sec:algebraic}

The common feature of algebraic CSLs is that the gap between bottom flat band(s) and higher band(s) closes at some points in the momentum space in a \textit{singular} manner, or, in the words of flat band theory, 
these \textit{singular} band touching points determine the class of the algebraic CSL.
By examining the eigenvector configuration or, equivalently, the effective Hamiltonian near a band crossing point, one can derive the generalized U(1) Gauss's law emerging there.
The ground state fluctuations  are essentially 
effective electric field fluctuations that obey a charge-free condition, in which the charge is defined via the generalized U(1) Gauss's law.
The statements above are already well-understood for conventional U(1) spin liquids like the pyrochlore (and kagome $\mathcal{N}\geq4$) Heisenberg models.
In this \paper\ we generalize it to other types of U(1) spin liquids with a simple algorithm to identify the Gauss's law.

\subsection{U(1) structure of the ground state manifold}

We first show that with  singular  band touching points,
the linear space of ground states has a U(1) structure.

As we have established in previous section,
in this case, there are local fluctuators encoded in $\mathcal{F}_I(\mathbf{R})$, 
and also loop fluctuators we denote abstractly as $\mathcal{F}_{\text{loop } 1}$, $\mathcal{F}_{\text{loop } 2}$ \textit{etc} that are linearly independent from the local ones.
Together they account for all states on the bottom flat bands and the additional states at the band touching points.

\begin{figure}
 \centering
 \includegraphics[width=0.5\textwidth]{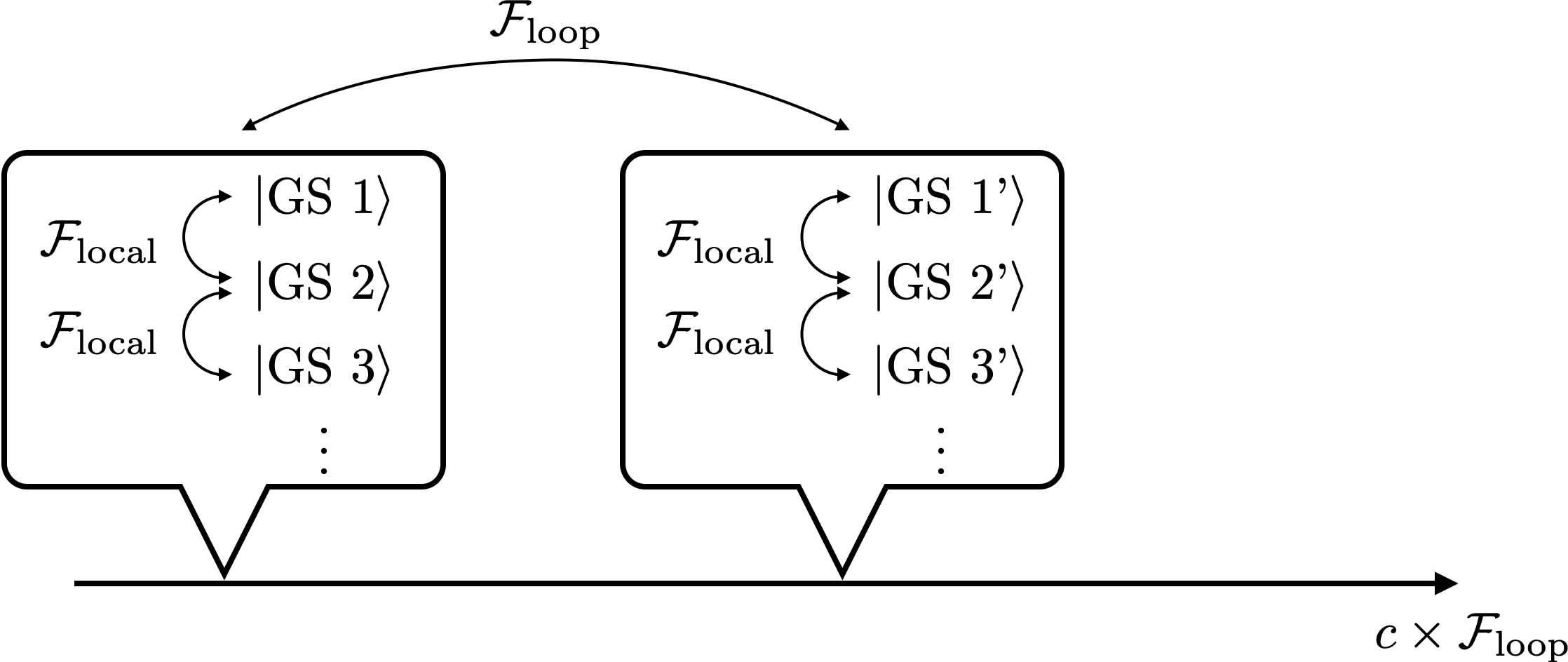}
 \caption{ $U(1)$ Structure of the ground states. States connected by local fluctuators $\mathcal{F}_{\text{local }}$ are in the same equivalence class, and different equivalence classes are connected via the non-local $\mathcal{F}_{\text{loop }}$ fluctuators.
 }
 \label{Fig_U1_structure}
\end{figure}

Hence the ground states can be divided into equivalence classes in the following sense:
two ground states are equivalent 
if and only if there are some local fluctuators that take one to the other.
Then, applying a loop fluctuator to a ground state takes it to another equivalence class. 
Note that, each loop fluctuator comes with a real coefficient $c$.
The equivalence classes hence have an uncompactified $U(1)$ (or $\mathbb{R}$) structure.
The loop fluctuators play the role similar to logical operator in topological orders, taking the ground state from one equivalence class  (a.k.a. superselection sector)  to another.
This is schematically shown in Fig.~\ref{Fig_U1_structure}.

Now the question is:
how to describe the $U(1)$ structure? 
It turns out that if the band-touching manifold is a point (or a few points),
then associated with each point one can derive a generalized Gauss's law by examining the eigenvector structure around the point. This will be the central result for the algebraic CSL classification. 
 
In more exotic cases, the band-touching manifold is not a point (or a few points) but a higher dimensional object (curves, membranes etc.). Then it is no longer possible to write down the long-wavelength physics as an expansion around a point and obtain a Gauss's law to capture all the physics -- because there are infinitely many gapless points elsewhere.
In what follows we will mostly focus on the former case of isolated touching point(s).

\subsection{Generalized Gauss's laws and their physics} 
\label{sec:general.Gauss.fracton}
 
While  the Maxwell U(1) gauge theory and its reincarnation in classical spin ice are well known~\cite{Huse2003, Hermele2004PhysRevB.69.064404, Ryzhkin2005, Henley05PRB, Castelnovo2008, Henley2010ARCMP}, the concept and consequences of \textit{generalized} U(1) gauge theories may be an unfamiliar topic to some readers. 
In this section we introduce the electrostatics of these new theories, 
since many of the algebraic CSLs are  described in this language. 
Given our focus on \textit{classical} spin liquids, we focus here on the classical electrostatic sector of the generalized Maxwell theory, without the magnetic fields which would introduce quantum dynamics.

The Gauss's law of Maxwell U(1) theory is written as 
\[
\label{EQN_maxwell_gauss}
\rho = \partial_\alpha E_\alpha\ .
\]
The spin liquid ground states are described by an electrostatic theory requiring the charge-free condition
\[
\rho = \partial_\alpha E_\alpha\ = 0\ ,
\]
to be satisfied everywhere on the lattice.
As the simplest \textit{Lorentz invariant} gauge theory,
the Maxwell U(1) gauge theory describes one of the fundamental forces of the universe as well as the emergent behavior of various many-body systems. 
Obviously, electric field fluctuations obeying the charge-free condition preserve the net  Noether  charge of the system
\[
\int \mathop{dv} \rho = 0\ .
\]

A difference between condensed matter systems and the universe is that the Lorentz symmetry, including continuous rotational symmetry of  space, 
can be broken in the former cases. 
This means the emergent theories describing solid-state systems need not have Lorentz invariance. 
Instead, only a lower set of symmetries (e.g. discrete rotational symmetry of the lattice or even less) 
need to be satisfied. 
Applying this principle to the CSLs,
means that one can write down generalized U(1) gauge theories and their Gauss's laws that do not necessarily respect Lorentz/rotational symmetry.

Some of the preeminent examples in recent years are the rank-2 symmetric U(1) gauge theories~\cite{PretkoPRB16,PretkoPRB17}.   
Here we briefly review the so-called scalar charged case~\cite{PretkoPRB16}.
The theory respects rotational symmetry of space but not the Lorentz symmetry.
Its electric field is a rank-2 symmetric tensor $E_{\alpha \beta}$,
which can be chosen to be traceless or not. 
Its (scalar) charge is defined as 
\[
\label{EQN_r2u1_gauss}
\rho = \partial_\alpha \partial_\beta E_{\alpha \beta} 
\ .
\]
One exotic consequence is the conservation of charge dipole or higher multipoples. 
For the example given above, 
the total electric dipole in the $\gamma$ spatial direction  
\[
\begin{split}
    \int \mathop{dv} r_\gamma \rho& = \int \mathop{dv} r_\gamma  \partial_\alpha \partial_\beta E_{\alpha \beta} \\
&= \int d\Sigma_{\alpha} r_{\gamma} \partial_{\beta} E_{\alpha \beta}  - \int \mathop{dv} \delta_{\gamma, \alpha}\partial_\beta E_{\alpha \beta}\\
& = \int d\Sigma_{\alpha} r_{\gamma} \partial_{\beta} E_{\alpha \beta}  - \int \mathop{dv} \partial_\beta E_{\gamma \beta}  \\
& = \int d\Sigma_{\alpha} r_{\gamma} \partial_{\beta} E_{\alpha \beta}\ ,
\end{split}
\]
where  $\int d\Sigma_{\alpha}$ denotes an integral over the boundary surface
normal to the component $\alpha$. 
This implies that the total dipole moment is entirely determined
by the value of the fields at the boundary of the system,
which further implies that it cannot be changed by any local rearrangement of the electric field in the bulk.
Thus any local dynamics must 
conserve the 
electric dipole,
with the
consequence that isolated charges cannot move,
in contrast to Maxwell U(1) gauge theory.
Such immobile charges are dubbed \textit{fractons}, which have received much theoretical attention in the past decade~(see e.g. Ref.~\onlinecite{Pretko-review2020} for review and references therein).

We can take a further step in the generalization \cite{BulmashArxiv2018}. 
We need two pieces of data to define a generalized electromagnetism: (1) the electric field and (2) the Gauss's laws that define the charges.
The electric field  does not need to be in the form of vector or tensor, since we do not enforce the rotational symmetry in the first place. 
Instead, we just label different components of the electric field as $E_i$, where $i = 1, 2, \dots, n_E$.
Correspondingly, the charges do  not need to be a scalar, vector, or tensor. 
Instead it can have several components labeled as $\rho_j$ where $j = 1, 2, \dots, n_c$.
Each component is defined via the Gauss's law as 
\[
\label{EQN_general_gauss_law}
\sum_{i = 1}^{n_E} D^j_i E_i = \rho_j\ .
\]
Here, $D^j_i$'s are linear differential operators.
In the case of Maxwell electromagnetism,  Gauss's law is explicitly written in Eq.~\eqref{EQN_maxwell_gauss},
and in the case of rank-2 U(1) symmetry gauge theory   it is written in Eq.~\eqref{EQN_r2u1_gauss}.
One can also write down any other choice of $D^j_i$ to define a new U(1) electromagnetism.

For a generalized gauge theory, 
the conserved quantities are
\[
\label{EQN_general_charge_conservation}
Q_{\{f\}} = \sum_{j= 1}^{n_c} \int \mathop{dv} f_j \rho_j\ ,
\]
for any set of functions $\{f_1, f_2, \dots, f_{n_c}\}$ that satisfy
\[
\label{EQN_general_charge_conserv}
\sum_{j= 1}^{n_c} \tilde{D}^j_i f_j = 0\ .
\]
Here, $\tilde{D}^j_i$ is a linear differential operator, related to ${D}^j_i$ by multiplying every term in ${D}^j_i$ that has $n$ derivatives by $(-1)^n$.
It is obvious that total charge conservation, i.e. $f_j = \text{constant}$ holds for any generalized Gauss's law.
But depending on the form of ${D}^j_i$, there can be other sets of $\{f_j\}$ that satisfy Eq.~\eqref{EQN_general_charge_conserv}. 
For instance, choosing $f_j=r_j$ would correspond to the dipole moment conservation in Eq.~\eqref{EQN_r2u1_gauss}.
The above generalization encapsulates new 
 conservation laws in the form of charge dipoles, multipoles, or combinations thereof.
Like the rank-2 symmetric U(1)  gauge theory,
such multipole conservation laws lead to immobility of isolated charge excitations, which are fractons.

Eqs.~(\ref{EQN_general_gauss_law})-(\ref{EQN_general_charge_conservation}) complete  the definition of electrostatics (i.e. the classical sector) of the generalized U(1) gauge theory.
We will show that the algebraic CSLs are described by the low energy effective theory, written here in the Hamiltonian form
\[
\mathcal{H} = \sum_{j = 1}^{n_c} \left( \sum_{i = 1}^{n_E} D^j_i E_i \right)^2 = \sum_{j = 1}^{n_c}  \rho_j{}^2\ ,
\]
where $E_i$ emerges from the spin degrees of freedom (see section~\ref{sec:Gauss-from-spins} for the detailed derivation).
The ground state fluctuations are then described by a generalized Gauss's law and the requirement that all charges vanish
\[
\sum_{i = 1}^{n_E} D^j_i E_i = 0\ .
\]

Given the definition of electric field and  charge (Eq.~\eqref{EQN_general_gauss_law}), it is also straightforward to write down the gauge transformations (more accurately speaking, gauge redundancy) and construct the magnetic field as objects invariant under these gauge transformations. 
The synthetic magnetic field encodes the fluctuations within the classical manifold of degenerate states and is necessary to describe the quantum spin liquid that originate from its `parent' CSL, see e.g. the well-known U(1) description of quantum spin ices~\cite{Hermele2004PhysRevB.69.064404}. 
This completes the construction of electromagnetism of the generalized U(1) gauge theory; 
interested readers can refer to Refs.~\cite{BulmashArxiv2018,GromovPhysRevX2019} for more details.

\subsection{Extracting Gauss's laws: one constrainer models}

\label{sec:Gauss-from-spins}

The generalized Gauss's laws introduced above provide a  
description of the ground state fluctuations in terms of the generalized charge-free condition in the corresponding U(1) theory.
Hence, the Gauss's law distinguishes different algebraic CSLs.

We will describe the general mathematical recipe to determine the Gauss's law in this  section, 
and then apply it to concrete examples in Sec.~\ref{sec:algebraic.csl.models}.
Since the only terms in the Hamiltonian are the constrainers,  
they must dictate the emergent Gauss's law.
In momentum space, 
FT-constrainers (i.e. the eigenvectors of the higher bands) describe the energetically costly spin configurations. 
Upon the inverse Fourier transform into real space, these become the (generalized) derivatives $D_i^j E_i$ (see Eq.~\eqref{EQN_general_gauss_law}) in the long-wave length limit, which turn out to be precisely the formulation of Gauss's law.

In real space, the Hamiltonian is given by the constrainer form Eq.~\eqref{eq:constr}. To lighten the notation,  we assume one constrainer in what follows:
\[
\mathcal{H} = \sum_{\mathbf{R}  \in \text{u.c.}}   \mathcal{C}(\mathbf{R}  )^2 \ .
\]
In Sec.~\ref{Sec_Constrainer_Hamiltonian},
we have analyzed the mathematical detail of this type of Hamiltonians. 
It has one dispersive top band and $N-1$ bottom flat bands,  
where $N$ is the number of sub-lattice sites  in a unit cell.
The Fourier transformed constrainer (FT-constrainer) $\bfT(\bfq)$  has $N$ components.

The Hamiltonian in momentum space is then represented by an $N\times N$ matrix in Eq.~\eqref{eq:Jab(q)}
\[
{J}_{ab} = T_{a}(\bfq) T_{b}^*(\bfq)\   .
\]  
The 
eigenvector of the top band is 
$\mathbf{T}(\bfq )$, 
and its eigenvalue (dispersion) is $\omega_\mathsf{top}(\bfq) = |\mathbf{T}(\bfq)|^2$.
The $N-1$ bottom bands are at energy $0$,
whose eigenvectors are those orthogonal to $\hat{\mathbf{T}}(\bfq ) \equiv \mathbf{T}(\bfq )/|\mathbf{T}(\bfq )|$.

Since we are studying the cases in which singular band-touching happens,
there must be one (or more) wavevector $\bfq_0$ where the dispersive band has zero eigenvalue: $\omega_\mathsf{top}(\bfq_0) = |\mathbf{T}(\bfq_0)|^2 = 0$.
At this point, all components of $\mathbf{T}(\bfq_0)$ are identically zero.
This is reflecting the singular nature of the band-touching point:
due to the non-smoothness of the eigenvector configuration around the singular gap-closing point,
the only way to write it down continuously is to have $\mathbf{T}(\bfq_0) = \mathbf{0}$.
If the band-touching point is non-singular, then such a requirement does not apply, and one can choose $\mathbf{T}(\bfq)$ in such a way as to be smooth and non-vanishing in the neighborhood of $\bfq_0$.

Expanding $\mathbf{T}$ around $\bfq_0$ for small $\mathbf{k} = \bfq - \bfq_0$, we  get 
\[
\tilde{\bfT}(\mathbf{k}) \equiv \bfT(\mathbf{k} + \bfq_0) \ .
\]
Note that by construction of the FT-constrainer $\bfT(\bfq)$ (Eq.~\eqref{eqn:FT_constrainer_def}), $q_x, q_y$ always appear in   exponential forms as $\exp{(i \mathbf{q}\cdot \mathbf{a}_{i,j})}$, 
we can then expand each component $\tilde{T}_a(\mathbf{k})$ as a polynomial of $ik_x, ik_y$ which satisfies $\tilde{T}_a(0, 0) = 0,\ \text{for}\ a= 1,\dots, N$. 
That is, there is no constant term in the polynomial, so the leading term must have finite powers of $k_x, k_y$.

The emergent Gauss's law is encoded in the algebraic form of the FT-constrainer $\tilde{\mathbf{T}}(k_x, k_y)$. 
Note that $\tilde{\mathbf{T}}(k_x, k_y)$ lives on the top band,
so it describes the spin configurations that cost  energy. 
That is, it encodes the generalized electric charge in terms of the spins $S_1,\ S_2,\ \dots,\ S_N$.

Before describing the most general scenario, 
let us look at a simple example. 
Consider a system with $N=2$ degrees of freedom per unit cell, and 
\[
\begin{pmatrix}
T_1 (\bfk)\\
T_2 (\bfk)\\
\end{pmatrix} = 
\begin{pmatrix}
-i k_x\\
-i k_y\\
\end{pmatrix}  + \mathcal{O}(k_x^2,k_y^2, k_x k_y)\ .
\]
Then the bottom band eigenvector $(\tilde{S}_1, \tilde{S}_2)$ satisfies
\[
\tilde{\bfT}^\ast \cdot (\tilde{S}_1, \tilde{S}_2) 
= i \mathbf{k} \cdot (\tilde{S}_1, \tilde{S}_2) = 0 \ .
\] 
Identifying the Fourier modes of the emergent electric field with the spins: $\tilde{\mathbf{E}} \equiv (\tilde{E}_1 , \tilde{E}_2) = ( \tilde{S}_1,  \tilde{S}_2 ) $,
this condition $ i \mathbf{k} \cdot \tilde{\mathbf{E}}(\mathbf{k)} = 0$ is exactly the Fourier transformed conventional $U(1)$ charge-free constraint in real space
\[
\partial_x E_x + \partial_y E_y = 0\ ,
\]
using $ik_x \rightarrow \partial_x$, $ik_y \rightarrow \partial_y$.
The long-wavelengh effective Hamiltonian  
is then formulated as
\[
\mathcal{H} = (\partial_x E_x + \partial_y E_y)^2\ ,
\]
in real space.  
This imposes exactly the two dimensional electrostatics of the Maxwell U(1) gauge theory, i.e., the electric field configuration has to obey charge-free condition at low energy.

Now let us formulate the   general description.
For each polynomial $\tilde{T}_a(k_x, k_y)$, we only need to keep the leading-order terms in $ik_x, ik_y$, 
since higher-order terms become negligibly small for sufficiently small $k_x, k_y$.
Suppose for a component $\tilde{T}^\ast_a(\bfk)$, the leading order term is of power $m_a\geq 1$, then it takes the general form
\[
\label{eq:T*(q)}
\tilde{T}^\ast_a = \sum_{j=0}^{m_a} c_{aj}^\ast (ik_x)^{j} (ik_y)^{m_a-j}\  .
\]
The emergent Gauss's law in momentum space is then written as 
\[
\label{EQN_Gauss_general_k}
\sum_{a=1}^N\left(\sum_{j=0}^{m_a} c_{aj}^\ast (ik_x)^{j} (ik_y)^{m_a-j} \tilde{S}_a (\mathbf{k}_0 + \mathbf{k})\right) = 0\ .
\] 

If the expansion is around a general   wavevector point $\mathbf{q}_0$,
then the $c_{ij}$'s can be complex. 
The Fourier mode of spin field $S_i$ is also complex. 
It is reconciled with the fact that the spins are real scalars by the constraint that
\[
\mathbf{T}^\ast(\mathbf{q}) =\mathbf{T}(-\mathbf{q})\ .
\]
This guarantees the Fourier mode expansion of the real scalar field is also real after taking into consideration of both $\bfq_0$ and $-\bfq_0$. 
This also means we also have to take into account of what happens at $-\mathbf{q}_0$.
We have 
\[
\mathbf{T}^\ast(\mathbf{k} - \mathbf{q}_0 ) =\mathbf{T}( - \mathbf{k} + \mathbf{q}_0 ) \ ,
\]
so that 
\[
\begin{split}
    T_a (\mathbf{k} - \mathbf{q}_0 )^\ast & = T_a( - \mathbf{k} + \mathbf{q}_0 )  \\
 & =\tilde{T}_a( - \mathbf{k})  \\
 & = \sum_{j=0}^{m_a} c_{aj} (ik_x)^{j} (ik_y)^{m_a-j} 
\end{split} 
\]
imposes the complex conjugated version of Eq.~\eqref{EQN_Gauss_general_k}.
We then have a \textit{complex} Gauss's law 
whose charge-free condition around $\mathbf{q}_0$ is 
\[
\label{EQN_Gauss_general_real}
\sum_{a=1}^N\left(\sum_{j=0}^{m_a} c^*_{aj} (\partial^{(\mathbf{q}_0)}_x)^{j} (\partial^{(\mathbf{q}_0)}_y)^{m_a-j} \tilde{S} \right) = 0\ .
\]
The Gauss's law at  $-\mathbf{q}_0$ is the complex conjugate of it, so we only need to consider one copy of them.

Let us elaborate on the meaning of the Gauss's law appearing at a general wavevector $\mathbf{q}_0$ in real space.
We first define the ``phase-shifted derivative'' $\partial^{(\mathbf{q}_0)}_\alpha$. 
For derivative in a general direction $\mathbf{a}$, we define
\[
\partial^{(\mathbf{q}_0)}_{\mathbf{a}} S(\mathbf{r})=
 S(\mathbf{r}) -e^{ i\mathbf{a} \cdot  \mathbf{q}_0} S(\mathbf{r}- \mathbf{a})
\] 
For example, for $\mathbf{q}_0 = (\pi, \pi)$ on a square lattice of lattice constant $1$,
we have 
\[
\partial^{(\pi, \pi)}_x S(\mathbf{r})=
 S(\mathbf{r}) + S(\mathbf{r}- (1,0)) \ ,
\]
which agrees with how we extract the soft mode from an anti-ferromagnetic background ~\cite{Haldane1983-PRL}.
More generally, $S(\mathbf{r}- \mathbf{a})$ does not have to be on the lattice site if we take a proper coarse-graining procedure, and $\partial^{(\mathbf{q}_0)}_{\mathbf{a}} S(\mathbf{r})$ is complex. 

The phase-shifted derivative $\partial^{(\mathbf{q}_0)}_\alpha$ is the correct spatial derivative from the expansion around general wavevector $\mathbf{q}_0$. 
When it acts on $S(\mathbf{r})$, it yields the correct Gauss's law in momentum space. For example,
\[
\begin{split}
  & \partial^{(\mathbf{q}_0)}_\mathbf{a} \left[ \int d \mathbf{k} \  \tilde{S}(\mathbf{q}_0 +\mathbf{k} ) e^{  i(\mathbf{q}_0 +\mathbf{k})\cdot \mathbf{r}  }\right] \\
= &  \int d \mathbf{k} \  \tilde{S}(\mathbf{q}_0 +\mathbf{k} ) e^{  i(\mathbf{q}_0 +\mathbf{k})\cdot \mathbf{r}  } \\
&\qquad -e^{i\mathbf{a} \cdot  \mathbf{q}_0}   \tilde{S}(\mathbf{q}_0 +\mathbf{k} ) e^{  i(\mathbf{q}_0 +\mathbf{k})\cdot( \mathbf{r} -  \mathbf{a}) } \\
= & \int d \mathbf{k} \ (1 - e^{i\mathbf{a} \cdot  \mathbf{q}_0}e^{ -i(\mathbf{q}_0 +\mathbf{k})\cdot \mathbf{a} } ) \tilde{S}(\mathbf{q}_0 +\mathbf{k} ) e^{  i(\mathbf{q}_0 +\mathbf{k})\cdot \mathbf{r}  }  \\
\simeq & \int d \mathbf{k} \  (i \mathbf{k}\cdot \mathbf{a} ) \tilde{S}(\mathbf{q}_0 +\mathbf{k} ) e^{  i(\mathbf{q}_0 +\mathbf{k})\cdot \mathbf{r}  }   \ .
\end{split}
\]
This again confirms the relation of $i k_\alpha \leftrightarrow \partial_\alpha$ (omitting some factors from lattice constants).
We see that here, although $S(\mathbf{r})$ is real, its phase-shifted derivative can be complex. So indeed the emergent Gauss's law (Eq.~\eqref{EQN_Gauss_general_real}) is  defined over complex fields. 
However, we did not double the number of DOFs or the constraints.
This is because we have 
\[
\partial^{(\mathbf{q}_0)}_{\mathbf{a}} S(\mathbf{r}) =\left[ \partial^{(-\mathbf{q}_0)}_{\mathbf{a}} S(\mathbf{r}) \right]^*\ ,
\]
so the other copy of Gauss's law at $-\mathbf{q}_0$, which contains shifted derivatives of the form $ \partial^{(-\mathbf{q}_0)}_{\mathbf{a}} S(\mathbf{r})$, is automatically obeyed when the original Gauss's law is. Therefore, nothing gets doubled.
Another equivalent point of view is that the DOFs and constraints around $\mathbf{q}_0$ and $-\mathbf{q}_0$ combine together to form the complex-valued field that obeys the complex Gauss's law. 
Because the complex Gauss's law has two constraints (one on the real component and one on the imaginary one), the counting of DOFs and constraints remain correctly unchanged.

Finally, once Eq.~\eqref{EQN_Gauss_general_real} is written down,
we can separate its real and imaginary components to form two copies of a real Gauss's law. 

A special situation -- which actually happens often -- is 
when the FT-contrainer is purely real, i.e. we have the condition $\tilde{\mathbf{T}}(\mathbf{k}) = \tilde{\mathbf{T}}^*(-\mathbf{k})$.
This happens if $\mathbf{q}_0$ is some high symmetry point so that $\mathbf{q}_0$ and $-\mathbf{q}_0$ are identified. 
For example, if $\mathbf{q}_0 = \mathbf{0}$, or their difference $\mathbf{q}_0- (-\mathbf{q}_0)$ is a reciprocal lattice vector ($\mathbf{q}_0$ is often on the BZ boundary in this case). 
Then we have all $c_{ij}$ real, and 
Eq.~\eqref{EQN_Gauss_general_real} (or equivalently, its charge-conjugate)
has the real space interpretation as the charge-free condition for a generalized Gauss's law
\[
\label{EQN_Gauss_general_real_2}
\sum_{i=a}^N\left(\sum_{j=0}^{m_a} c_{aj}  (\partial_x)^{j} (\partial_y)^{m_a-j} \tilde{S}_a \right) \equiv D_a^{(m_a)} \tilde{S}_a= 0\ ,
\]
where we have defined a generalized differential operator $D_a^{(m_a)}$ of order $m_a\geq 1$ on site $a$. 
The effective long-wavelength Hamiltonian is then 
\[
\label{EQN_Gauss_general_real4}
\mathcal{H} = \left[\sum_{a=1}^N\left(\sum_{j=0}^{m_a} c_{aj}  (\partial_x)^{j} (\partial_y)^{m_a-j} S_i \right)\right]^2  = (D_a^{(m_a)} \tilde{S}_a)^2 
\]
in real space.
Note that the number of sublattice sites in a unit cell $N$ is not necessarily the number of components of the electric field.
The equation ~\eqref{EQN_Gauss_general_real_2} needs to be regrouped in terms of different $D_a^{(m_a)}$'s. 
We will see plenty of examples later.

\begin{table*}[th]
\caption{Some common algebraic classes of CLS
\label{TABLE_algebraic_class}
} 
\begin{tabular}{  c @{\hskip 10pt} c @{\hskip 10pt} c @{\hskip 10pt} c} 
	\toprule
	 $\mathbf{T}(\mathbf{k})$ & Gauss's law & specturm of $\mathbf{J}$  & examples\\[3pt] \midrule 
	$ (ik_x, ik_y)$ 
	&\begin{tabular}{@{}c@{}} 
	Maxwell $U(1)$ \\ [0.5em]
	$\partial_\alpha E_\alpha = 0$
	\end{tabular}
	&\raisebox{-.5\height}{\includegraphics[width=0.1\textwidth]{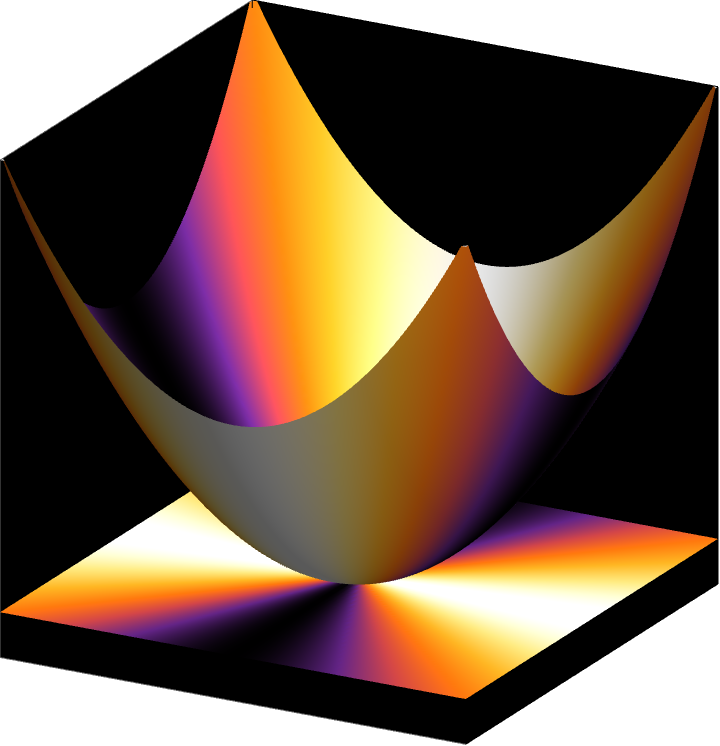}}
	& \begin{tabular}{@{}c@{}} 
    checkerboard~\cite{Moessner1998PhysRevB.58.12049}, Sec.~\ref{subsubsec_checkerboard}\\ 
    kagome (large-$\mathcal{N}$ limit)~\cite{Moessner1998PhysRevB.58.12049,Garanin1999PhysRevB.59.443}, Sec.~\ref{subsubsec_kagome}\\
    pyrochlore~\cite{MC_pyro_PRL,Moessner1998PhysRevB.58.12049}, Sec.~\ref{subsubsec_pyro} \\ 
	honeycomb-snowflake model  ~\cite{Benton21PRL}, Sec.~\ref{Sec_BM_classification_application} 
	\end{tabular}
	 \\[30pt]
	$ (-k_x^2 + k_y^2, -2k_xk_y)$ 
	&\begin{tabular}{@{}c@{}} 
	scalar rank-2 $U(1)$  : \\ [0.5em]
	$\partial_\alpha \partial_\beta E_{\alpha \beta}   = 0$,\\[0.5em]
	$\mathbf{E} = \begin{pmatrix}E_{xx} & E_{xy} \\ E_{xy} & -E_{xx}\end{pmatrix}$
	\end{tabular}
	&\raisebox{-.5\height}{\includegraphics[width=0.1\textwidth]{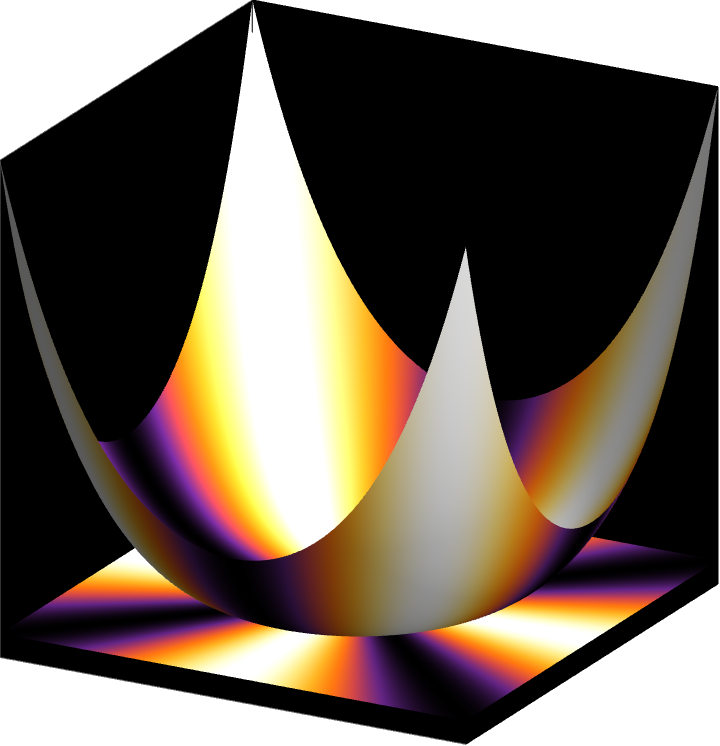}}
	&  
    honeycomb-snowflake model  ~\cite{Benton21PRL}, Sec.~\ref{Sec_BM_classification_application} 
	\\[30pt]
    \begin{tabular}{@{}c@{}} 
	$ ( k_x , k_y,0 )$ \\ [0.5em]
	and $ ( 0, k_x , k_y  )$ 
	\end{tabular}
	&\begin{tabular}{@{}c@{}} 
	vector rank-2 $U(1)$  : \\ [0.5em]
	$\partial_\alpha  E_{\alpha \beta}   = 0$,\\[0.5em]
	$\mathbf{E} = \begin{pmatrix}E_{xx} & E_{xy} \\ E_{xy} &  E_{yy}\end{pmatrix}$
	\end{tabular}
	&\raisebox{-.5\height}{\includegraphics[width=0.1\textwidth]{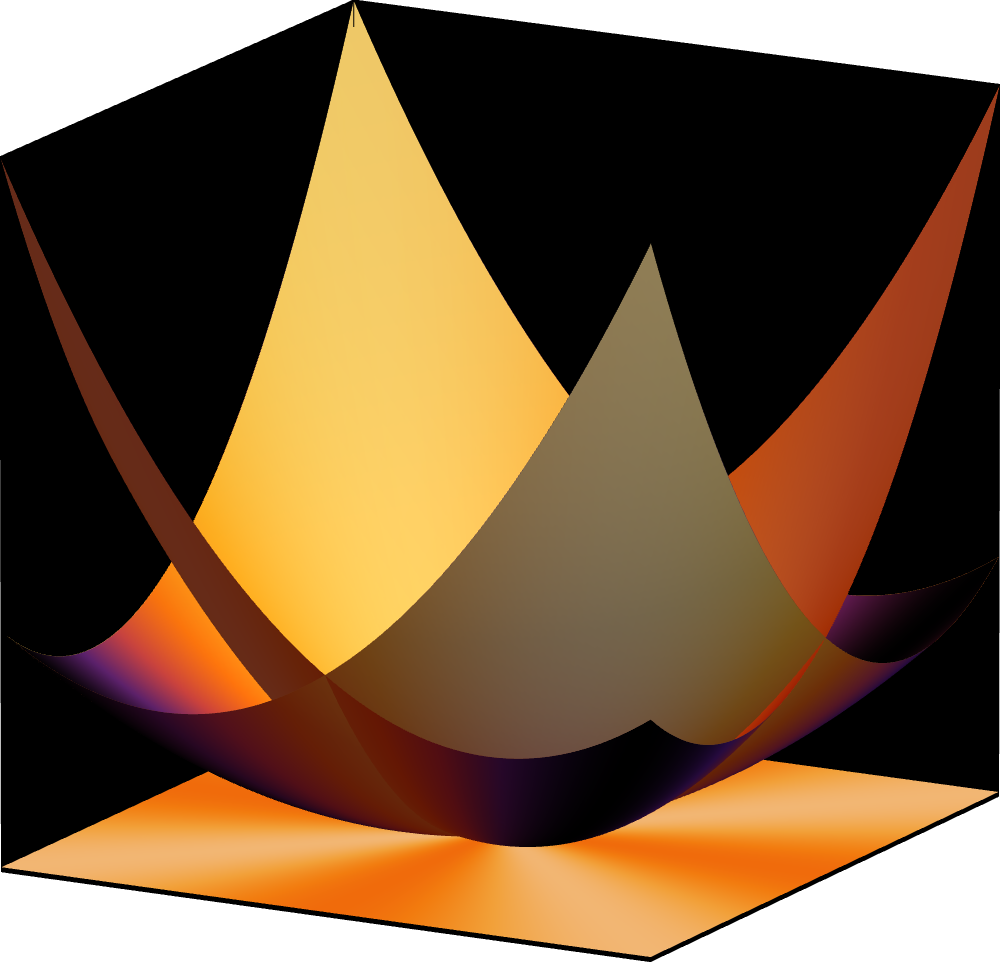}}
	& \begin{tabular}{@{}c@{}} 
	Breathing pyrochlore model ~\cite{Yan20PRL}  \\ [0.5em]
	(when generalized to 3D)  
	\end{tabular} 	\\[30pt]
    Equation~\eqref{EQN_T_Anisotropic_HS}
	&\begin{tabular}{@{}c@{}} 
	anisotropic $U(1)$: \\ [0.5em]
	$  3\partial_y E_1 + \frac{3}{4} (\partial_x^2  - 3 \partial_y^2) E_2=0$
	\end{tabular}
	&\raisebox{-.5\height}{\includegraphics[width=0.1\textwidth]{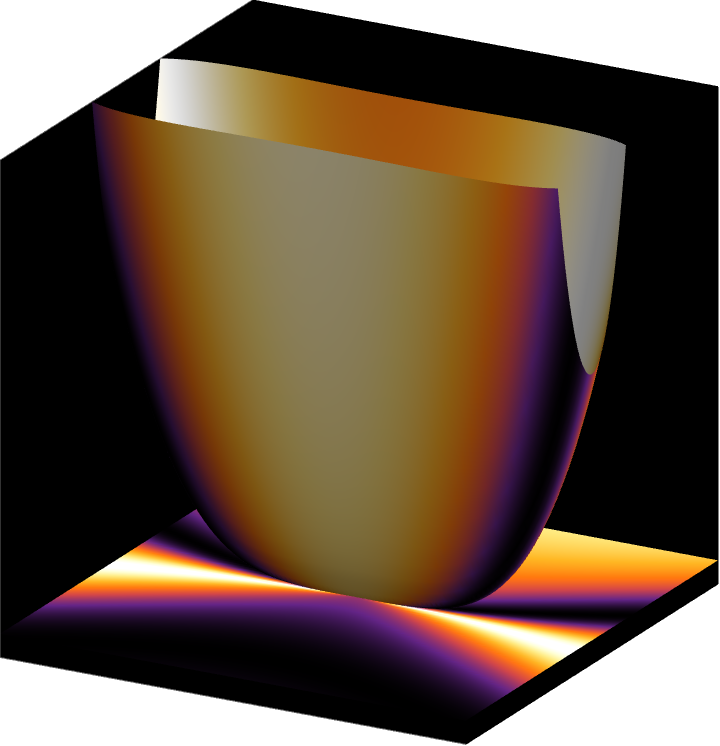}}
	&  \begin{tabular}{@{}c@{}} 
	Anisotropic honeycomb-snowflake \\
	 Model, Sec.~\ref{Sec_anisotropic_snowflake}
	\end{tabular}
	\\ 
	\bottomrule
	\end{tabular}
\end{table*}

\subsection{Extracting the Gauss's laws: multiple constrainer models}
\label{sec:multi_constrainer}

We now discuss the physics when there are multiple constrainers per unit cell. 
In this case, the Hamiltonian is in its most general form (repeating Eq.~\eqref{EQN_abstract_Ham_general})
\[\label{EQN_abstract_Ham_general_2}
\begin{split}
 \mathcal{H} & = \sum_{\mathbf{R} \in \text{u.c.}}\sum_{I=1}^M   [\mathcal{C}_I(\mathbf{R} )]^2 \\
&= \sum_{\mathbf{R} \in \text{u.c.}}\sum_{I=1}^M\left[
\sum_{\mathbf{r}} \mathbf{S}(\mathbf{r})\cdot \mathbf{C}_I(\mathbf{R},\mathbf{r})
\right]^2 .   
\end{split}\nonumber
\]
There are $M$ FT-constrainers $\mathbf{T}_{1}, \mathbf{T}_{2},\dots, \mathbf{T}_\mathsf{M}$.
At a general momentum $\mathbf{q}$, these FT-constrainers span the space for eigenvectors for the higher dispersive bands. 
However, different FT-constrainers are not necessarily orthogonal to each other, and each FT-constrainer is not necessarily the  eigenvector of a certain band.

In this case, there are two possible ways to close the gap. 
The first way is the same as the single constrainer case, i.e., 
one (or several) of the FT-constrainers vanishes at $\mathbf{q}_0$.
The second is when a subset of the FT-constrainers become linearly dependent, so that the dimension of the linear space they span (i.e. the number of the non-flat higher bands) decreases. 

To extract the Gauss's law, the core idea is the same as before: we would like to know the eigenvector configuration on the higher dispersive band in the vicinity of the $\mathbf{q}$-points where it becomes gapless. 
However, more care is needed since the FT-constrainers themselves are not necessarily the eigenvectors we look for.
To find the eigenvector, one has to make sure that the orthogonality condition is satisfied.
This is just an exercise in linear algebra.

Let us use the case of two FT constrainers $\mathbf{T}_\mathsf{1,2} (\bfq)$ as an example. 
In the first case, when one of the constrainers vanishes, let us assume $\mathbf{T}_{1} (\mathbf{0}) = \mathbf{0}$ without loss of generality. 
We then have $\mathbf{T}_{1} (\mathbf{k})$ as a vector polynomial Taylor expansion in powers of $k_{\alpha}$, and we keep only the leading order term in each of its components. 
The Gauss's law should be extracted using
\[
\label{EQN_T1_project_out}
\tilde{\mathbf{T}}( \mathbf{k}) = \mathbf{T}_{1}  ( \mathbf{k}) -\frac{\mathbf{T}_{2}(\mathbf{0} )}{|\mathbf{T}_{2}(\mathbf{0} )|^2} \left[\mathbf{T}^\ast_{1}  ( \mathbf{k}) \cdot \mathbf{T}_{2}(\mathbf{0} ) \right] \  .
\]

Here, the second term on the right hand side is to project out the part of $\mathbf{T}_{1}$ that is along the direction of $\mathbf{T}_{2}$,
so that the rest, $\tilde{\mathbf{T}}$, is orthogonal to $\mathbf{T}_{2}$. 
Since $\tilde{\mathbf{T}}$ is still in the space spanned by the FT-constrainers, it is then guaranteed to be the eigenvector of the band that becomes gapless at $\mathbf{0}$.
We can use $\mathbf{T}_{2}(\mathbf{0})$ instead of $\mathbf{T}_{2}(\mathbf{k})$ because only the leading order term needs to be kept. 

In the second case mentioned above, the FT-constrainers $\mathbf{T}_{1}$ and $\mathbf{T}_{2}$ become linearly dependent at $\mathbf{q}=\mathbf{0}$. 
Let us separate   $\mathbf{T}_{1}$ via
\[
\mathbf{T}_{1}(\mathbf{k)}= \mathbf{T}_{1}(\mathbf{0})
+ \delta\mathbf{T}_{1}(\mathbf{k})\ .
\]
So we know 
\[
\delta\mathbf{T}_{1}(\mathbf{0}) = 0\ ,
\]
and its Taylor expansion is some polynomial of $k_{\alpha}$ for each of its components.
The Gauss's law can then be extracted via 
\[
\label{EQN_delta_T1_project_out}
\tilde{\mathbf{T}}( \mathbf{k}) = \delta\mathbf{T}_{1}  ( \mathbf{k}) -\frac{\mathbf{T}_{2}(\mathbf{0} )}{|\mathbf{T}_{2}(\mathbf{0} )|^2} \left[\delta\mathbf{T}^\ast_{1}  ( \mathbf{k}) \cdot \mathbf{T}_{2}(\mathbf{0} ) \right] \  .
\]

The above considerations can be generalized to the case of more constrainers. 
In each case, suppose we need to do Taylor expansion on $\mathbf{T}_{1}$ or $\delta\mathbf{T}_{1}$ , 
then we should first find an orthognal basis of the linear space spanned by $\mathbf{T}_{2},\dots,\mathbf{T}_{M}$. 
Let us denote the unit vectors of this basis by $\mathbf{T}'_{2},\dots,\mathbf{T}'_{M}$,
then Eq.~\eqref{EQN_T1_project_out} should be replaced by
\[ 
\tilde{\mathbf{T}}( \mathbf{k}) = \mathbf{T}_{1}  ( \mathbf{k}) -\sum_{I=2}^M \mathbf{T}'_{I}(\mathbf{0} )  \left[\mathbf{T}^\ast_{1}  ( \mathbf{k}) \cdot \mathbf{T}'_{I}(\mathbf{0} ) \right]  \ ,
\]
 and Eq.~\eqref{EQN_delta_T1_project_out} should be replaced by
\[ 
\tilde{\mathbf{T}}( \mathbf{k}) = \delta\mathbf{T}_{1}  ( \mathbf{k}) -\sum_{I=2}^M \mathbf{T}'_{I}(\mathbf{0} )  \left[\delta\mathbf{T}^\ast_{1}  ( \mathbf{k}) \cdot \mathbf{T}'_{I}(\mathbf{0} ) \right] \  .
\]

\subsection{Transitions between different algebraic CSLs  }

We can classify different algebraic CSLs by examining their gap-closing points. Specifically, two algebraic CSLs belong to the same class if one can smoothly transform the constrainer Hamiltonian and the Gauss's law of one CSL into that of the other, without encountering singular processes that involve merging, splitting, or lifting any of these points. On the other hand, two algebraic CSLs are considered distinct if they have a different number of gap-closing points or if their associated Gauss's laws involve a different number of effective electric field degrees of freedom or a different order of $\partial_x$ and $\partial_y$. It is impossible to make these gap-closing points identical without going through certain singular transitions.

By identifying the emergent Gauss's law with the structure of the gap-closing point,
we can also study the transition between different algebraic CSLs as merging/splitting of the gapless points on the bottom flat band.

The simplest structure of the band-touching point is the one associated with the (complex) Maxwell Gauss's law, shown in the first row of Table.~\ref{TABLE_algebraic_class}.
Let us call it the \textit{basic band-touching point}.
Other band-touching points corresponding to more exotic Gauss's laws can often be obtained by merging some of the basic band touching point. 

Often, the scenario is the following (see e.g. \cite{Benton21PRL}).
We start with an algebraic CSL with only basic band-touching points  in  momentum space.
By tuning some parameters of the Hamiltonian, 
the positions
 of the basic band-touching points can be changed,
or new basic band-touching points can emerge when  higher bands  come down to zero energy.
When the parameters are tuned to certain critical values,
several basic band-touching points can merge  to form a new band-touching point.
The new band-touching point is then described by a different generalized Gauss's law.

For readers familiar with topological band theory,
this scenario is very similar to the knowledge that the Weyl point is the ``basic'' gap-closing point containing   divergent Berry curvature   at the singularity, 
and merging a few Weyl points together generates other types of gap-closings. 
In fact, the basic band-touching point in the spectrum of the CSL is exactly equivalent to two merged Weyl cones.

From the perspective of  the effective theory, this tells us that by taking a few copies of Maxwell electrostatics and tuning them to a critical point,
one can obtain more general forms of U(1) electrostatics. 

A summary of the important points in 
the classification of algebraic CSL is shown in Table.~\ref{TABLE_flat_band_theory}.
In Sec.~\ref{Sec_BM_classification_application},
we will see concretely how  transitions between
algebraic CSLs happen in the case of the honeycomb-snowflake model.

\section{Algebraic CSL models}
\label{sec:algebraic.csl.models}
In this section, we will analyze many old and new examples of algebraic CSLs using our classification scheme as well as tools from flat band theory introduced in Sec.~\ref{SubSec.flat.band.theory}. A survey of various CSL models and prior studies, all fitting within the present classification, can be found  Table.~\ref{table:summary.of.models}.

\subsection{Checkerboard, kagome, and pyrochlore AFM}
\label{sec:square_ice}
To understand how the classification scheme works on concrete examples, 
let us first apply it to the checkerboard~\cite{Moessner1998PhysRevB.58.12049}, kagome~\cite{Moessner1998PhysRevB.58.12049,Garanin1999PhysRevB.59.443}, and pyrochlore~\cite{MC_pyro_PRL,Moessner1998PhysRevB.58.12049} antiferromagnets in the large-$\mathcal{N}$  limit.
These models, due to their geometric frustration, 
were the first ones discovered to host spin liquids,
and are perhaps the most familiar to readers.

\subsubsection{Emergent Gauss's law from the checkerboard AFM}
\label{subsubsec_checkerboard}

\begin{figure*}[ht]
 \centering
 \includegraphics[width=0.95\textwidth]{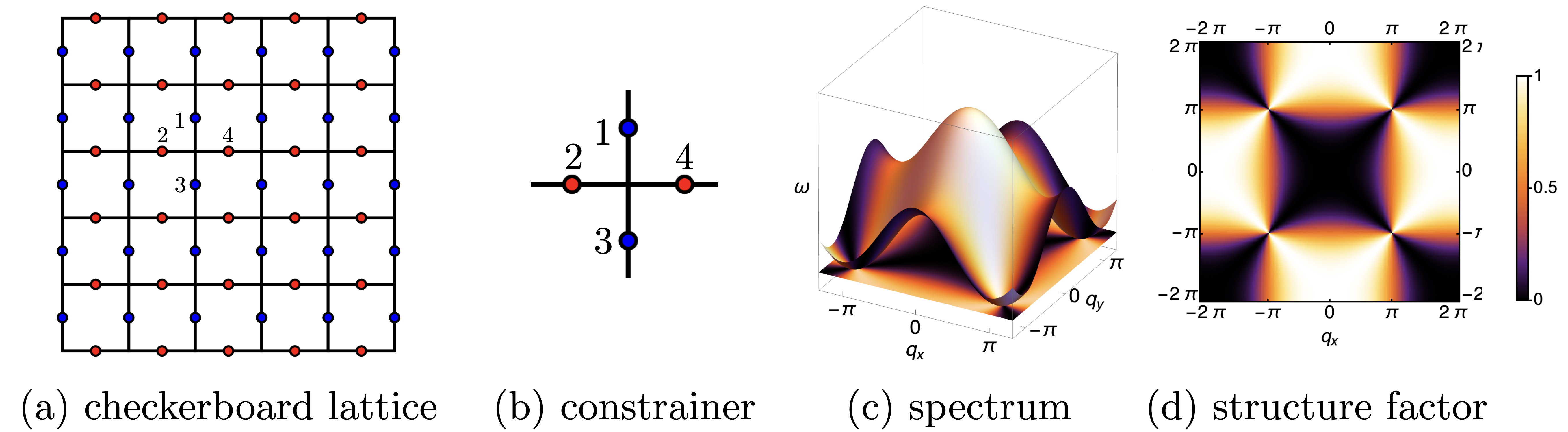}
 \caption{ (a) Checkerboard lattice.
 (b) Constrainer of the checkerboard model. Classical spins are arranged on the edges of a square lattice, with ground states defined by the constraint that the sum of spins on each vertex   must vanish (Eqs.~(\ref{EQN_checkerboard_ham})-(\ref{EQN_checkerboard_constrainer})).
 (c) Spectrum $\omega({\bf q})$ that arises from diagonalizing
 the Hamiltonian Eq.~\eqref{EQN_J_Square_ice}. There is one  flat band  at the bottom of the spectrum and a dispersive upper band with gap-closing points between them.
 (d) Spin structure factor showing pinch points at the position of gap-closing points.
 }
 \label{Fig_Square_Ice_all}
\end{figure*}

Let us demonstrate our classification scheme with   the checkerboard lattice model.
The model is illustrated in Fig.~\ref{Fig_Square_Ice_all}(a).
The spins sit on the edges of the square lattice, and the constrainer Hamiltonian is
\[\label{EQN_checkerboard_ham}
\mathcal{H}_\mathsf{CB} 
= \sum_{\text{all star}}  (S_1 + S_2 +S_3 + S_4 )^2 \equiv
\sum_{\mathbf{R} \in \text{all star}}
[\calC_\mathsf{CB}(\bfR)]^2\ .
\]

\begin{figure}[ht]
 \centering
 \includegraphics[width=\columnwidth]{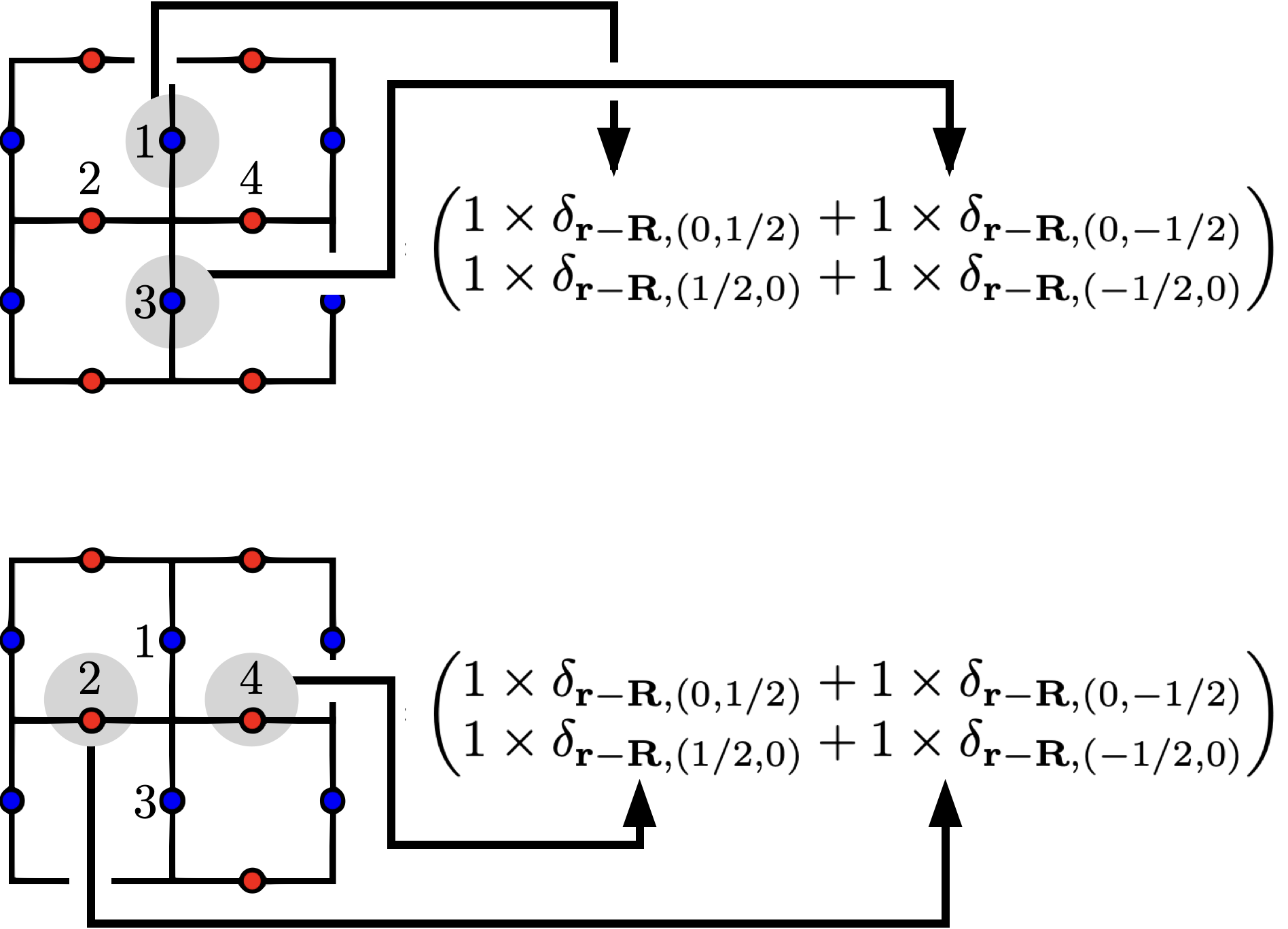}
 \caption{How to write down the vector form constrainer $\bfC_\mathsf{CB}(\bfR,\mathbf{r})$ (Eq.~\eqref{EQN_checkerboard_constrainer}) from its real space image (Fig.~\ref{Fig_Square_Ice_all}(b)).
 \label{Fig_Square_Ice_proj}  
 } 
\end{figure}

Note that there are $N=2$ inequivalent sites in the periodic unit cell. 
Without loss of generality, we can take the spin on sites $1, 4$ in Fig.~\ref{Fig_Square_Ice_all}(b) to be the first and second sublattice DOFs in one unit cell, respectively. 
In this convention, the spins on site $2$ and $3$ are related by lattice translation to the other two sites: 
the spin on site $2$ is a second sublattice DOF in the unit cell to the left,
and the spin on site $3$ is a first sublattice DOF in the unit cell below.
Therefore, the constrainer is  (see Fig.~\ref{Fig_Square_Ice_proj} on how each spin maps to each term in the constrainer)
\[\label{EQN_checkerboard_constrainer}
\begin{split}
 &\bfC_\mathsf{CB} (\bfR,\mathbf{r})\\
 &= \begin{pmatrix}
1\times \delta_{\mathbf{r} - \bfR,(0,1/2)} + 1\times \delta_{\mathbf{r} - \bfR,(0,-1/2)} \\
1\times \delta_{\mathbf{r} - \bfR,(1/2,0)} + 1\times \delta_{\mathbf{r} - \bfR,(-1/2,0)} 
\end{pmatrix}\ .   
\end{split} 
\]

The FT-constrainer is then
\[
\mathbf{T}_\mathsf{CB} (\bfq)
=
\begin{pmatrix}
e^{-iq_y/2} +e^{iq_y/2}   \\
e^{ - iq_x/2} +e^{iq_x/2}   \\
\end{pmatrix}\ .
\]
The Hamiltonian in momentum space is 
\[ 
\label{EQN_J_Square_ice}
({J}_\mathsf{CB})^{\alpha\beta} = ({T}_\mathsf{CB})^\alpha[({T}_\mathsf{CB} )^\beta{}]^*\ .
\] 
Its spectrum is illustrated in Fig.~\ref{Fig_Square_Ice_all}(c).
We see that it has gapless points at $\bfq = (\pm \pi , \pm \pi )$.
We can expand the FT-constrainer around $\bfq = ( \pi ,  \pi )$ to get (upon adding an overall factor $-i$)
\[
\tilde{\mathbf{T}}_\mathsf{CB} (\bfk)
=
\begin{pmatrix}
i k_y  \\
 i k_x \\
\end{pmatrix} \ .
\]
This gives us ground state constraints
\[
\partial_y S_1 + \partial_x S_2 = 0\  ,
\]
which is exactly the expected Maxwell U(1) Gauss's law upon identifying the spin sites with the components of the electric field: $E_x \equiv S_2, \; E_y \equiv S_1$.
The charge-free Gauss's law is shown as pinch poings around these gapless points in the equal-time spin structure factor (Fig.~\ref{Fig_Square_Ice_all}(d)).

Finally we note that  when writing down $\bfC_\mathsf{CB}(\bfR,\mathbf{r})$, we made the ``gauge choice'' equivalent to treating two sublattice sites to be at their physical locations in the unit cell. One can also use other   gauge choice (for example, assuming they are at the same position in the unit cell) as long as the complex phase factor is taken care of.

We also note that, on the checkerboard lattice, if the constrainer is symmetric regarding inversion about the center of the constrainer (the vertex of the lattice),
the spectrum is guaranteed to be gapless at $(\pm \pi, \pm \pi)$. 
Such constrainers include the one we used above, and also more generalized ones containing spins on sites farther from the vertex.

The argument, which works for the checkerboard lattice (but not all other lattices), is the following. 
For the first sublattice sites, if the constrainer involves a spin at site $\mathbf{a}_{1,1}= (r'_x, r'_y)$ relative to its center set at $\bfR = \mathbf{0}$ (the vertex) with coefficient $c_{1,1}$, then it also involves a spin at site $-\mathbf{a}_{1,1}$, with the same coefficient for the second spin.
So the first element of the constrainer must have a pair of terms in the form of 
\[
[C_\mathsf{CB} (\mathbf{0},\mathbf{r})]_1 = 
c_{1,1}( \delta_{\mathbf{r},\mathbf{a}_{1,1}} +   \delta_{\mathbf{r},-\mathbf{a}_{1,1}} ) + \dots .
\]
Note that, due to the symmetry, any term in the constrainer  appears in the form above.
Hence, we know the first component of the FT-constrainer must look like
\[
[T_\mathsf{CB}  (\bfq)]_1 =
2c_{1,1} \cos({\bfq \cdot \mathbf{a}_{1,1}} )+ 2c_{1,2} \cos({\bfq \cdot \mathbf{a}_{1,2}} ) + \dots .
\]
Since the vector $\mathbf{a}_{1,1}$, pointing from the lattice vertex to the sublattice site  on the checkerboard lattice, must be 
\[
\mathbf{a}_{1,1} = n_x \hat{\mathbf{x}} +  (n_y+ \frac{1}{2} )\hat{\mathbf{y}} \ ,
\]
where $n_{x,y}$ are integers, 
the term $\cos({\bfq \cdot \mathbf{a}_{1,1}} )$ is guaranteed to vanish for 
\[
\bfq =( \pm\pi, \pm \pi)\ .
\]
This applies to any other terms in $\mathbf{T}_\mathsf{CB}  (\bfq)$, so the FT-constrainer must vanish at $\bfq =( \pm\pi, \pm \pi)$, at which point the spectral gap between the dispersive top band and the bottom flat band closes.

We hence conclude that given the checkerboard lattice crystalline symmetry, and the properly-chosen action of the constrainer under the  crystalline symmetry, 
existence of gapless points in the spectrum is guaranteed, i.e.,
the algebraic CSL is protected by symmetry.

Such analysis can be generalized to  all crystalline symmetries and their associated constrainer behaviors. 
Given the proper combination of them,
the band touching points are protected and the CSL has to be an algebraic CSL. 
A systematic examination of all crystalline symmetries and constrainer behaviors is achievable, but lies beyond the scope of this work.

\subsubsection{Emergent Gauss's law from the kagome AFM}
\label{subsubsec_kagome}

\begin{figure*}[ht]
 \centering
 \includegraphics[width=0.9\textwidth]{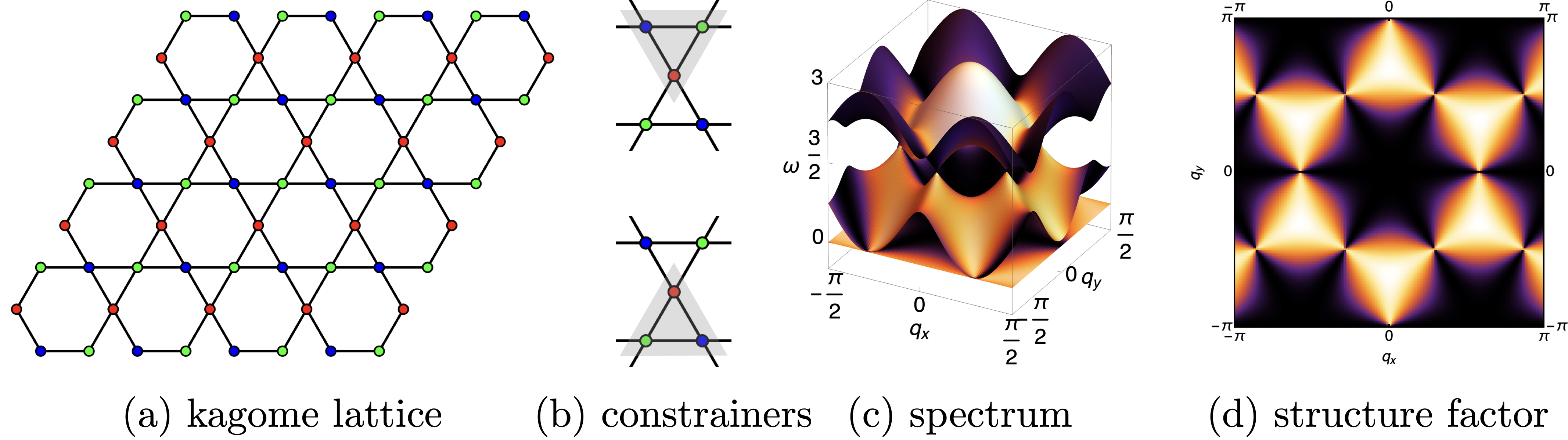}
 \caption{  (a) Kagome lattice.
 (b) Two constrainers of the kagome model shown  in shaded regions. Classical spins are arranged on the edges of a square lattice, with ground states defined by the constraint that the sum of spins on each vertex   must vanish (Eqs.~(\ref{eqn:kagomeAFM_2})-(\ref{EQN_Kagome_AFM_constrainer_2})).
 (c) Spectrum $\omega({\bf q})$ that arises from diagonalizing
 the Hamiltonian Eq.~\eqref{eqn:kagomeAFM_2}. There are one  flat band  at the bottom of the spectrum and two dispersive upper bands with gap-closing points between them.
 (d) Spin structure factor showing pinch points at the position of gap-closing points.
 }
 \label{Fig_Kagome_AFM_lattice_proj}
\end{figure*}

\begin{figure}[ht]
 \centering
 \includegraphics[width=0.7\columnwidth]{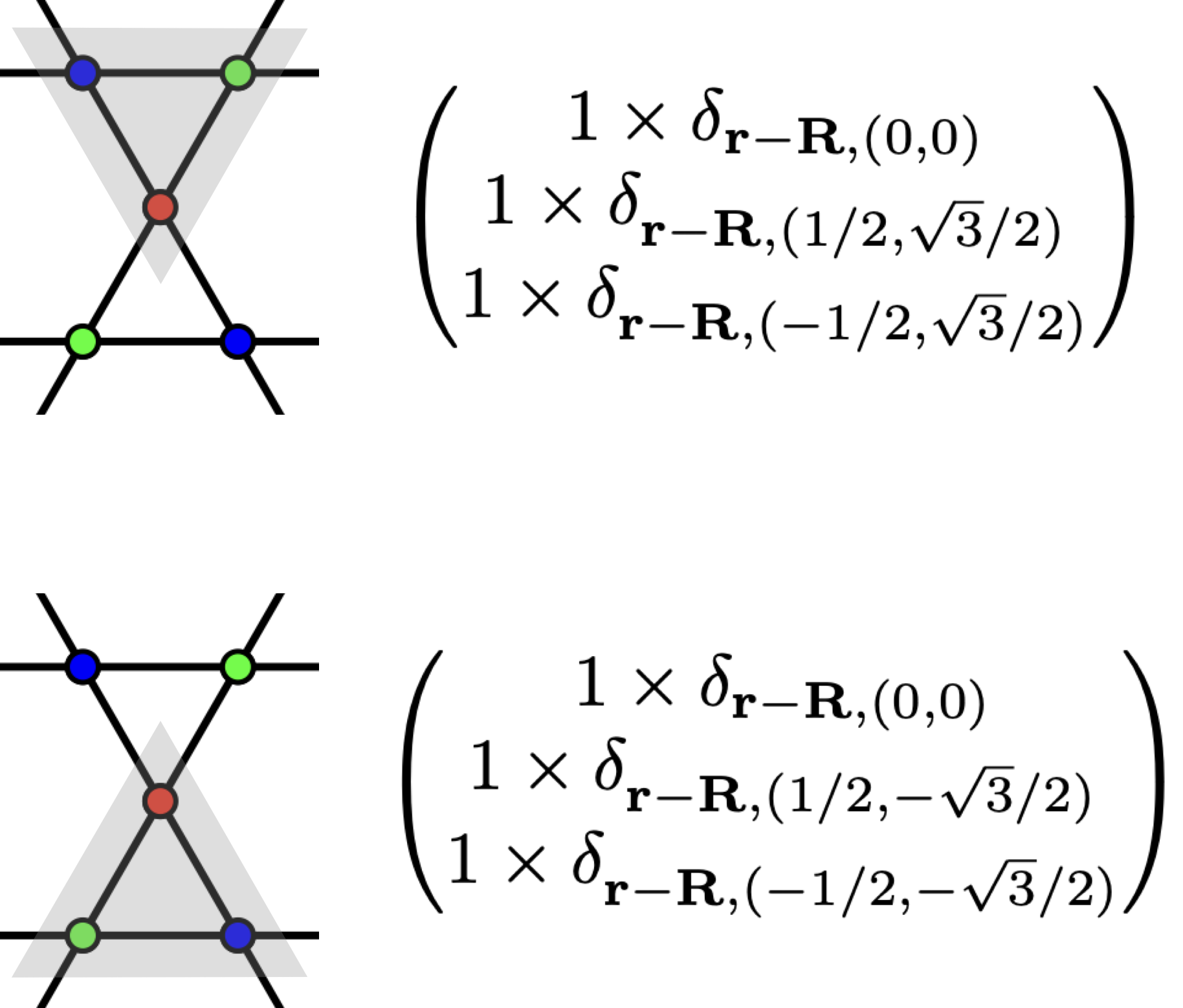}
 \caption{ The two constrainers colored in shaded regions, and their corresponding vectors (Eqs.~(\ref{EQN_Kagome_AFM_constrainer_1},\ref{EQN_Kagome_AFM_constrainer_2})) in the model of Eq.~\eqref{eqn:kagomeAFM_2}.
 }
 \label{Fig_Kagome_AFM_lattice_proj_eq}
\end{figure}

Next we discuss the kagome lattice model with AFM interactions (Fig.~\ref{Fig_Kagome_AFM_lattice_proj}(a)), which we have already introduced in Sec.~\ref{SubSec.flat.band.theory} in the context of the flat band theory. 
The Hamiltonian contains two constrainers, as shown in Fig.~\ref{Fig_Kagome_AFM_lattice_proj}(b), which we repeat here:
\[
\label{eqn:kagomeAFM_2}
\begin{split}
   \mathcal{H}_\mathsf{KGM} & =  \sum_{\langle i,j\rangle} S_i S_j + 2 \sum_i S_i^2 \\ 
   & = \sum_\bigtriangleup\left( \sum_{i \in \bigtriangleup} S_i\right)^2 + \sum_\bigtriangledown\left( \sum_{i \in \bigtriangledown} S_i\right)^2 \\
& \equiv  \sum_{\mathbf{R}  \text{ for } \bigtriangleup }
[\calC_\mathsf{KGM1}(\bfR)]^2 +\sum_{\mathbf{R} \text{ for } \bigtriangledown }
[\calC_\mathsf{KGM2}(\bfR)]^2 \ . 
\end{split}
\]

The two constrainers, written in  vector form, are
\begin{align}
\label{EQN_Kagome_AFM_constrainer_1}
\bfC_\mathsf{KGM1} (\bfR,\mathbf{r})=  & \begin{pmatrix}
1\times \delta_{\mathbf{r} - \bfR,(0,0)} \\
1\times \delta_{\mathbf{r} - \bfR,(1/2,\sqrt{3}/2)} \\
1\times \delta_{\mathbf{r} - \bfR,(-1/2,\sqrt{3}/2)} \\
\end{pmatrix}\ ,\\
\label{EQN_Kagome_AFM_constrainer_2}
\bfC_\mathsf{KGM2} (\bfR,\mathbf{r}) =  & \begin{pmatrix}
1\times \delta_{\mathbf{r} - \bfR,(0,0)} \\
1\times \delta_{\mathbf{r} - \bfR,(1/2,-\sqrt{3}/2)} \\
1\times \delta_{\mathbf{r} - \bfR,(-1/2,-\sqrt{3}/2)} \\
\end{pmatrix}
\ .
\end{align}
The FT-constrainers are 
\begin{align}
\mathbf{T}_\mathsf{KGM1} (\bfq)&
=
\begin{pmatrix}
 1 \\   e^{   i (-q_x / 2 -\sqrt{3}q_y/2)} \\ e^{  i ( q_x / 2 - \sqrt{3}q_y/2)} 
\end{pmatrix}\ ,\\
\mathbf{T}_\mathsf{KGM2} (\bfq)&
=
\begin{pmatrix}
 1 \\   
 e^{  i (-q_x / 2 + \sqrt{3}q_y/2)} \\ 
 e^{   i (q_x / 2 + \sqrt{3}q_y/2)} 
\end{pmatrix}\ .
\end{align}
Since there are two constrainers,  there is one flat bottom band and two upper dispersive bands at a general momentum $\mathbf{q}$.
However,  at $\mathbf{q} = \mathbf{0}$, the two constrainers become linearly dependent,
\[
\mathbf{T}_\mathsf{KGM1}  ( \mathbf{0})  =  \mathbf{T}_\mathsf{KGM2}  ( \mathbf{0}) \propto \frac{1}{\sqrt{3}}(1,1,1)^\text{T} \equiv \mathbf{T}^0\ ,
\]
which means a gap closing happens there, as shown in the spectrum in Fig.~\ref{Fig_Kagome_ice}. 
Hence we have to expand $\mathbf{T}_\mathsf{KGM1}$ around $\mathbf{q} = \mathbf{0}$, and take its component perpendicular to $\mathbf{T}^0$, which is
\[
\begin{split}
 &  \tilde{\mathbf{T}}( \mathbf{k}) = \mathbf{T}_\mathsf{KGM1}  ( \mathbf{k}) - (\mathbf{T}_\mathsf{KGM1}  ( \mathbf{k}) \cdot \mathbf{T}^0 )\mathbf{T}^0 \\
  =&  \frac{i}{6}\begin{pmatrix}
  2\sqrt{3} k_y \\
 -3 k_x - \sqrt{3}k_y   \\
 3 k_x - \sqrt{3}k_y  
\end{pmatrix}  
\end{split}\ .
\]
Expanding $\mathbf{T}_\mathsf{KGM2}$  and extracting its perpendicular component gets the same result. Fourier-transforming the constrainer to the real space as in Section~\ref{sec:Gauss-from-spins}, we obtain Gauss's law in the form of Maxwell's U(1) theory:
\[
\begin{split}
&3\partial_x (-  S_2 +   S_3)
+ \sqrt{3} \partial_y (2  S_1 - S_2 -S_3 ) \\
&\equiv \partial_x E_x + \partial_y E_y  = 0
\end{split}
\]

Note that the number of sublattice sites does not necessarily need to equal to the number of components of the electric field. 
Here, the DOF $(S_1 + S_2 + S_3)/\sqrt{3}$ is not involved in the low energy physics. 
It is instead relevant to the third band on top whose eigenvector is $ \mathbf{T}^0$.

The same physics can also be obtained by analyzing the bottom band eigenvector and fluctuator, which has been discussed in Sec.~\ref{SubSec.flat.band.theory}. 
However, in general it is easier to use the higher dispersive bands because their eigenvectors can be obtained analytically as shown here.

\subsubsection{Emergent Gauss's law from the pyrochlore AFM  model}
\label{subsubsec_pyro}

\begin{figure*}[ht]
 \centering
\includegraphics[width=0.95\textwidth]{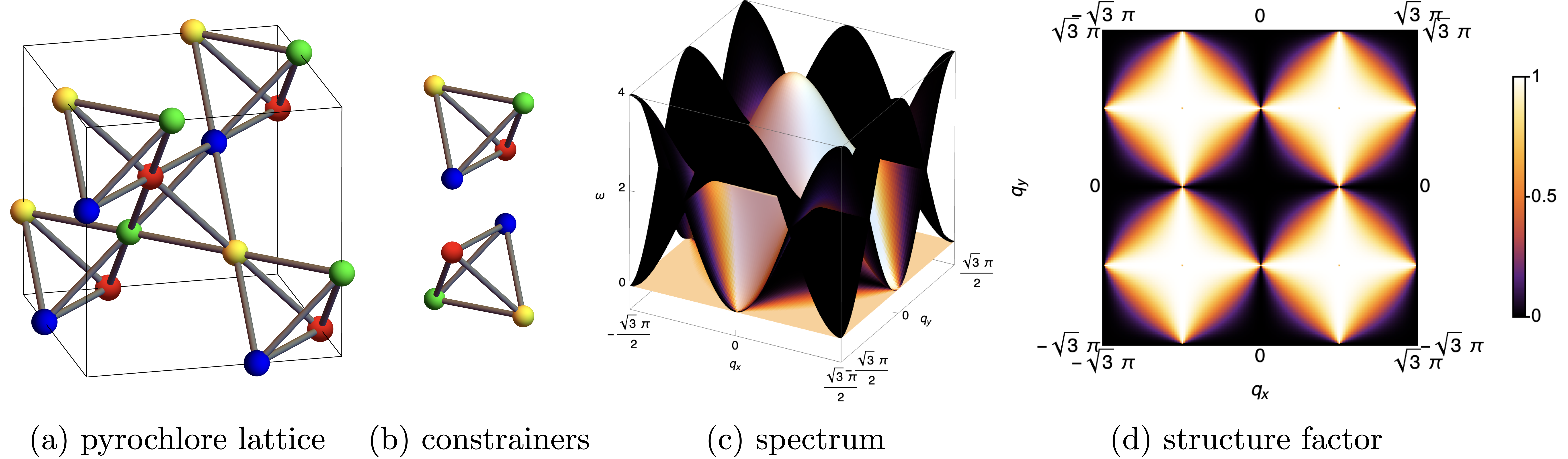}
 \caption{(a) Pyrochlore lattice.
 (b) Two constrainers of the Pyrochlore model corresponding to Eqs.~(\ref{EQN_pyro_AFM_constrainer_1},\ref{EQN_pyro_AFM_constrainer_2}).
 (c) Spectrum   from diagonalizing the Hamiltonian (Eq.~\eqref{eqn:pyrochlore_AFM}) in momentum space on plane $q_z = 0$. There are degenerate two flat bands at the bottom of the spectrum and two  dispersive upper band.
 The gap-closing point's eigenvector configuration (Eqs.~(\ref{EQN_pyro_AFM_FTconstrainer_1},\ref{EQN_pyro_AFM_FTconstrainer_2})) encodes the Gauss's law (Eq.~\eqref{EQN_pyro_gauss_law}). 
 (d) Equal-time spin structure factor on plane $q_z = 0$. 
 \label{Fig_pyro_AFM_lattice}}
\end{figure*}

The third model we review is   the Pyrochlore AFM  Model. 
The lattice is  a network of tetrahedra, shown in Fig.~\ref{Fig_pyro_AFM_lattice}(a).
Its Hamiltonian also contains two constrainers (Fig.~\ref{Fig_pyro_AFM_lattice}(b)), written as
\[
\label{eqn:pyrochlore_AFM}
\begin{split}
   \mathcal{H}_\mathsf{PC} &=  \sum_{\langle i,j\rangle} S_i S_j + 2 \sum_i S_i^2 \\
   &= \sum_\text{up-tet.}\left( \sum_{i \in \text{tet.}} S_i\right)^2 + \sum_\text{down-tet.}\left( \sum_{i \in \text{tet.}} S_i\right)^2 \\
& \equiv  \sum_{\mathbf{R}  \text{ for up-tet.} }
[\calC_\mathsf{PC1}(\bfR)]^2 +\sum_{\mathbf{R} \text{ for down-tet.}}
[\calC_\mathsf{PC2}(\bfR)]^2 \ . 
\end{split}
\]
The treatment is very similar to that of the kagome AFM model. For completeness, let us write down all the steps again.

The two constrainers, written in the vector form, are
\begin{align}
\label{EQN_pyro_AFM_constrainer_1}
\bfC_\mathsf{PC1} (\bfR,\mathbf{r})=  & \begin{pmatrix}
1\times \delta_{\mathbf{r} - \bfR,\mathbf{0}} \\
1\times \delta_{\mathbf{r} - \bfR,\mathbf{a}_1} \\
1\times \delta_{\mathbf{r} - \bfR,\mathbf{a}_2} \\
1\times \delta_{\mathbf{r} - \bfR,\mathbf{a}_3} \\
\end{pmatrix}\ ,\\
\label{EQN_pyro_AFM_constrainer_2}
\bfC_\mathsf{PC2} (\bfR,\mathbf{r}) =  & \begin{pmatrix}
1\times \delta_{\mathbf{r} - \bfR,\mathbf{0}} \\
1\times \delta_{\mathbf{r} - \bfR,-\mathbf{a}_1} \\
1\times \delta_{\mathbf{r} - \bfR,-\mathbf{a}_2} \\
1\times \delta_{\mathbf{r} - \bfR,-\mathbf{a}_3} \\
\end{pmatrix}
\ .
\end{align}
where $\mathbf{a}_i$'s are along the edges of the tetrahedron:
\begin{align}
    \mathbf{a}_1   =& (1,1,0)\ , \\
    \mathbf{a}_2   =& (1,0,1)\ ,\\
    \mathbf{a}_3   =&(0,1,1)\ .\\
\end{align}
The FT-constrainers are 
\begin{align}
\label{EQN_pyro_AFM_FTconstrainer_1}
\mathbf{T}_\mathsf{PC1} (\bfq)&
=(
 1 ,
 e^{  - i \mathbf{q} \cdot \mathbf{a}_1} ,
 e^{  - i \mathbf{q} \cdot \mathbf{a}_2} ,
 e^{  - i \mathbf{q} \cdot \mathbf{a}_3},
)^\text{T}\ ,\\
\label{EQN_pyro_AFM_FTconstrainer_2}
\mathbf{T}_\mathsf{PC2} (\bfq)&
=(
 1 ,
 e^{    i \mathbf{q} \cdot \mathbf{a}_1} ,
 e^{    i \mathbf{q} \cdot \mathbf{a}_2} ,
 e^{    i \mathbf{q} \cdot \mathbf{a}_3},
)^\text{T}\ .
\end{align}
Again, since there are two constrainers,  there are  two flat bottom bands and two higher dispersive bands at a general momentum $\mathbf{q}$.
However,  at $\mathbf{q} = \mathbf{0}$, the two constrainers become linearly dependent,
\[
\mathbf{T}_\mathsf{PC1}  ( \mathbf{0})  =  \mathbf{T}_\mathsf{PC2}  ( \mathbf{0}) \propto \frac{1}{2}(1,1,1,1)^\text{T} \equiv \mathbf{T}^0\ ,
\]
which means a gap closing happens there (ref. spectrum in  Fig.~\ref{Fig_pyro_AFM_lattice}(c)). 
Thus we   expand $\mathbf{T}_\mathsf{PC1}$ around $\mathbf{q} = \mathbf{0}$, and take its component perpendicular to $\mathbf{T}^0$, which is
\[
\begin{split}
  \tilde{\mathbf{T}}( \mathbf{k})& = \mathbf{T}_\mathsf{PC1}  ( \mathbf{k}) - (\mathbf{T}_\mathsf{PC1}  ( \mathbf{k}) \cdot \mathbf{T}^0 )\mathbf{T}^0\\
&= \frac{i}{2}\begin{pmatrix}
   q_x + q_y + q_z\\
   -q_x - q_y + q_z \\
   -q_x + q_y - q_z \\
   +q_x - q_y - q_z \\
\end{pmatrix}  \ .
\end{split} 
\] 
This  yields the Gauss's law  of 3D Maxwell U(1):
\[
\label{EQN_pyro_gauss_law}
\begin{split}
  & \partial_x (S_1 -   S_2  -    S_3 +    S_4)
+\partial_y (S_1 -   S_2  +    S_3 -    S_4)\\
& \qquad +\partial_z (S_1 +  S_2  -    S_3 -    S_4)\\
& \equiv \partial_x E_x + \partial_y E_y + \partial_z E_z = 0  \ .
\end{split}
\] 
The gapless points also appears at $\mathbf{q} = (\pm \sqrt{3}\pi/2,0 , 0)$ and its cubic rotations. 
At these points, the equal-time spin correlation, shown in Fig.~\ref{Fig_pyro_AFM_lattice}(d),
exhibits the two-fold pinch points (2FPP), which is the canonical hallmark of the emergent U(1) electrostatics physics. 
We note that, although the spectrum is also gapless at $\mathbf{q} = (0,0 , 0)$, the equal-time spin correlation does not observe a pinch point there. But this is merely due to the cancellation of intensity when all spin correlation channels are summed together.

\subsection{Honeycomb-snowflake model}
\label{Sec_BM_classification_application}

\subsubsection{Emergent Gauss's laws from the honeycomb-snowflake model}

\begin{figure}[ht!]
 \centering
 \includegraphics[width=0.45\textwidth]{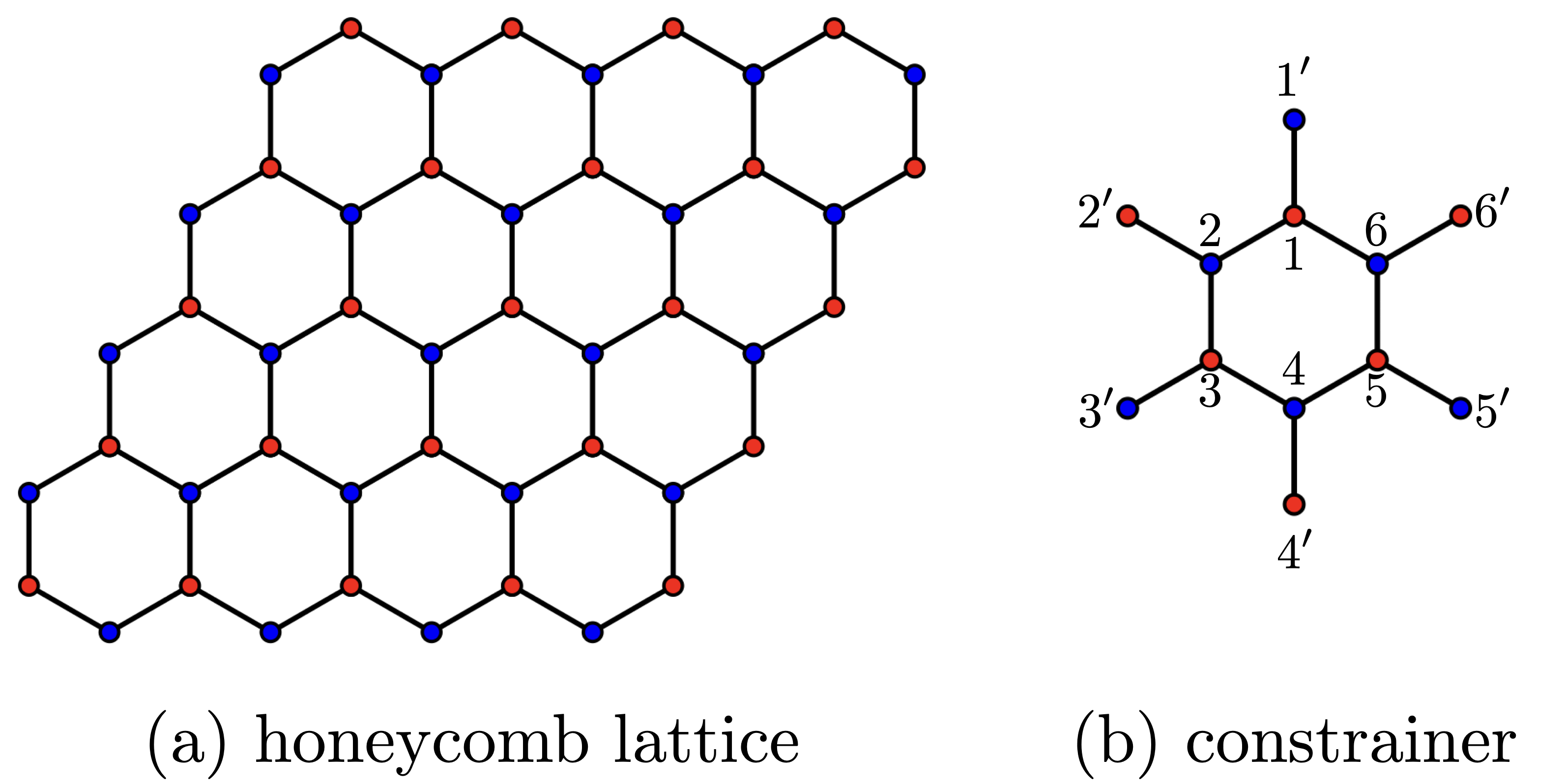}
 \caption{  (a) Honeycomb lattice for the Honeycomb-snowflake  model (Eq.~\eqref{eq:Hhex2}) introduced in \cite{Benton21PRL}.
 (b) The constrainer of the model (Eq.~\eqref{eq:Mhex2}). 
 This figure is a replication of Fig.~\ref{fig:BMmodel}, reproduced here for convenience.
 \label{fig:BMmodel2}}
\end{figure}

Now let us apply the classification algorithm to the honeycomb-snowflake model \cite{Benton21PRL}, which we introduced in Sec.~\ref{Sec_BM_Honeycomb_model}.

The model  is defined on the honeycomb lattice, with a spin on each site, which we treat as a scalar within the large-$\mathcal{N}$
approximation. 
The Hamiltonian is 
\[ 
\mathcal{H}_\mathsf{HS}=
\frac{J}{2} \sum_{\mathbf{R} \in \text{u.c.}} \   [\mathcal{C} ^{  \gamma}(\mathbf{R} )]^2\ . 
\label{eq:Hhex2}
\]
The sum  of $\mathbf{R}$ is  taken over all unit cells, which is best visualized as hexagonal plaquettes.
The constrainer $\mathcal{C}^{  \gamma}(\mathbf{R} )$ defined on the   hexagons contains  weighted sums 
of spins around each hexagon shown in Fig.~\ref{fig:BMmodel2}:
\[ 
\mathcal{C}_{\mathsf{HS},\alpha}^{  \gamma}(\mathbf{R} ) 
= \sum_{i  = 1  }^6 {S}^\alpha_i + \gamma \sum_{j = 1'  }^{6'}{S}^\alpha_j \ .
\label{eq:Mhex2} 
\]

\begin{figure}[ht!]
 \centering
 \includegraphics[width=\columnwidth]{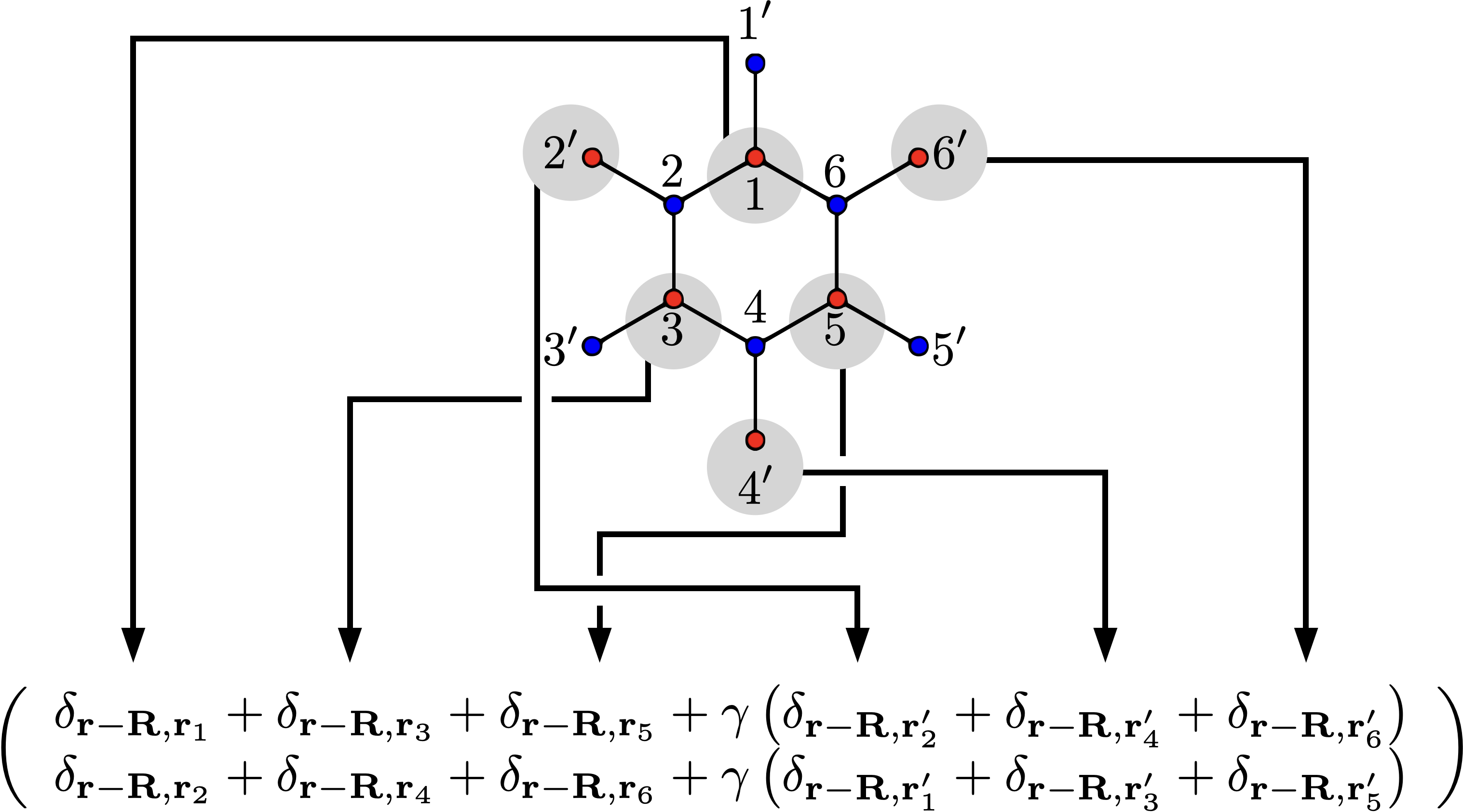}
 \caption{How to write down the vector form constrainer $\bfC^\gamma_\mathsf{HS}(\bfR,\mathbf{r})$ (Eq.~\eqref{EQN_honey_snow_constrainer_vector}) from its real space image (Fig.~\ref{fig:BMmodel2}(b)).
 \label{Fig_honey_snow_constrainer}}
\end{figure}

The constrainer reads
\[
\label{EQN_honey_snow_constrainer_vector}
\begin{split}
       \bfC^\gamma_\mathsf{HS}(\bfR,\mathbf{r})    
   =&
\begin{pmatrix}
 \delta_{\mathbf{r} - \bfR,\mathbf{r}_1} + \delta_{\mathbf{r} - \bfR,\mathbf{r}_3} + \delta_{\mathbf{r} - \bfR,\mathbf{r}_5}   
  \\
 \delta_{\mathbf{r} - \bfR,\mathbf{r}_2} + \delta_{\mathbf{r} - \bfR,\mathbf{r}_4} + \delta_{\mathbf{r} - \bfR,\mathbf{r}_6}   
\end{pmatrix} \\
 &+\gamma \begin{pmatrix}
 \delta_{\mathbf{r} - \bfR,\mathbf{r}_2'} + \delta_{\mathbf{r} - \bfR,\mathbf{r}_4'} + \delta_{\mathbf{r} - \bfR,\mathbf{r}_6'} 
 \\
\delta_{\mathbf{r} - \bfR,\mathbf{r}_1'} + \delta_{\mathbf{r} - \bfR,\mathbf{r}_3'} + \delta_{\mathbf{r} - \bfR,\mathbf{r}_5'} 
\end{pmatrix}\ .
\end{split}
\]
Here, $\bfr_j$ is the vector from the center of 
the snowflake to the corresponding site $j$ labeled in Fig.~\ref{fig:BMmodel2}(b).
Figure~\ref{Fig_honey_snow_constrainer} shows how the first element of the constrainer is constructed by going over all the first sub-lattice sites in the adjacent unit cells.

\begin{figure*}[ht]
 \centering
 \includegraphics[width=0.7\textwidth]{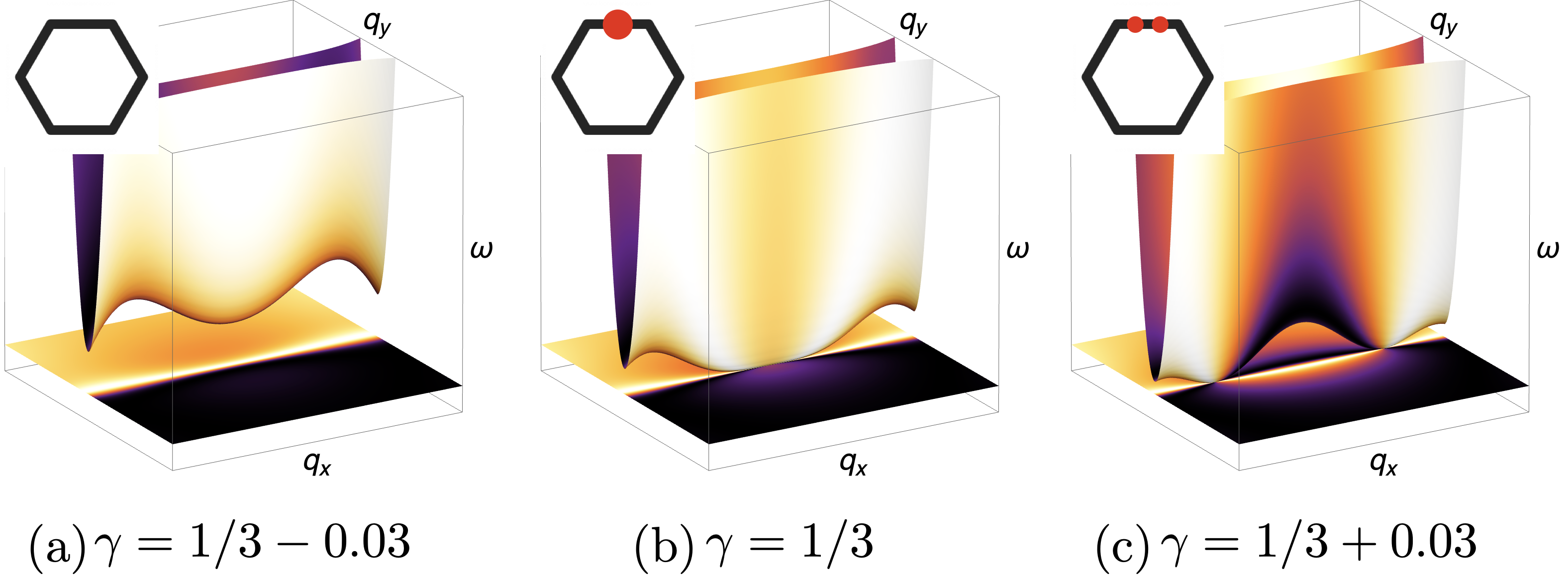}
 \caption{ 
 The transition between different algebraic CSLs as the emergence  and splitting of the gap-closing points. 
 This figures shows one such transition in the  honeycomb-snowflake model around $\gamma = 1/3$.
 The three plots are the zoomed-in view of the spectrum  at the center of the BZ edge. The insets on the top left corner show  the position of gap-closing points in the BZ (actual distance is exaggerated for better visibility).
  (a) At $\gamma = 1/3-0.03$, there is no gap-closing there, but a higher dispersive band moves down to approach the bottom flat band.
 (b) At $\gamma = 1/3$, a new gap-closing point appears as the higher dispersive band touches the bottom flat band. 
 (c) As $\gamma$   increases to  $\gamma = 1/3+0.03$, the gap-closing point split into two and moves toward the conner of BZ. 
 }
 \label{Fig_BM_emerge}
\end{figure*}

\begin{figure*}[ht]
 \centering
 \includegraphics[width=0.7\textwidth]{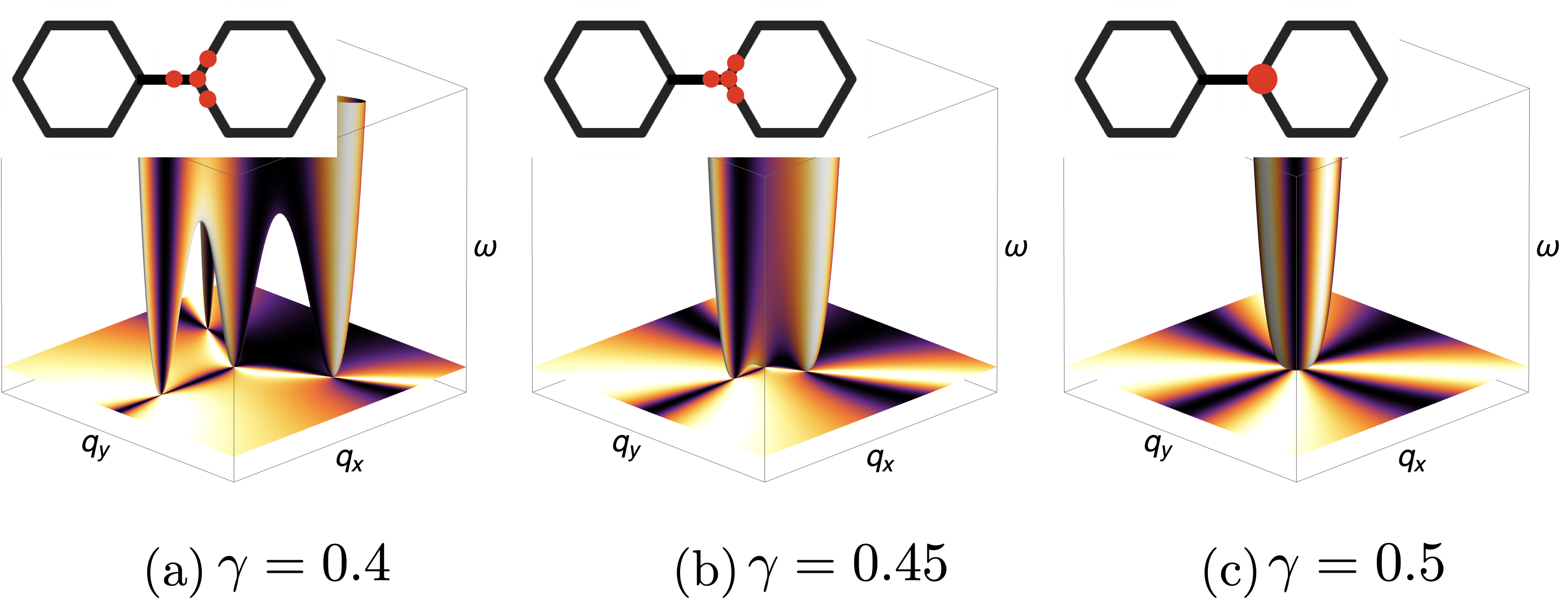}
 \caption{ The transition between different algebraic CSLs as the merging of the gap-closing points. 
 This figures shows one such transition in the  honeycomb-snowflake model around $\gamma = 1/2$.
 The three plots are the zoomed-in view of the spectrum structure at the corner   of the BZ. The insets on the top left corner show  the position of gap-closing points in the BZ (actual distance is exaggerated for better visibility).
 (a) 
 At $\gamma  = 0.4$, there are four  gap-closing points, each associated with the Maxwell Gauss's law. 
 (b) 
 As $\gamma $ gradually grows to $\gamma  = 0.45$, the four gap-closing points move close to each other.
 (c)
 At the critical value $\gamma = 0.5$, the four points merge together to form a new type of gap-closing points, which is associated with the  rank-2 U(1) Gauss's law.
 }
 \label{Fig_BM_merge}
\end{figure*}
The FT-constrainer is then obtained by Fourier transforming $\bfC^\gamma_\mathsf{HS}(\bfR,\mathbf{r})$,
\[
\label{eq:T(q)-honeycomb}
\begin{split}
     \mathbf{T}^\gamma_\mathsf{HS} ( \bfq)  
     =& 
\begin{pmatrix}
 e^{-i\bfq \cdot \bfr_1}  +  e^{-i\bfq \cdot \bfr_3}  + e^{-i\bfq \cdot \bfr_5}  
  \\
 e^{-i\bfq \cdot \bfr_2}  +  e^{-i\bfq \cdot \bfr_4}  + e^{-i\bfq \cdot \bfr_6}  
    \\
\end{pmatrix}\\
& +  \gamma 
\begin{pmatrix}
  e^{-i\bfq \cdot \bfr_2'}  +  e^{-i\bfq \cdot \bfr_4'}  + e^{-i\bfq \cdot \bfr_6'}  \\
  e^{-i\bfq \cdot \bfr_1'}  +  e^{-i\bfq \cdot \bfr_3'}  + e^{-i\bfq \cdot \bfr_5'}   \\
\end{pmatrix} \ .
\end{split} 
\]
And the Hamiltonian in momentum space is 
\[
\label{EQN_BM_Hamiltonian_J}
[J^\gamma_\mathsf{HS}]_{ab} = [{T}^\gamma_\mathsf{HS}]_a[{T}^\gamma_\mathsf{HS}]_b^*\ .
\]

The spectrum structure is plotted in Fig.~\ref{fig:honeycomb_bands} for different values of $\gamma$.
It has two bands.
The top band always undergoes a gap closing at wavevector
\[
\bfq_0 = (\frac{4\pi}{3\sqrt{3}},0)\ .
\]

Let us now examine the physics for small $\bfk = \bfq - \bfq_0$ for two cases: $\gamma  = 0$ and $\gamma = 1/2$.

When $\gamma = 0$, we have, at leading order, the FT-constrainer
\[
\tilde{\mathbf{T}}^0_\mathsf{HS} (\bfk ) = \mathbf{T}^{0}_\mathsf{HS}(\bfq_0 + \bfk )
= \frac{3}{2}
\begin{pmatrix}
   -k_x + i k_y  \\
   -k_x - i k_y  \\
\end{pmatrix}\ .
\]

So the spin fluctuations around the ground state satisfy the constraint
\[
\tilde{\mathbf{T}}^0_\mathsf{HS} (\bfk  )\cdot \tilde{ \mathbf{S} }
=\frac{3}{2} \left[( -k_x + i k_y )\tilde{S}_1 + (-k_x - i k_y )\tilde{S}_2 \right]= 0
\ .
\]
Reorganizing the DOFs, we have 
\[
ik_x i( \tilde{S}_1 +  \tilde{  S_2}) + ik_y (\tilde{S}_1 -\tilde{S}_2) = 0,
\]
which upon Fourier tranformation into the real space yields a Maxwell Gauss's law 
\[
\partial_\alpha E_\alpha = 0
\label{eq:honeycombgauss}
\]
which acts on a complex electric field
${\bf E}=(i(\tilde{S}_1+\tilde{S}_2), 
\tilde{S}_1-\tilde{S}_2)$.
Note here that because of the
phase shift arising from expanding around
finite momentum ${\bf q_0}$, $\tilde{S}_a$
are themselves complex in real space and
therefore $E_x$ is not purely imaginary and
$E_y$ is not pureley real.

If we separate Eq. (\ref{eq:honeycombgauss})
into real and imaginary parts we may consider it as two real Gauss's laws.
States satisfying these two Gauss's laws are also guaranteed to satisy the Gauss's laws that would be obtained from an expansion around the singular band touching at ${\bf q}=-{\bf q}_0$, due to the property $\tilde{S}_a({\bf q})^{\ast}=\tilde{S}_a(-{\bf q})$.
We thus have two real Gauss's laws in total, which is to be expected, as there are
two singular band touchings per BZ.

When $\gamma = 1/2$, on the other hand, we obtain the FT-constrainer
\[
\tilde{\mathbf{T}}^{1/2}_\mathsf{HS} (\bfk ) = \mathbf{T}^{1/2}_\mathsf{HS} (\bfq_0 + \bfk )
= \frac{9}{8}
\begin{pmatrix}
   k_x^2 -  2i k_x k_y  - k_y^2  \\
    k_x^2 + 2i k_x k_y  - k_y^2\\
\end{pmatrix}\ .
\]
We then have the emergent Gauss's law in the form
\[
\begin{split}
&\tilde{\mathbf{T}}^{1/2}_\mathsf{HS} (\bfk  )\cdot \tilde{ \mathbf{S} } = 0 \\ 
&=\frac{9}{8} \left[(  k_x^2 -  2i k_x k_y  - k_y^2 )\tilde{S}_1 + ( k_x^2 + 2i k_x k_y  - k_y^2)\tilde{S}_2 \right]
\ ,    
\end{split}
\]
which we rewrite as
\[
\left[-(ik_x)^2 + (ik_y)^2\right](\tilde{S}_1+\tilde{S}_2)
+ (i k_x)( ik_y) 2i(\tilde{S}_1-\tilde{S}_2) = 0
\ .
\]
Again this is a complex Gauss's law.
If we identify a traceless, symmetric complex matrix to be 
\[
\mathbf{E}
=
\begin{pmatrix}
  E_{xx} & E_{xy}  \\
    E_{xy} & - E_{xx}\\
\end{pmatrix}
\equiv 
\begin{pmatrix}
 - (\tilde{ S}_1+\tilde{ S}_2) & i(\tilde{ S}_1-\tilde{ S}_2)  \\
  i(\tilde{ S}_1-\tilde{ S}_2)  & (\tilde{ S}_1+\tilde{ S}_2)\\
\end{pmatrix}
\ ,
\]
then the Gauss's law becomes 
\[
\label{EQN_BM_r2u1_gauss}
\partial_\alpha \partial_\beta E_{\alpha \beta} = 0
\ ,
\]
which is a (complex) realization of the electrostatics for a symmetric rank-2 U(1) gauge theory.
 
Breaking the complex Gauss's law into real
and imaginary parts, we obtain two
real Gauss's laws.
As before, these also take care of the
constraints arising from the band touching at  $-{\bf q}_0$, and the presence of two Gauss's laws agrees with the presence
of two band touchings in the BZ.

\subsubsection{Transition between   algebraic CSLs}

Let us now study the transition between different algebraic CSLs in the honeycomb-snowflake model. 
We will study the transitions near two critical points: $\gamma = 1/3$ and $\gamma = 1/2$.

For the critical point $\gamma = 1/3$, a  new band touching point emerges at the mid point of the BZ boundary, as illustrated in Fig.~\ref{Fig_BM_emerge}(b). 
This happens, as the top band gradually moves down, when the top band touches the bottom band, as $\gamma\rightarrow 1/3^-$.
The new band touching then splits into two band touching points as $\gamma$ increases above $1/3$, see Fig.~\ref{Fig_BM_emerge}(c).
Each single band touching point is associated with a Maxwell's U(1) Gauss's law.

For  $1/3 < \gamma < 1/2$, 
there are band touching points on the BZ corner and boundary. 
Each single band touching point is associated with a Maxwell U(1) Gauss's law, as just shown. 
This can also be seen from Fig.~\ref{fig:honeycomb_bands}, where the structure factor (defined in Eq.~\eqref{eq:S(q)}) on each band touching point exhibits the characteristic two-fold pinch point.

As $\gamma$ increases and approaches $1/2$, 
three band touching points on the BZ boundary move toward the fourth one on the BZ corner, as illustrated in Fig.~\ref{Fig_BM_merge}.
At the critical point $\gamma = 1/2$, the four points merge together (see Fig.~\ref{Fig_BM_merge}c), and form a new band touching point with a different structure: one associated with the rank-2 U(1) Gauss's law shown in Eq.~\eqref{EQN_BM_r2u1_gauss}.

The lesson we learn here is that the transition between different algebraic CSLs can be understood as the emergence/disappearance and merging/splitting of the band touching points in their spectrum.
Mathematically, such transitions are described in the same way as in topological band theory, and much prior knowledge can be borrowed to understand transitions of algebraic CSLs.
This will be a topic for future study.

\subsubsection{Symmetry and topological protection of the gapless points}
In the honeycomb-snowflake model these gap closings are symmetry- and topologically protected, provided that the Hamiltonian respects inversion symmetry.
This is because inversion symmetry requires the two components of the constrainer to obey
\[
[C^\gamma_\mathsf{HS}]_1(\bfR,\mathbf{r}) = [C^\gamma_\mathsf{HS}]_2(\bfR,-\mathbf{r})\ .
\]
Hence, 
the components in the FT-constrainer ${\bf T}({\bf q})$ are related by
\begin{eqnarray}
\label{eq:T(q)-honeycomb2}
[{ T}^{\gamma}_\mathsf{HS}]{}_1({\bf q}) = [{ T}^{\gamma}_\mathsf{HS}]{}^{\ast}_2({\bf q})\ .
\end{eqnarray}

Combining this with the normalization of the eigenvector implies $\hat{\bf T}({\bf q})$ can always be written in the form
\begin{eqnarray}
\hat{\bf T}^{\gamma}_\mathsf{HS}({\bf q})=
\frac{1}{\sqrt{2}}
\begin{pmatrix}
\exp(i \phi({\bf q})) \\
\exp(-i \phi({\bf q}))
\end{pmatrix}\ .
\end{eqnarray}
and can thus be represented by a point on the unit circle $\phi({\bf q})$.

We can thereby define a winding number of the vector field 
\[
\mathbf{v} ({\bf q})= (\cos \phi({\bf q}), \sin \phi({\bf q})) 
\]
around closed paths in reciprocal space:
\begin{eqnarray}
Q_C=\frac{-1}{\pi} \oint_C d{\bf q}\cdot\left(
v_2 \nabla_{\bf q} v_1
\right) \qquad \in \ \mathbb{Z}\ .
\end{eqnarray}
The topologically stable gap closing points correspond to vortices of 
${\bf v}({\bf q})$ with a finite integer winding number for closed paths encircling them.
The gapless points are thus topologically stable and cannot be removed by small changes to the ground state constraint, provided that inversion symmetry is maintained.

Let us now revisit the  ground state phase diagram of the honeycomb-snowflake model shown in  Fig.~\ref{fig:honeycomb_bands}.
Transitions between distinct CSLs occur as $\gamma$ is varied, via pair creation/annihilation of vortices, or by coalescence of vortices with like winding number.
The various CSLs have a distinct arrangement of singularities (known as pinch points)
in their spin correlation functions, affirming their distinctive nature.
At $\gamma=0$ the model is an algebraic CSL with gap closings and corresponding pinch points at the Brillouin zone corners (K points) (Fig.~\ref{fig:honeycomb_bands} (a)). 
On increasing $\gamma$ this remains the case until $\gamma=1/3$, at which point pairs of oppositely charged vortices nucleate at the M points of the BZ
(Fig. \ref{fig:honeycomb_bands} (b) and Fig.~\ref{Fig_BM_emerge}(b-c)),  in addition to the existing pinch points at the $K$ wavevector. 
This leads to a new CSL with 8 pinch points per BZ instead of only 2, with all pinch points on the zone boundaries (the points shared by several adjacent Brillouin zones are included only once in this count). 
As $\gamma$ is further increased the vortices formed at the M points migrate towards the K points, such that three vortices of one charge converge on one of the opposite charge
(Fig. \ref{fig:honeycomb_bands} (c) and Fig.~\ref{Fig_BM_merge}(c)).
This leads to the formation of vortices with winding number $+/-2$ at the K points when $\gamma=1/2$, and four-fold pinch points in the spin structure factor
(Fig. \ref{fig:honeycomb_bands}(d) and Fig.~\ref{Fig_BM_merge}(c)).
This is indicative of a spin liquid described by a higher-rank U(1) gauge theory 
\cite{Prem18PRB, Yan20PRL} in Eq.~\eqref{EQN_BM_r2u1_gauss}, as explained earlier.
On increasing $\gamma$ further the vortices at the zone corners separate again and the system enters a new CSL, with 8 pinch points per BZ but now with 6 of them in the interior of the BZ rather than on the boundary
(Fig. \ref{fig:honeycomb_bands} (l,p)).
 
For negative $\gamma$, the story is similar, and the readers can refer to the original paper ~\cite{Benton21PRL} for more detail.

\subsection{Anisotropic U(1) CSL }
\label{Sec_anisotropic_snowflake}

The honeycomb-snowflake model with $\gamma=1/2$ provides a simple
example of a classical spin liquid with an isotropic
Gauss's law, as shown in Ref.~\onlinecite{Benton21PRL}.
Here, we propose a simple model exhibiting a  spin liquid described by an anisotropic Gauss's law, and demonstrate its
nature using the algebraic classification from
Section \ref{sec:algebraic}.

\begin{figure}[ht!]
    \centering
    \includegraphics[width=\columnwidth]{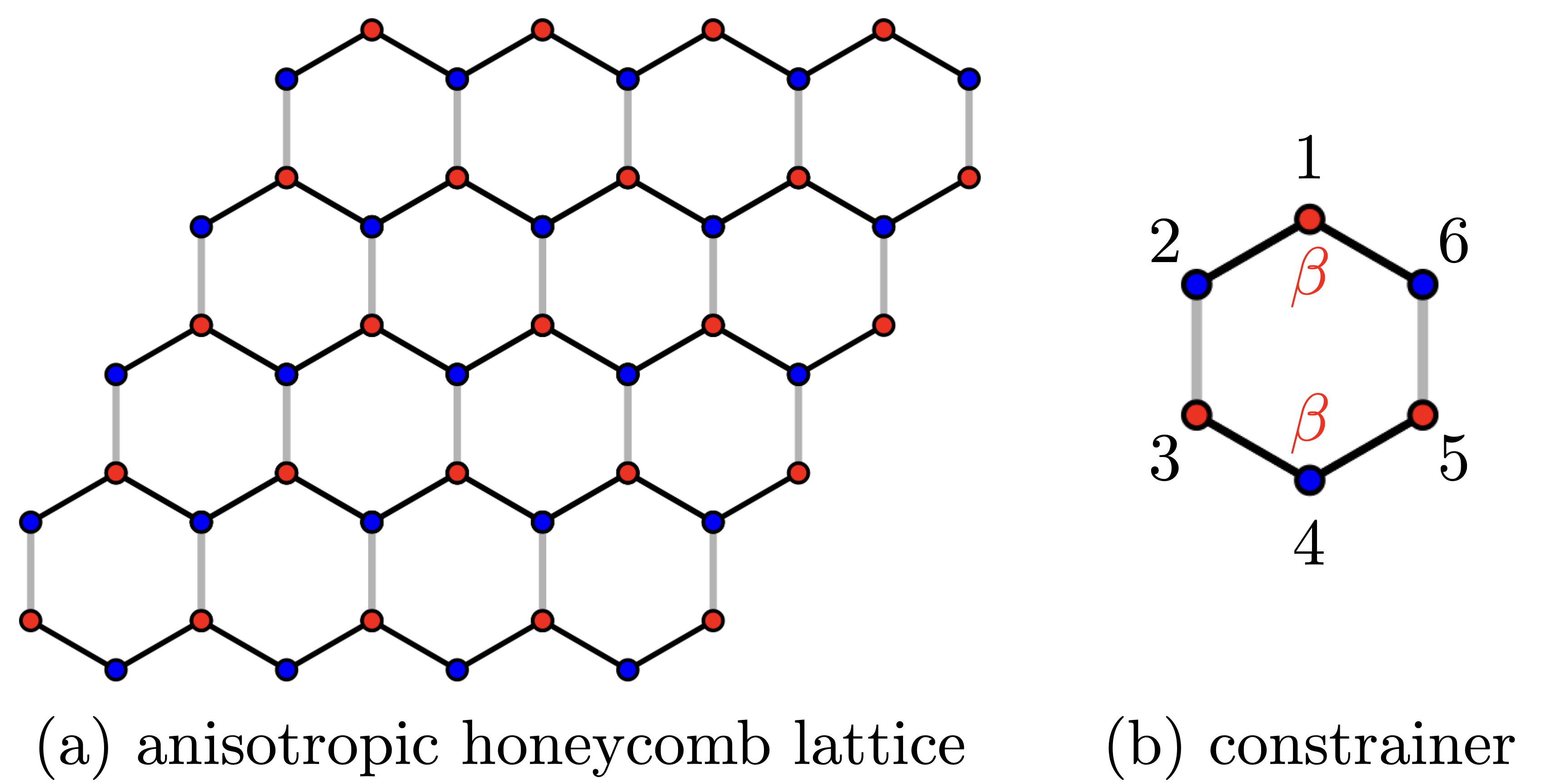}
    \caption{ (a) The  honeycomb lattice with explicit lattice symmetry breaking. 
    (b) Constrainer of the anisotropic honeycomb lattice.
    Tuning from $\beta=1$ to $\beta=2$ tunes
    from the ordinary U(1) CSL from \cite{Rehn17PRL} to anisotropic  U(1) CSL.}
    \label{fig:typeII_constraint}
\end{figure}

The model can be considered as a generalisation 
of the honeycomb-snowflake model with explicit lattice symmetry breaking.
Specifically, we take the honeycomb-snowflake model with $\gamma=0$
such that the constraint on each hexagon only involves
spins belonging to that hexagon.
We then adjust the contribution of each spin to the constraint according to a new parameter $\beta$,
such that spins at the top and bottom of the hexagon contribute 
to the ground state constraint with weight $\beta$ and the others
with weight 1: 
\[
\label{EQN_constrainer_aniso_HS}
\mathcal{C}_{\sf hex, \beta}= \beta (S_{1, \alpha}+S_{4, \alpha}) + (S_{2, \alpha}+S_{3, \alpha}+S_{4, \alpha}+S_{5, \alpha})=0\ ,
\]
with the sites numbered around each hexagon as shown in Fig.~\ref{fig:typeII_constraint}(b).
The case $\beta=1$ corresponds to the isotropic honeycomb model from Ref.~\onlinecite{Rehn16PRL}.

Upon increasing $\beta$ from $\beta=1$, gap closing 
points migrate along the Brillouin zone boundaries
normal to the $q_y$ axis. At $\beta=2$ they merge
at the M point of the Brillouin zone $({\bf q}={\bf q}_M)$. 
At this merging point,
we can expand the FT-constrainer ${\bf T}({\bf q})$
around ${\bf q}={\bf q}_M$. Here, similarly to Section \ref{sec:square_ice}, we use a gauge in which we reference the spins to the position of the centre of
their unit cell, rather than their physical position on the lattice. This leads us to:
\[
\label{EQN_T_Anisotropic_HS}
{\bf T}({\bf q}_M + {\bf k})
=
\begin{pmatrix}
3i k_y + \frac{3}{4} (-k_x^2 + 3 k_y^2) \\[1em]
e^{-i2\pi/3}\left[-3i k_y + \frac{3}{4} (-k_x^2 + 3 k_y^2)\right]
\end{pmatrix}\ .
\]

The dispersion $\omega({\bf q})$
is anisotropic around the band touching, having a form:
\[ 
\begin{split}
 & \omega({\bf q}_M + {\bf k})  =\mathbf{T}^\ast({\bf q}) \cdot \mathbf{T}({\bf q}) \\
  =&   18 k_y^2 + \frac{9}{8} (k_x^4 -6 k_x^2 k_y^2 +9 k_y^4)  \ .
\end{split} 
\]

To obtain the Gauss law we  use  $\sum T^\ast_a({\bf q})\cdot \tilde{S}_a ({\bf q})=0$.
By adding a phase $\tilde{S}_2' = \tilde{S}_2 e^{-i2\pi/3}$, 
we have  
\[
3i k_y (\tilde{S}_1 ({\bf q})- \tilde{S}_2' ({\bf q}))
+ \frac{3}{4} (-k_x^2 + 3 k_y^2)
(\tilde{S}_1 ({\bf q}) + \tilde{S}_2' ({\bf q})) =0,
\]
and therefore a real space Gauss' law:
\begin{eqnarray}
 3\partial_y E_1 + \frac{3}{4} (\partial_x^2  - 3 \partial_y^2) E_2=0,
\end{eqnarray} 
where we have identified the  electric field components with the suitable combination of the fluctuating spin variables $\tilde{S}_i$.

We found the conserved quantities to be the following. First, the obvious one is the net charge conservation:
\[
    Q_1 = \int \mathop{dv}\rho\ .
\] 
We can also look for other conservation laws defined by a suitably chosen function $f(x,y)$ in the integrand:
\[
\begin{split}
 Q_2 & =  \int \mathop{dv} f \rho  \\
& =  \int \mathop{dv}\left[ - f (\partial_x^2 + 3 \partial_y^2) E_1 + f \partial_y E_2 \right]  \\
& =  \int \mathop{dv} \left[ -E_1 (\partial_x^2 + 3 \partial_y^2) f - E_2 \partial_y f \right]\ ,    
\end{split} 
\]
after integration by parts. 
To make sure this is zero, we need to choose $f$ such that the following two conditions are simultaneously satisfied:
\begin{align}
(\partial_x^2 + 3 \partial_y^2) f = & 0 \ ,\\
\partial_y f =& 0.
\end{align} 
The solutions are
\[
f_1  =a_0 +  a_1 x,  
\]
for any choice of real numbers $a_i$.
Hence we deduce that the second conserved quantity is the charge dipole in the $x$ direction:
\[
Q_2  =  \int \mathop{dv} x \rho \ .
\]
The charge 
therefore has reduced mobility as in fracton theories~\cite{Pretko-review2020}: it is immobile in $x$ direction but can move in the transverse (here, $y$) direction. 

\begin{figure*}[ht]
    \centering
    \subfloat{\includegraphics[width=0.8\textwidth]{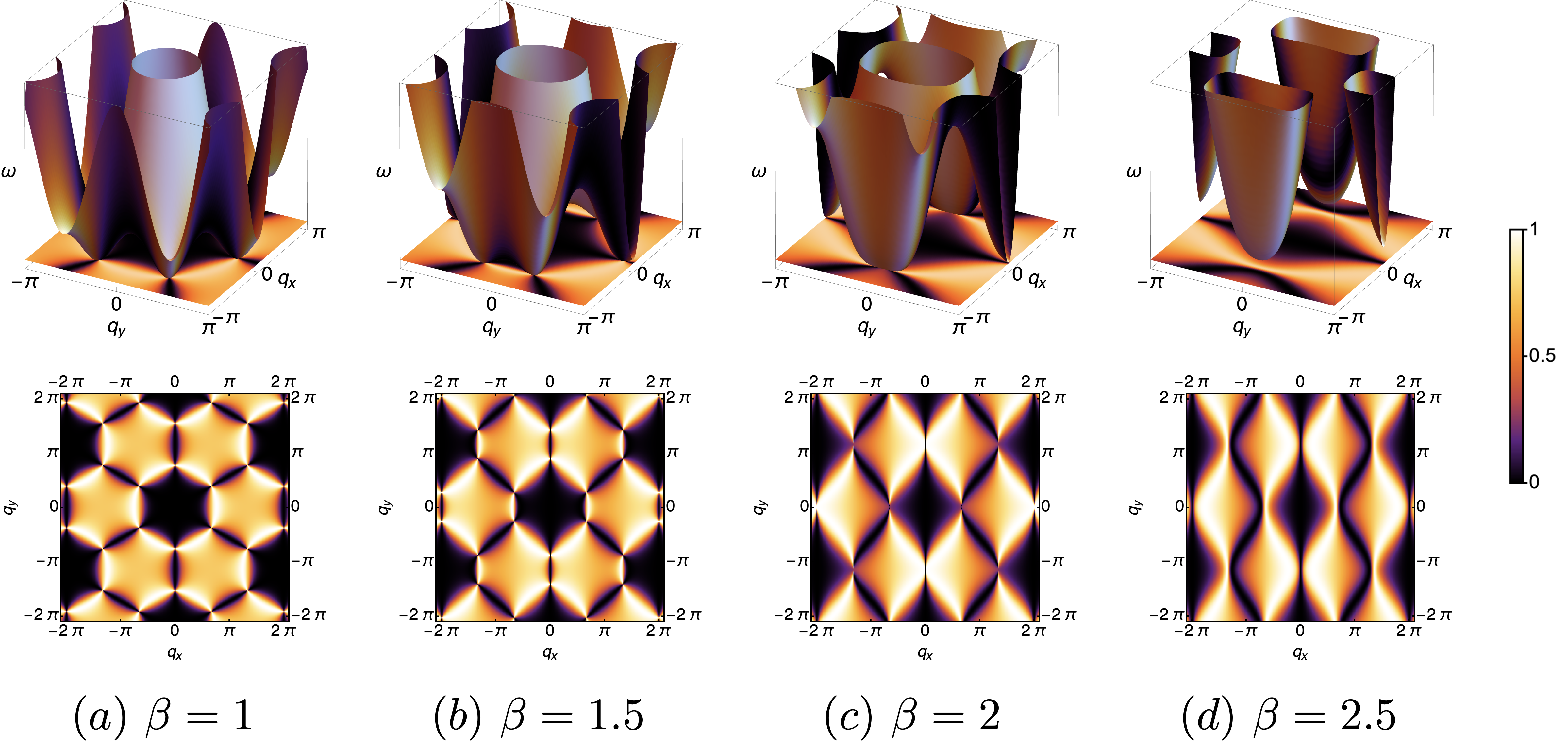}}
    \caption{Structure factor S({\bf q}) for the anisotropic honeycomb model defined by the constrainer Eq.~\eqref{EQN_constrainer_aniso_HS}. 
    (a) At $\beta = 1$, the quadratic band touching is visible as standard pinch points in the structure factor.
    (b) 
    As $\beta$ is increased, two gap-closing points migrate along the Brillouin zone boundary toward the M point (edge center of the  Brillouin zone). 
    (c) At $\beta = 2$, two gap-closing point merge at the M point of the Brillouin zone, creating a parabolic pinch point singularity.
    (d) For $\beta>2$ the system enters
    a trivial paramagnetic phase with smooth spin correlations throughout the Brillouin zone.}
    \label{fig:type2sq}
\end{figure*}

This model can actually be viewed as a 2D cut of the generalized U(1) gauge theory for Haah's code (before Higgsing). 
Its structure factor is essentially identical to that proposed in Ref.~\onlinecite{Hart2022PhysRevB},
featuring pinch points with parabolic contours.
The evolution of the spin correlations on tuning through the critical point is shown in Fig. \ref{fig:type2sq}.

It is worth noting that despite the anisotropic Gauss's law,
this spin liquid is not a ``Type II'' fracton phase \cite{Hart2022PhysRevB}, 
which requires an infinite number of conservation laws. Here
we have only a finite number. 
The parabolic pinch points should therefore be understood as
a signature of an anisotropic Gauss's law, and not necessarily as 
a signature of ``Type II'' fracton phases, as proposed in Ref.~\onlinecite{Hart2022PhysRevB}.

This spin liquid occurs at the special point of
parameter space $\beta=2$.
For $\beta>2$, there are no band touchings in the Brillouin Zone and the momentum space correlations
are smooth.
The model for $\beta>2$ connects smoothly to
the $\beta \to \infty$ limit, which is a trivial
paramagnet in which a pair of spins is coupled within each unit cell, but there is no inter-unit-cell coupling.
Hence, the anisotropic  spin liquid  occurs at the transition point between a Coulomb phase and a short-range correlated trivial paramagnet.

\subsection{Higher-dimensional gapless manifolds: pinch lines \textit{etc.}}
\label{subsec:pinch.line.CSL}

The discovery of various forms of algebraic spin liquid based on gapless points
of the Hamiltonian ${\bf J}({\bf q})$ in Eq.~\eqref{eq:Jab(q)} leads to a natural question: are there spin liquids associated to nodal lines of ${\bf J}({\bf q})$ and, if so, what are their properties?
The connection between gapless points in ${\bf J}({\bf q})$ and pinch points in the spin structure factor $S({\bf q})$ (see Eq.~\ref{eq:S(q)}) is suggestive of a generalization to nodal lines, i.e. the extension
of pinch point singularities along lines of reciprocal space, namely pinch lines.

Such features have previously been found in a classical spin liquid based on an
anisotropic spin Hamiltonian \cite{Benton16NComms}, and indeed a soft spin treatment
of this model finds nodal lines in the dispersion attached to the flat bands at the bottom of the spectrum.
The spin liquid in \cite{Benton16NComms} thus establishes one example of a nodal line spin liquid.

Here we present a new, simple model, of a nodal line spin liquid with isotropic spin interactions, based on the concept of the symmetry protected topological phases, which we have previously applied to other algebraic CSL.

To motivate the construction, we consider once more the honeycomb-snowflake model (see Section \ref{Sec_BM_Honeycomb_model} and earlier subsection~\ref{Sec_BM_classification_application} for the definition of this model). The Fourier transformed constrainer ${\bf T}({\bf q})$ has two components, listed in Eqs.~\eqref{eq:T(q)-honeycomb} and \eqref{eq:T(q)-honeycomb2}, corresponding to the two sites per unit cell and obeys the relation $T_1({\bf q}) = T_2({\bf q})^{\ast}$ due to inversion symmetry.
When normalised, as done in Eq.~\eqref{eq:T(q)-honeycomb2}, this means that ${\bf T}({\bf q})$ lies on the unit circle and its evolution in reciprocal space can support stable vortices corresponding to the nontrivial homotopy classes $\pi_1(S^1)$ of the phase $\phi(\mathbf{q})$.
The singularities in the center of such momentum-space vortices correspond to band touchings in ${\bf J}({\bf q})$ and pinch points
in the equal-time spin structure factor $S({\bf q})$ (see Eq.~\eqref{eq:S(q)}). These singularities are then protected in the sense that they cannot be removed by small changes to the ground state constraint which respect inversion symmetry.
The presence of these vortices arose directly from a two site unit cell and a symmetry constraining ${\bf T}({\bf q})$ onto the unit circle.

In two dimensions vortices are point like, but in three dimensions they are line like. 
The above considerations lead us to expect that a classical spin liquid with two sites and one constraint per unit cell, and inversion symmetry should support pinch lines.

\begin{figure}
    \centering
    \subfloat{\includegraphics[width=\columnwidth]{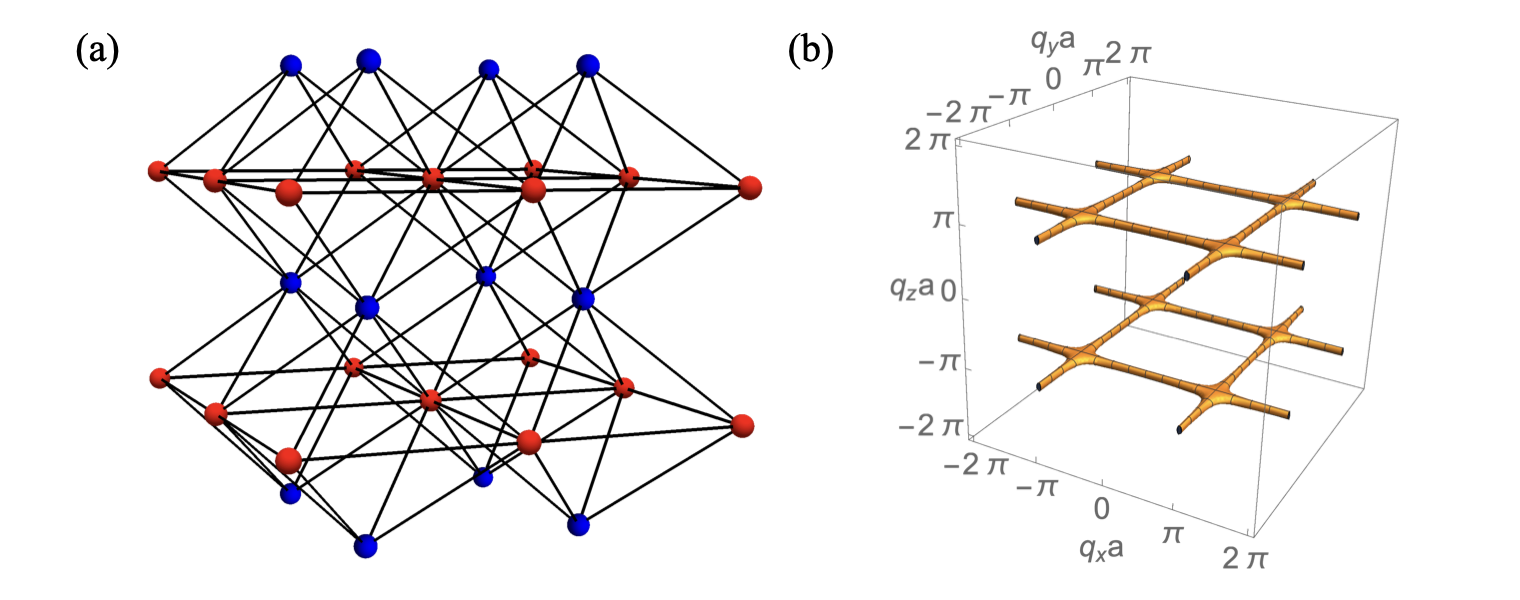}}
    \caption{(a) Frustrated lattice composed of octahedra which share edges in the $xy$ plane and connect via vertices in the $z$ direction. There is a two site unit cell, with inequivalent sites here indicated in red and blue. Defining a local constraint on the octahedra leads to a classical spin liquid with  nodal lines in ${\bf J}({\bf q})$ and hence pinch lines in $S({\bf q})$.
    (b) Location of nodal lines for the model defined in Eq. (\ref{eq:octahedra_model}). The nodal lines appear at wavevectors 
    ${\bf q}=({q_x, \pi, \pi})$, ${\bf q}=({\pi,q_y, \pi})$ and equivalent, creating a network of nodal lines along the edges of the $q_z=\pm\pi$ faces of the Brillouin zone.}
    \label{fig:octahedra_nodal_lines}
\end{figure}

One such example is found on the lattice shown in Fig. \ref{fig:octahedra_nodal_lines}(a). This lattice is formed from octahedral units which share edges in the $xy$ plane and join at vertices in the $z$ direction.
There are two sites per unit cell, indicated in red and blue in Fig. \ref{fig:octahedra_nodal_lines}(a).
We write down a Hamiltonian on this lattice, as a sum over octahedra:
\begin{eqnarray}
\mathcal{H}_{\sf oct.} = \sum_{\rm oct.} \left( \sum_{i \in {\rm oct.}} {\bf S}_i \right)^2\ .
\label{eq:octahedra_model}
\end{eqnarray}

The resulting soft-spin dispersion $\omega({\bf q})$ has two bands, a lower flat band and
a dispersive upper band. The upper band meets the flat band along the edges of the $q_z = \pm \pi$ faces of the Brillouin zone, i.e. along 
${\bf q}=({q_x, \pi, \pi})$, ${\bf q}=({\pi,q_y, \pi})$ and equivalent directions.
The location of the nodal lines is illustrated in Fig. \ref{fig:octahedra_nodal_lines}(b).

\begin{figure*}
    \centering
    \subfloat{\includegraphics[width=0.95\textwidth]{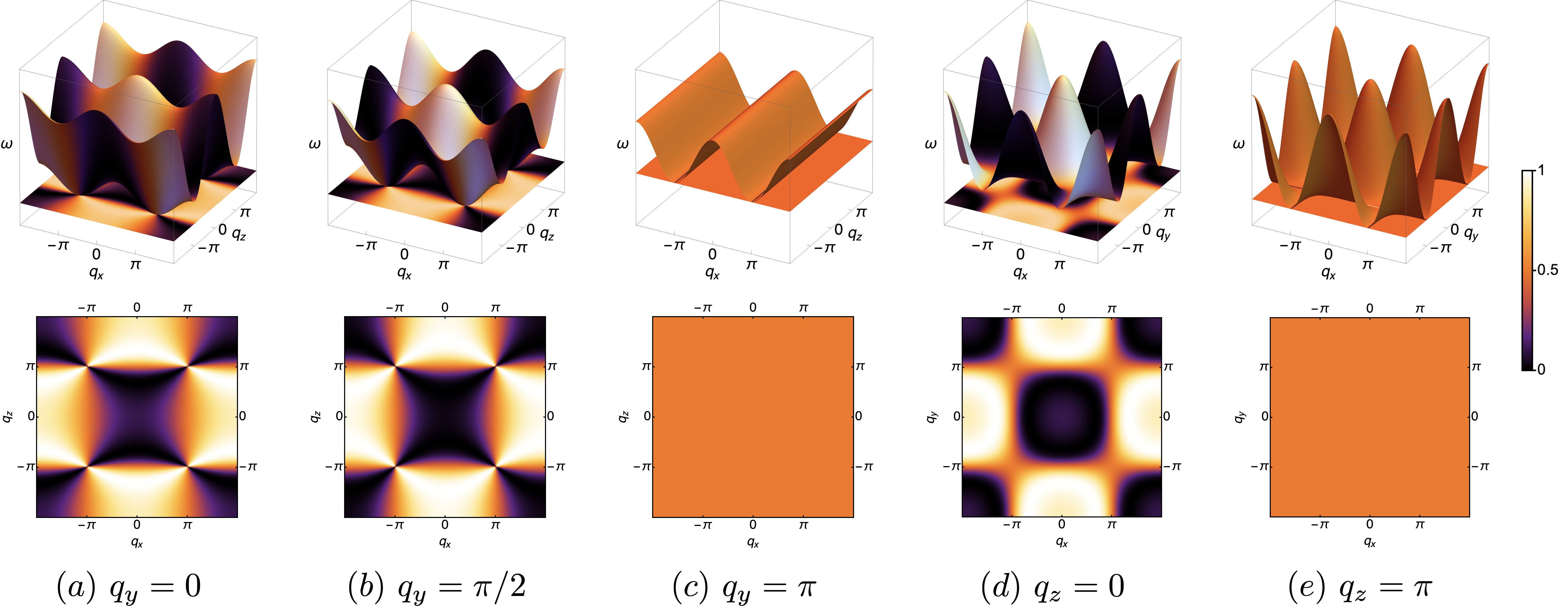}}
    \caption{  Spectrum and S({\bf q}) for the octahedral nodal line model (Eq.~\eqref{eq:octahedra_model}), taking cross sections at a series of fixed values of $q_y$ and $q_z$, cutting through the nodal lines at different points (see Fig.~\ref{fig:octahedra_nodal_lines}). 
    A pinch point is present at ${\bf q}=({\pm\pi,q_y, \pm\pi})$ for all values of
    $q_y$, thus forming an extended, line-like singularity: a pinch line.
    }
    \label{fig:sq_pinch_lines}
\end{figure*}

The structure factor $S({\bf q})$ for the moodel is depicted in Fig. \ref{fig:sq_pinch_lines}, as a series of cuts at fixed values of $q_y$.
These cuts intersect the pinch lines along the lines ${\bf q}=(\pm \pi, q_y, \pm \pi)$, and thus four pinch points are visible in each panel where the plane cuts the pinch line.

We thus establish a simple  model for a spin liquid with pinch line 
singularities. 
Based on the topological considerations outlined above,
pinch lines should be common to inversion symmetric three-dimensional 
classical spin liquids with two sites and one constraint per unit cell.

\section{Fragile topological CSL classification: eigenvector homotopy} 

\label{sec.topological.csl}

\subsection{The topological classification  }

Next we discuss the other category of classical spin liquids: the fragile topological CSLs with short-range spin correlations (the meaning of the qualifier `fragile' will be explained later in  section~\ref{sec:fragility}). 
A fundamental difference between this category and the algebraic one is that in fragile topological CSLs there is no  band-touching between the higher bands the bottom flat one(s).
Instead, the bottom flat bands are gapped from all other bands in the spectrum.
In real space, this means that all $L_x L_y$ local fluctuators (due to translation symmetry on an $L_x\times L_y$ lattice) are linearly independent and form a complete basis, 
thus accounting for all the ground states in the flat band.
The absence of  band-touching points 
means 
there are no emergent Gauss's laws describing the CSLs. 
For the same reason, the spin correlations decay exponentially instead of algebraically with distance.

In this category, we can still ask the question about the classification of fragile topological CSL models. 
More precisely, we consider two CSL systems  $A$, $B$ that have the same number of DOFs per unit cell, and same number of gapped flat bands,
and ask if it is possible to adiabatically tune CSL $A$ into $B$, while \textit{keeping the system in a CSL state} (i.e., maintaining the flatness of bottom bands)?
In terms of constrainers, 
this is to ask if we can smoothly change the $L_x L_y$  constrainers $\mathcal{C}_A(\mathbf{R})$'s into $\mathcal{C}_B(\mathbf{R})$'s without making them  linearly dependent at some point (for simplicity we use the one-constrainer Hamiltonian but its generalization is straightforward).
Although all of  $\mathcal{C}_A(\mathbf{R})$'s are linearly independent (i.e. the corresponding $\mathbf{T}(\bf{q})$ never vanishes and there is no band touching point), and so are the $\mathcal{C}_B(\mathbf{R})$'s,
in the process of tuning 
we may have to go through a  boundary point in parameter space whose constrainer $\mathbf{C}_X(\mathbf{R})$'s are not linearly independent anymore.
In the spectrum of the Hamiltonian, this would manifest itself in a gap closing. 
If such an intermediate gapless point is unavoidable, then we say that the two CSLs $A$ and $B$ belong to distinct equivalence classes. If, on the contrary, an adiabatic tuning of the constrainers from $\mathcal{C}_A(\mathbf{R})$ into $\mathcal{C}_B(\mathbf{R})$ is possible without closing the spectral gap, we identify the two CSLs as belonging to the same equivalence class.

The reason we 
make this distinction 
is because given the short-ranged spin correlations, one may naively expect all CSLs in this category to be equivalent to a \textit{trivial paramagnet}. 
The trivial paramagnet is defined as systems where spins only interact within a unit cell,
and there is no inter-unit cell couplings.
Using the breathing kagome lattice as an example, a trivial paramagnet is given by a Hamiltonian which constrains the spins only on the up-pointing triangles
\[
\label{eqn:BRkagomeAFM_trivial}
   \mathcal{H}_\mathsf{B-KGM}  = \sum_\bigtriangleup\left( \sum_{i \in \bigtriangleup} S_i\right)^2 \ .  
\]
{As we see here, it is  a model with two DOFs  in the unit cell freely fluctuating, while  the other DOF completely frozen to be zero.
More importantly, there is no inter-unit cell coupling, so the higher dispersive band has a constant eigenvector $T(\mathbf{q})=(1,1,1)/\sqrt{3}$.

Given another breathing kagome model with one constrainer per unit cell one can ask the following question:
if we keep the one-constrainer form of the Hamiltonian but change the constrainer smoothly to tune the model from the trivial model, Eq. (\ref{eqn:BRkagomeAFM_trivial})
to the new model, 
can this procedure happen without closing the gap in the spectrum at any step?
We will see later an example 
of a FT-CSL CSL which can be shown by such an argument to \textit{not} be equivalent to a trivial paramagnet (Section \ref{sec.kagoem.star}).

As one may expect, if adiabatic transitions between two CSLs are obstructed, 
there must be some mathematical quantities that distinguishes them.
The idea is very similar to the notion of  Chern insulators in band theory, wherein two theories with different Chern numbers  cannot be adiabatically transformed into each other by tuning the Hamiltonian, without the gap closing.
The classification can be further enriched by symmetry:
while there are paths to    deform $\mathcal{C}_A$ to $\mathcal{C}_B$ without closing gaps,
it is only possible to do so when the path   breaks
a symmetry. 
In such a symmetry-enforced scenario, the two states $A$ and $B$ are still considered to be different.

We will show that up to the equivalence of adiabatic connection, the CSLs can then be divided into different topological classes. 
What is the topological quantity that distinguishes the different topological classes? 
Since the bottom band eigenvector is globally well-defined (in mathematical terms, it is a section of a trivial vector bundle), the band always has zero Chern number, so that is not the quantity we look for. 
Instead, we found that the fragile topological CSLs are classified by the homotopy class of the bottom band \textit{eigenvector subspace configuration} on the torus of the BZ. 
When there is only one bottom band, 
the eigenvector subspace is simply the 
$N$-component unit vector modulo an overall phase.
When there is more then one bottom band,
the eigenvector subspace is a higher dimensional subspace of the total space of all the eigenvectors. We now consider these two cases in more detail.

\subsubsection{One bottom band}

Let us consider the simplest case first: a  2D model with $N$ spins per unit cell, and one bottom flat band in the spectrum of its Hamiltonian (generalizations to 3D exist, but we shall primarily focus on 2D models in what follows). 
The flat band has normalized eigenvector configuration, which we refer to as the fluctuator, see Eq.~\eqref{eqn:FT_constrainer_def}:
\[
\hat{\mathbf{B}}(\mathbf{q})=  {\mathbf{B}}(\mathbf{q})/|\mathbf{B} (\mathbf{q})|\ .
\]
The fact that the bottom band is gapped from the other bands means $\hat{\mathbf{B}}(\mathbf{q})$ is well-defined and non-vanishing everywhere in the momentum space.

At a fixed wavevector $\mathbf{q}$, $\hat{\mathbf{B}}(\mathbf{q})$ and $e^{i\theta}\hat{\mathbf{B}}(\mathbf{q})$
correspond to the same physical spin fluctuation.
Therefore, the physical configuration space for $\hat{\mathbf{B}}(\mathbf{q})$ is the complex projective space $\CP^{N-1}$.
Often, we have additional inversion or time reversal symmetries that constrain $\hat{\mathbf{B}}(\mathbf{q})$ to be real, in which case the physical configuration space is the real projective space $\RP^{N-1}$.
From now on,  we take $\hat{\mathbf{B}}(\mathbf{q})$ to denote a ray in the target space $\CP^{N-1}$ or $\RP^{N-1}$. 

Now $\hat{\mathbf{B}}(\mathbf{q})$ defines a map from the torus of the (two-dimensional) BZ to the space of $\CP^{N-1}$ or $\RP^{N-1}$:
\[
\hat{\mathbf{B}}(\mathbf{q}): {T}^2 \rightarrow \CP^{N-1} (\text{or }\RP^{N-1});\  \mathbf{q} \mapsto \hat{\mathbf{B}}(\mathbf{q})\ .
\]
The equivalence classes of such maps are classified by the relative homotopy group $[{T}^2, \CP^{N-1} ]$ or $[{T}^2, \RP^{N-1} ]$.

The homotopy classes are the topological quantities that distinguish different fragile topological CSLs.
Without closing the gap, i.e., having ${\mathbf{B}}(\mathbf{q})$ vanishing and $\hat{\mathbf{B}}(\mathbf{q})$ ill-defined at some momentum point,
the homotopy class cannot be changed. 
Hence two fragile topological CSLs of different homotopy classes cannot be adiabatically turned into each other without closing the gap.
Obviously, the comparison of homotopy classes is only sensible when $\hat{\mathbf{B}}(\mathbf{q})$'s have the same number of components. 
This indicates that the topology classification is a fragile concept (hence our use of the term \textit{fragile} topological CSL), which we will explain in more detail below in subsection~\ref{sec:fragility}.

In general, the homotopy group $[{T}^2, X ]$ is not easy to compute. 
However, if the target manifold $X$ is simply connected (i.e., $\pi_1(X) = 0$ and path-connected), then we have the homotopy group isomorphic to the second homotopy group of $X$:
\[
[{T}^2, X ] = \pi_2 (X) = H_2 (X)\ .
\]
Since $\pi_1 (\CP^{N-1})=0$ for any $N-1 \ge 1$, we have
\[
\label{eqn:homoT2RPn}
[{T}^2, \CP^{N-1} ] = \pi_2 (\CP^{N-1} ) = \mathbb{Z}\ .
\]
This is the homotopy class for complex eigenvector  $\hat{\mathbf{B}}(\mathbf{q})$.
That is, in general, the homotopy classes are labeled by an integer number in $\mathbb{Z}$.

The homotopy class for $[{T}^2, \RP^{N-1} ]$ is more complicated, since $ \RP^{N-1}$ is not simply connected: $\pi_1 (\RP^1)=\mathbb{Z}$, and $\pi_1 (\RP^{N-1})=\mathbb{Z}_2$ for $N-1\ge 2$. 
They are in principle calculable, but there is no simple, universal formula \cite{janich1987topological,bechtluft1999global}.

However, in some scenarios, we can consistently assign directions to the $\RP^{N-1}$ eigenvectors smoothly over  momentum space, without encountering any inconsistencies with  the BZ periodic boundary condition.  
Then, we can  treat the eigenvectors as unit vectors pointing on the $S^{N-1}$ sphere. 
For $N-1\ge 2$, $\pi_1 (S^{N-1})=0$ so $S^{N-1}$ is simply connected. 
In this case we have 
\[
\label{eqn:homoT2Sn}
[{T}^2, S^{N-1} ] = 
\pi_2 (S^{N-1} ) = \begin{cases}
 \mathbb{Z} & \text{if $N-1 = 2$}\\
 0 & \text{if $N-1 \ge 3$}
 \end{cases} \ .
\]
We see that here the only non-trivial case is when the model has $N=3$ degrees of freedom. The integer homotopy invariant is then nothing but the skyrmion number on the 2-torus. 
Given the eigenvector configuration, this skyrmion number $n_\mathsf{sk}$ can be computed by 

\[
Q_\mathsf{sk}=\frac{1}{4 \pi} \int_{\rm BZ} \mathrm{d}^2{\bf q}\  \hat{\mathbf{B}}({\bf q}) \cdot
\left(
\frac{\partial \hat{\mathbf{B}}({\bf q})}{\partial q_x} 
\times 
\frac{\partial \hat{\mathbf{B}}({\bf q})}{\partial q_y}
\right)\  .
\]

We mention in passing that 
here the skyrmion number should not be confused with the winding of effective magnetic field for a two level Hamiltonian  $\mathbf{J}(\qq)=\mathbf{B}(\qq)\cdot \hat{\boldsymbol{\sigma}}$. 
In that case, the skyrmion configuration makes it impossible to smoothly define the phase of a band,
which is equivalent to the statement of a nontrivial Chern number of the bottom band. 
By contrast, the skyrmion characterizing  the band eigenvector in a CSL has a fundamentally different physical meaning.
In fact, when a band's eigenvector is well defined in the BZ (so we can talk about its skyrmion number to begin with),  there is no problem in smoothly defining the phase of the band at all wavevectors -- it is then a section of a trivial vector bundle, with zero Chern number. 
In particular,  exact flat bands with finite range of interactions have been shown to be alway have zero Chern number ~\cite{Chen2014}.

\subsubsection{$N-1$ bottom bands}
Another equally simple case is when we have a single constrainer in the Hamiltonian Eq.~\eqref{eq:H_constrainer}, resulting in
$N-1$ bottom bands and one disperstive top band separated by a gap. 
In this case we can  examine the homotopy for eigenvector $\hat{\mathbf{T}}(\mathbf{q})$ of the top band instead. 
All the analysis in the previous sub-sub-section carries over by replacing $\hat{\mathbf{B}}(\mathbf{q})$ with $\hat{\mathbf{T}}(\mathbf{q})$.

Equations (\ref{eqn:homoT2RPn})-(\ref{eqn:homoT2Sn}) applied to a single bottom band or a single top band
explicitly tell us the possible homotopy classes 
of the corresponding cases.
In Sec.~\ref{sec.kagoem.star}, we will present a concrete microscopic spin model which exhibits several  topological classes $[{T}^2, S^{2} ] = \mathbb{Z}$ and  transitions between them, as the Hamiltonian is tuned.

\subsubsection{Other cases}
\label{sec.homotopy.several.band}
The more complicated situation is to have 
$N-M$ degenerate bottom bands where $1< M < N-1$. 
In this case, the target space is not a ray in $\CP^{N-1}$ or $\RP^{N-1}$, but a projective plane (for two bottom bands), 
or generally the projective $(N-M)$--dimensional subspace in $\CP^{N-1}$ (or $\RP^{N-1}$) generated by the $N-M$ eigenvectors $\hat{\mathbf{B}}^1(\mathbf{q}), \dots, \hat{\mathbf{B}}^{N-M}(\mathbf{q})$. 
These homotopy classes are in principle calculable, though we are not aware of a simple closed-form expression.

Additionally, our analysis generalize to three dimension by computing the homotopy classes of $[{T}^3, \CP^{N-1} ]$,  $[{T}^3, \RP^{N-1} ]$ and $[{T}^3, S^{N-1} ]$. 
One   example is the Hopf invariant $[{T}^3, S^2 ]=\mathbb{Z}$ \cite{Moore2008PhysRevLett}, whose computation method is known \cite{PhysRevLett.51.2250}.

\subsection{Properties of the fragile topological CSLs}

We now discuss the general properties of the fragile topological CSLs.
For concreteness, we use the one  bottom flat band (or equivalently one top band)  cases for demonstration.
Our discussion is straightforward to generalize to multi-band cases.

\subsubsection{Transition between   homotopy classes}

The homotopy equivalence class remains unchanged
upon adiabatically tuning the Hamiltonian while keeping the bottom bands flat and gapped.
The changing of topological class thus requires  the gap between the higher bands and bottom bands to close, 
so that the bottom band eigenvector configuration can go through singular changes at the gap closing point.

From this point of view, 
while all CSLs' Hamiltonians are fine-tuned, 
the fragile topological CSLs are the more common ones over the entire parameter space. 
The algebraic CSLs, with the spectral gap closing, require additional tuning and denote the critical boundaries between different fragile topological CSLs, or higher order critical points where the critical boundaries intersect.
In this sense, the algebraic spin liquids are more fine-tuned than the topological CSLs. 
A schematic phase diagram indicating both types of spin liquids is shown in Fig.\ref{fig:Fig_overview}.

\subsubsection{Eigenvector homotopy is fragile  }
\label{sec:fragility}

The homotopy class is a ``fragile'' topological quantity (see e.g. discussion in Refs.~\cite{Kitaev2009,song2020twisted,peri2020experimental}), in that
upon adding a new spin DOF per unit cell to the model,
the previous non-trivial homotopy class of the $N$-component eigenvector configuration  may become trivial as a $(N+1)$-component eigenvector configuration. 

Let us demonstrate this using the following model.
Consider the original model with $N$ sublattice sites and only one constrainer $\mathcal{C}$:
\[
\mathcal{H}_\mathsf{0} = \sum_{\bfR \in \text{u.c.} }\mathcal{C}_\mathsf{0}(\bfR)^2\ .
\]
Such a model has $N-1$ degenerate flat bands and $1$ higher dispersive band.
The top band has eigenvector $\mathbf{T}_\mathsf{0}(\bfq)$ obtained by Fourier transforming the vector $\bfC_\mathsf{0}(\mathbf{r}, \bfR)$.

We now add a new DOF $S_{N+1}$ to the system, and introduce a parameter $\gamma$ to tune the interactions. 
The new Hamiltonian is
\[
\mathcal{H}(\gamma) = \sum_{\bfR \in \text{u.c.} }\left[(1-\gamma)\mathcal{C}_\mathsf{0}(\bfR) + \gamma S_{N+1}(\bfR) \right]^2\ ,
\]
which   has one higher dispersive band and $N$ bottom flat bands.
The corresponding constrainer in its vector form is then
\[
\bfC( \gamma, \mathbf{r}, \bfR) =  \left[ (1-\gamma) \bfC_\mathsf{0}(\mathbf{r}, \bfR), \gamma \right]\   ,
\]
and the FT-transformed constrainer is 
\[
{\mathbf{T}}( \gamma, \mathbf{q})= \left[
(1-\gamma)  \mathbf{T}_\mathsf{0}(\mathbf{r}, \bfR),    \gamma \right]\ ,
\]

Now, we adiabatically tune the Hamiltonian parametrized by $\gamma$ going from $0$ to $1$.

Note that the norm square of ${\mathbf{T}}( \gamma, \mathbf{q})$ is always positive everywhere,
\[
|{\mathbf{T}}( \gamma, \mathbf{q})|^2 = (1-\gamma)^2 |{\mathbf{T}}^0|^2 + \gamma^2 > 0\ ,
\]
so the gap between the bottom band and the higher bands never closes. 
However, at the end of this adiabatic tuning, the eigenvector becomes 
\[
{\mathbf{T}}( 1, \mathbf{q})  = (0,\dots,0, 1)\ ,
\]
which belongs to the trivial homotopy class for the $N+1$ band model.

By the above argument, the $N$ band model of any homotopy class can be adiabatically changed to the trivial homotopy class of the $N+1$ band model.
We can also join two such processes together to adiabatically change between the two different homotopy classes of the $N$ band model without closing any gap:
\[
\begin{split}
  \mathbf{T}(\gamma=0,\qq) &= ({\mathbf{T}_\text{A}} \text{ of homotopy class A}, 0) \\
  \rightarrow 
   \mathbf{T}(\gamma=\frac{1}{2},\qq) &= (0, \dots, 0 , 1 )\\
\rightarrow  \mathbf{T}(\gamma=1,\qq) &=({\mathbf{T}_\text{B}} \text{ of homotopy class B}, 0)\ .
 \end{split}
\]

We would like to stress that such a construction is not possible without introducing the new DOF. 
That is to say, the homotopy class of the eigenvector configuration is a fragile topological quantity only when allowing for arbitrary `padding' the unit cell with new DOFs.
However, when restricted to the original degrees of freedom,
the homotopy class is well-defined and can only be changed via spectral gap closing.

\subsubsection{Absence of algebraic boundary correlations}
\label{subsubsec:no_boundary}

One may naturally wonder whether fragile topological CSLs have a notion of bulk-boundary correspondence, in analogy with topological insulators. Specifically, given the association between gapless points in the band structure of ${\bf J}({\bf q})$ and algebraic correlations of a CSLs, one may have hoped that   topological  CSLs would host gapless edge states of ${\bf J}({\bf q})$, and therefore algebraic correlations at their boundaries.

However, as we have argued above that topological CSLs are generically fragile in nature; and since fragile topology does not guarantee gapless boundary modes \cite{Po2018}, the scenario of algebraic boundary correlations is not realized.
Fragile topological CSLs with open boundary conditions will generally have short ranged correlations on the edge, as well as in the bulk.

It is known that fragile topology can have an
associated bulk-boundary correspondence in the presence of specially chosen twisted boundary conditions \cite{Song2020}. 
However, the naturalness of such twisted boundary conditions in a CSL is doubtful,   so we do not pursue this topic further in this work, and leave its investigation for the future studies.
How the fragile topological CSLs manifest its non-trivial topology in experimentally measurable quantities is still an open and important question.

\section{Kagome-star model : a series of FT-CSLs}
\label{sec.kagoem.star} 

\begin{figure}[ht!]
 \centering
 \includegraphics[width=\columnwidth]{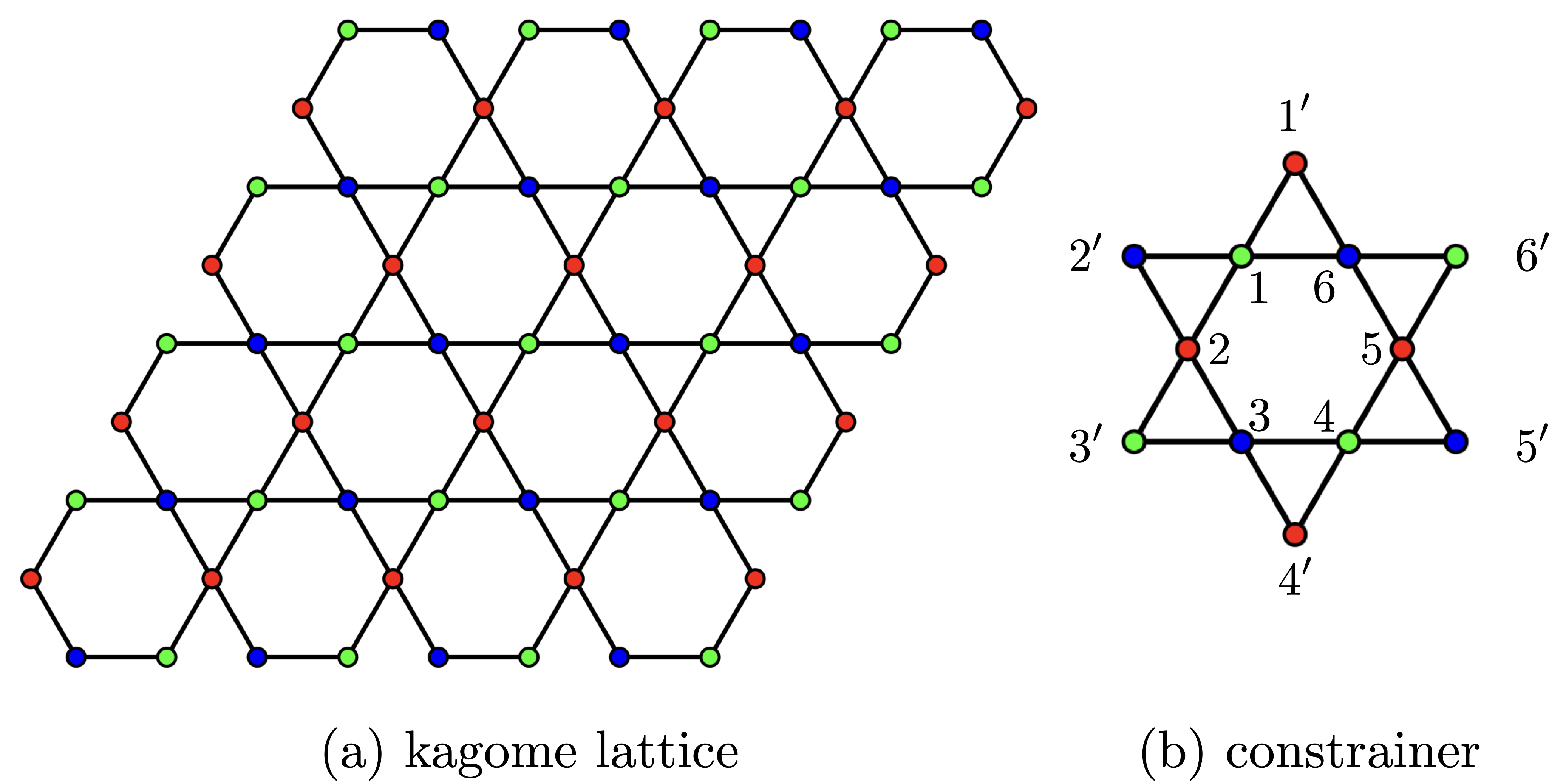}
 \caption{
 (a) The kagome lattice.
 (b) The  constrainer of the kagome-star model.   
 The constraint is defined on each hexagon of the lattice, with the spins on the interior of the hexagon $(1-6)$ contributing to the constraint with coefficient $1$, and the spins connected to the exterior $(1'-6')$ contributing with coefficient $\zeta$ (Eq. (\ref{eq:Pstar})-(\ref{eq:constraint_star})).
 }
 \label{fig:BRSM_model}
\end{figure}

In this section we introduce a generalization of the kagome-hexagon model \cite{Rehn17PRL} introduced earlier in Eq.~\eqref{eq:HRSM},  which is used to demonstrate the application of our 
scheme of fragile topological CSL and   establishes the possibility of transitions between distinct fragile topological CSLs.
We refer to this generalized model as the kagome-star model.
The Hamiltonian is:
\[ 
\mathcal{H}_{\mathsf{KS}}  
= \frac{J}{2} \sum_{{\bf R} \in {\rm hexagons}} \sum_{\alpha=x,y,z} \left[ \mathcal{C}_{\mathsf{KS}, \alpha} ({\bf R}, \zeta) 
\right]^2\ ,
\label{eq:Hstar}
\]
where $\zeta$ is a dimensionless (real) tuning 
parameter and
\begin{eqnarray}
\mathcal{C}_{\mathsf{KS}, \alpha} ({\bf R}, \zeta) 
= \sum_{i  = 1}^{6} S_i^{\alpha}
+ \zeta \sum_{j  = 1'}^{6'} S_{i'}^{\alpha}\ .
\label{eq:Pstar}
\end{eqnarray}
The two contributions to the constrainer $\mathcal{C}_{{\sf KS} \alpha}$ in Eq. (\ref{eq:Pstar}) are illustrated in Fig. \ref{fig:BRSM_model}.
The first sum over $i$ is a sum over spins $1, \dots, 6$
belonging to the interior of the hexagon centred at ${\bf R}$.
The second sum over $j$ is a sum over spins $1', \dots, 6'$
connected to the exterior of the hexagon, forming the points of a six-pointed star.
The ground states are those which satisfy the constraints
\begin{eqnarray}
\mathcal{C}_{\mathsf{KS}, \alpha}  ({\bf R}, \zeta)=0   f
\quad \forall \ \  {\bf R}, \alpha\ .
\label{eq:constraint_star}
\end{eqnarray}

Since the three components $\alpha = x, y, z$ are identical and decouple from each other,
we can focus on one copy of them now and drop the index $\alpha$.
There is one star motif - and hence one constrainer - per unit cell.
The Fourier transformed constrainer is:
\[
\begin{split}
   {\bf T}_{\zeta} ({\bf q})
=&
\begin{pmatrix}
\cos(\sqrt{3} q_x) \\
\cos\left( -\frac{\sqrt{3}}{2} q_x + \frac{3}{2} q_y \right)  \\
\cos\left( -\frac{\sqrt{3}}{2} q_x - \frac{3}{2} q_y \right)  
\end{pmatrix}\\
&+  \zeta \begin{pmatrix}
  \cos(3 q_y)\\
  \cos(-  \frac{3\sqrt{3}}{2} q_x - \frac{3}{2} q_y) \\
  \cos( \frac{3\sqrt{3}}{2} q_x - \frac{3}{2} q_y) 
\end{pmatrix}
\ . 
\end{split} 
\]

For $\zeta=0$, this model reduces to the Kagome-Hexagon model \cite{Rehn17PRL}, in which ${\bf T}_{ 0} ({\bf q})$ 
is well-defined and non-zero for all ${\bf q}$.
Correspondingly, the soft spin band structure is gapped
everywhere, with two flat bands at the bottom of the
spectrum separated from one top band.

The top band is topologically non-trivial, as can be seen through calculating the momentum space skyrmion number associated to $T_{\zeta} ({\bf q})$:
\[
Q_\mathsf{sk}(\zeta) = \frac{1}{4 \pi} \int_{\rm EBZ}
\mathrm{d}^2{\bf q}  \ \ \hat{\bf T}_{\zeta}({\bf q}) \cdot \left(
\frac{\hat{\bf T}_{\zeta}({\bf q})}{\partial q_x} \times
\frac{ \hat{\bf T}_{\zeta}({\bf q})}{\partial q_y}
\right)\ ,
\]
where $\hat{\bf T}({\bf q})={\bf T}(\bf q)/|{\bf T}({\bf q})|$
is the normalized constrainer.
The integral is taken over the extended Brillouin zone (EBZ), corresponding to the periodicity of ${\bf T}({\bf q})$, which  due to the relative phase of different sites within the unit cell has double the period of the primitive Brillouin zone.

Provided that inversion symmetry is maintained and no further sites are added to the unit cell, $Q_\mathsf{sk} \in \mathbb{Z}$ takes integer values which can only be changed by tuning the model through a gapless point,
as we discussed in detail in Sec.~\ref{sec.topological.csl}.
The evolution of $Q_\mathsf{sk}$ with increasing $\zeta$ is shown 
in Fig. \ref{fig:Qsk}.

\begin{figure}
\centering
\includegraphics[width=\columnwidth]{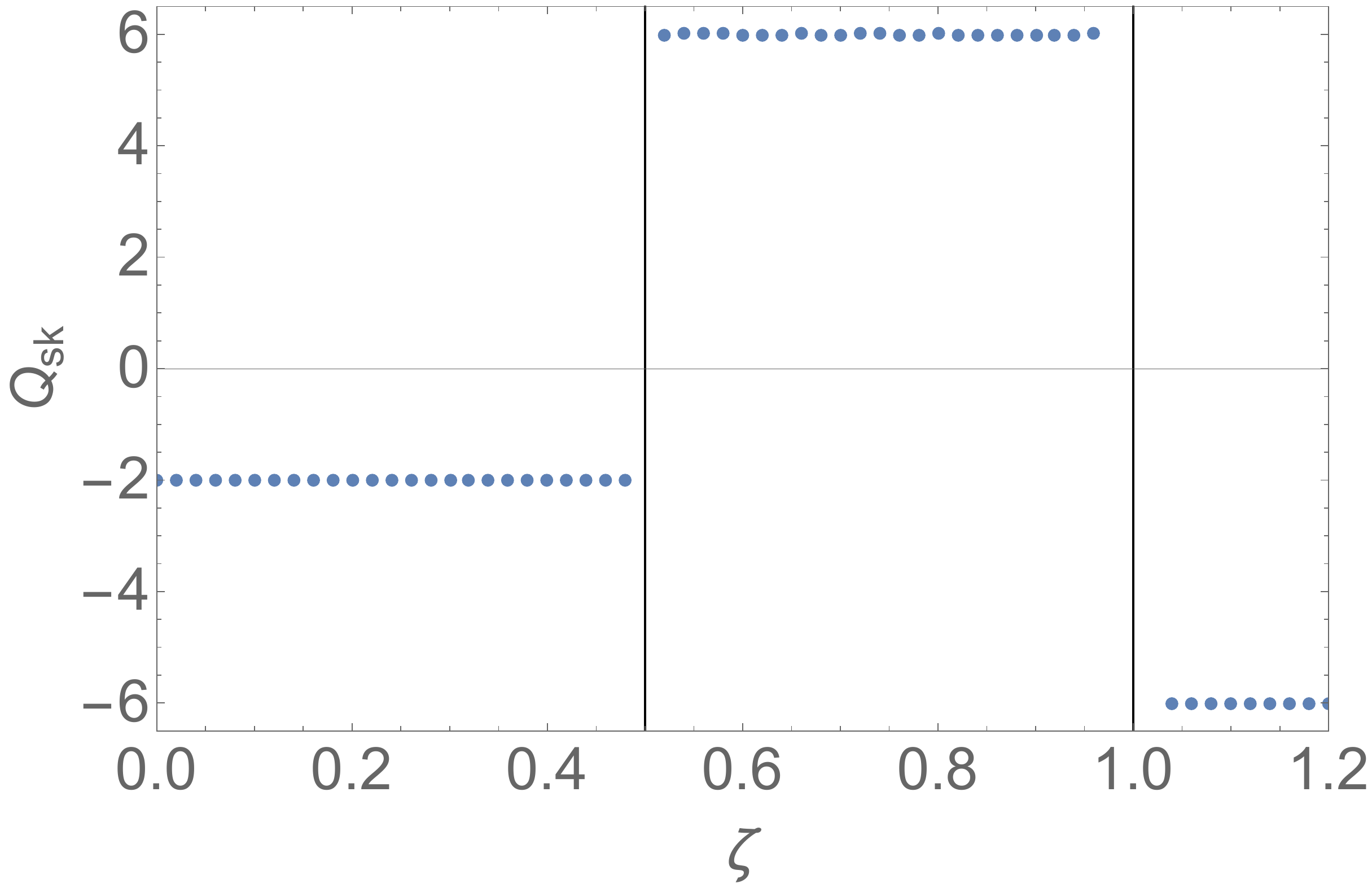}
\caption{Evolution of momentum space Skyrmion number,
$Q_{sk}$, as a function of the tuning parameter $\zeta$
in The kagome-star model.
Jumps in $Q_\mathsf{sk}$ at $\zeta=1/2$ and $\zeta=1$ reveal 
zero temperature transitions between distinct, short
range correlated, CSLs.
Algebraic CSLs emerge at the boundaries.
}
\label{fig:Qsk}
\end{figure}

\begin{figure*}
\centering
\includegraphics[width=0.95\textwidth]{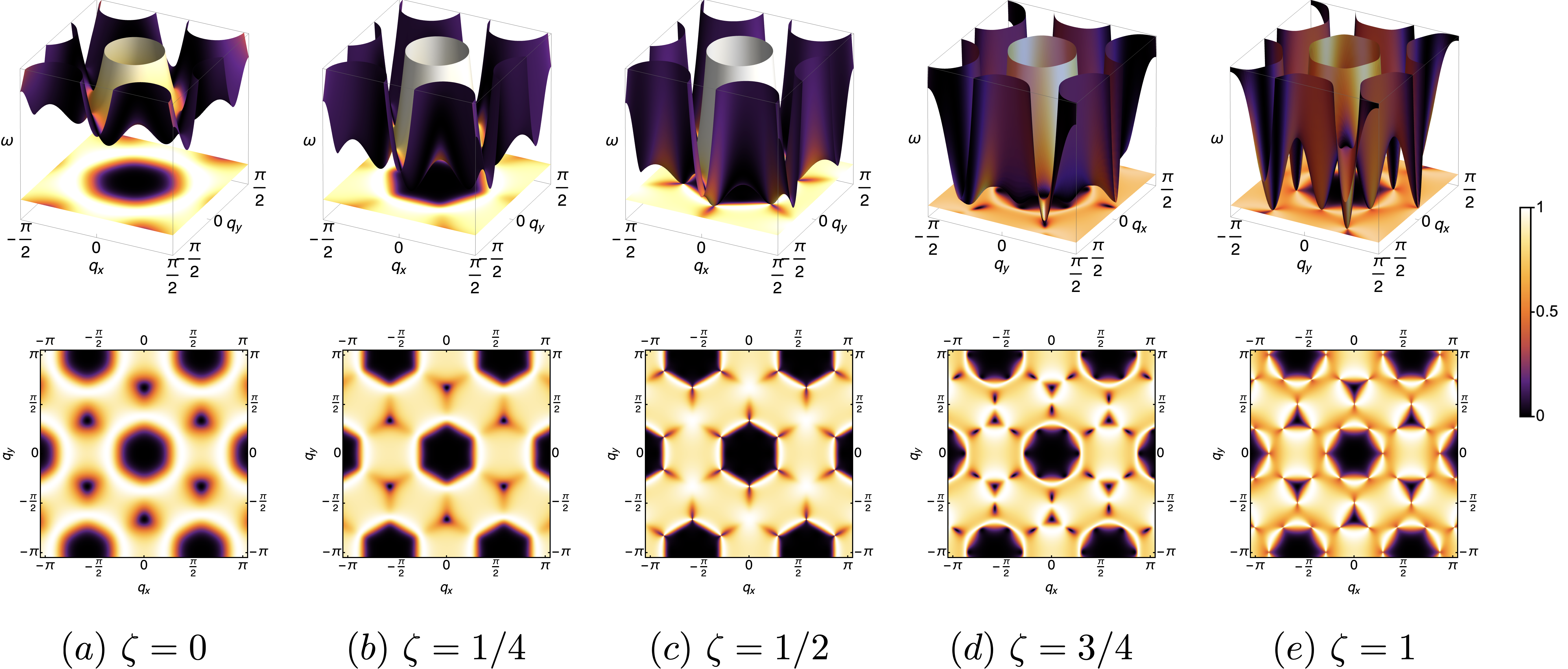}
\caption{
Evolution of spectrum and structure factor 
$S({\bf q})$, as a function of the tuning parameter $\zeta$
in the kagome-star model (Eq.~\eqref{eq:Hstar}).
$S({\bf q})$ is smooth throughout momentum space for generic values of $\zeta$ (panels (a), (b), (d)), indicating the 
short range correlations of the fragile topological CSLs.
At the boundaries between these fragile topological CSLs
algebraic CSLs appear, with gap-closing points in the spectrum and  pinch point singularities in $S({\bf q})$ (panels (c), (e)).
\label{fig:SqKag}
}
\end{figure*}

\begin{figure*}
\centering
\includegraphics[width=0.75\textwidth]{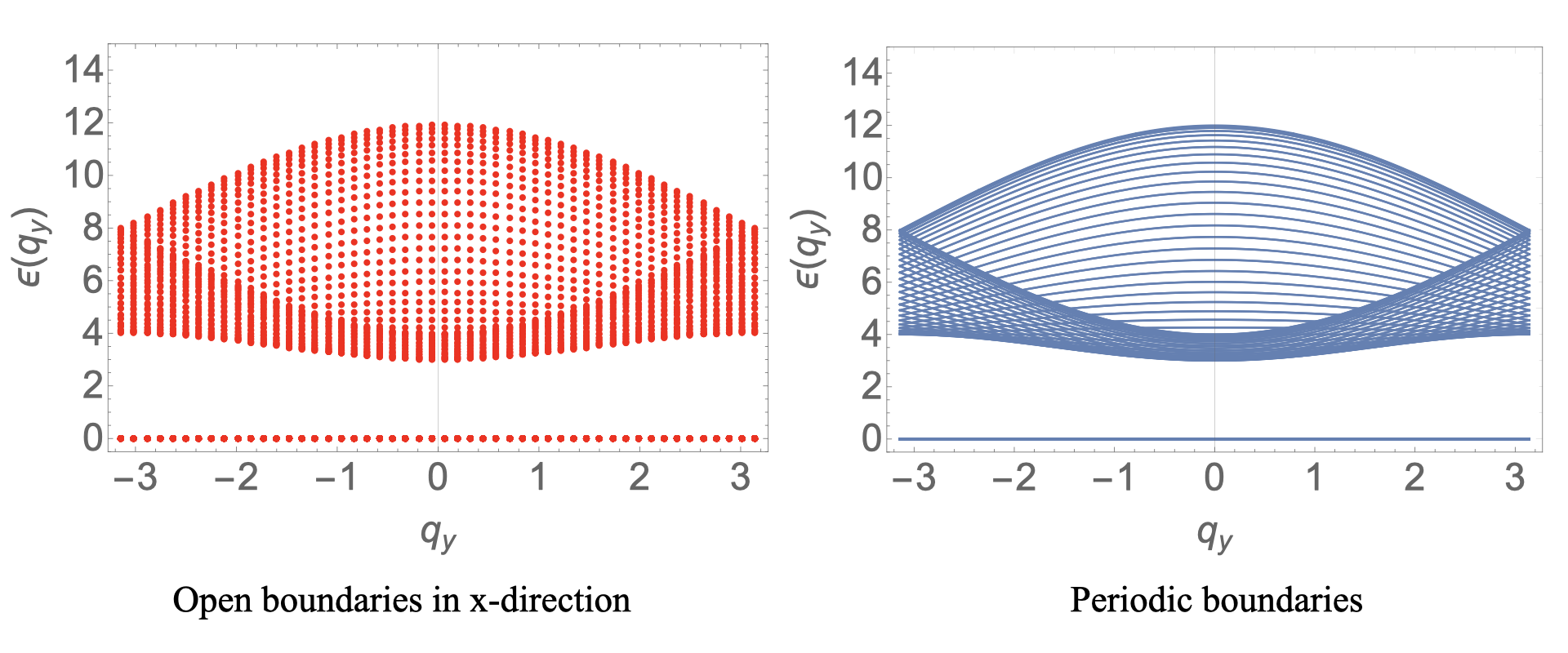}
\caption{
Soft spin spectrum of the kagome-star model at $\zeta=0$ (Kagome-Hexagon model) with open boundary conditions in one direction and with fully periodic boundary conditions. 
There is little difference between the spectra and no additional gapless points, in accordance with the fragile topological nature of the CSL and absence of bulk-boundary correspondence.
}
\label{fig:no_boundary}
\end{figure*}

The Skyrmion number jumps discontinously at $\zeta=1/2$ and $\zeta=1$. These changes in $Q_{sk}$ indicate zero temperature
transitions between distinct CSLs, all with short range correlations, but distinguished by the homotopy of the momentum space constrainer.
At the  boundaries between the fragile topological CSLs in parameter space, 
the soft spin dispersion has gapless points. Based on our discussion above, this implies the emergence of algebraic CSLs at the boundaries.
Indeed, calculating the equal-time spin structure factor $S({\bf q})$, defined in Eq.~\eqref{eq:S(q)} as a function of $\zeta$ reveals that pinch point singularities appear precisely at $\zeta=1/2$ and $\zeta=1$
 (Fig.~\ref{fig:SqKag}) demonstrating the emergence of algebraic classical spin liquids where fragile topological classical spin liquids meet.

As discussed in \ref{subsubsec:no_boundary}, the fragile topological nature of the CSLs does not guarantee bulk-boundary correspondence, meaning that there no additional gapless points arising at the edge with open boundary conditions, and that the correlations remain short ranged up to the edge of the lattice.
This is demonstrated in Fig. \ref{fig:no_boundary} where we plot the soft spin dispersions with open boundary conditions in one direction and with fully periodic boundary conditions,
for $\zeta=0$. 
There is no additional gap closing at the boundary and hence no algeberaic boundary correlations, underscoring the fragility of the topology underlying the short-range correlated CSLs.

We thus establish the kagome-star model as exemplifying a series of distinct fragile topological CSLs,
with algebraic CSLs at the boundary between them.

Another example based on the modified kagome-star model can be found in the companion article Ref.~\onlinecite{yan2023ArxiVclassification}, which provides a richer phase diagram of different FT-CSL phases and their algebraic CSL boundaries. 

\section{Aspects of application} 
\label{sec.app.to.exp}

\begin{table*}[ht!]
\caption{Survey of known CSL models. See respective references for detailed definition of the models.}
 \centering
\begin{tabular}{  c @{\hskip 10pt} c @{\hskip 10pt} c @{\hskip 10pt} c} 
    \toprule
    Category & Model lattice & Gauss's law/homotopy & Reference\\ \midrule 
    \multirow{11}{*}{Algebraic CSLs} 
    & pyrochlore    & Maxwell U(1) &  Ref.~\onlinecite{MC_pyro_PRL,Moessner1998PhysRevB.58.12049}  and Sec.~\ref{subsubsec_pyro} \\   
    & kagome (large-$\mathcal{N}$ limit)    & Maxwell U(1) &  Ref.~\onlinecite{Moessner1998PhysRevB.58.12049,Garanin1999PhysRevB.59.443}  and Sec.~\ref{subsubsec_kagome} \\
    & checkerboard    & Maxwell U(1) &  Ref.~\onlinecite{Moessner1998PhysRevB.58.12049} and Sec.~\ref{subsubsec_checkerboard}  \\    
    & honeycomb  & Maxwell  U(1) &  Ref.~\onlinecite{Rehn16PRL}  \\ 
    & breathing pyrochlore  & rank-2 U(1) &  Ref.~\onlinecite{Yan20PRL}  \\   
    & pyrochlore  & pinch line  &  Ref.~\onlinecite{Benton16NComms}  \\  
    & checkerboard   &  various &  Ref.~\onlinecite{davier2023combined}  \\  
    & kagome  &  various   &  Ref.~\onlinecite{davier2023combined}  \\  
    & honeycomb-snowflake &  various    &  Ref.~\onlinecite{Benton21PRL} and Sec.~\ref{Sec_BM_classification_application}  \\ 
    & honeycomb  & anisotropic U(1)  &   Sec.~\ref{Sec_anisotropic_snowflake}  \\ 
    & octahedral   &    pinch line  &  Sec.~\ref{subsec:pinch.line.CSL} \\ 
    & octahedral   &   various  &  Ref.~\onlinecite{Benton21PRL} \\ 
    \midrule
    \multirow{3}{*}{FT-CSLs}  
    & kagome    & skyrmion &  Ref.~\onlinecite{Rehn17PRL}  \\  
    & kagome-square    & Sec.~\ref{sec.homotopy.several.band}   &  Ref.~\onlinecite{Rehn17PRL}  \\   
    & centered pyrochlore  &  Sec.~\ref{sec.homotopy.several.band}  &  Ref.~\onlinecite{Nutakki2023PhysRevResearch,nutakki2023arxivCPyro}  \\  
    \midrule
    \multirow{3}{*}{Both}  
    & kagome    &   various Gauss's laws and skyrmion &  Ref.~\onlinecite{yan2023ArxiVclassification}   \\
    &  kagome-square    &   Maxwell Gauss's laws  and $[T^2, \mathbb{R}P^3]$ &  Ref.~\onlinecite{Gembe2023arxiv}  \\
    & kagome-star     &  various Gauss's laws and skyrmion &  Ref.~\onlinecite{davier2023combined} and Sec.~\ref{sec.kagoem.star}    \\
    \bottomrule
\end{tabular} 
\label{table:summary.of.models}
\end{table*}

\subsection{Self-consistent Gaussian approximation and lattices with symmetry-inequivalent sites}

Our classification formalism employs a scheme for soft spins in constrainer Hamiltonians,  i.e., each spin component is a real scalar free of any non-linear constraints, and the Hamiltonian is written as a sum of squared linear constrainers. 
Adding to the discussion when introducing this formalism, we comment on aspects of what happens
in the siutation when the spins have hard constraints (particularly Heisenberg   spin with the hard constraint $\mathbf{S}^2 = 1$), and the bare Hamiltonian does not take the constrainer form.
First, note that  the  constrainer Hamiltonian of soft spins is largely equivalent to the Self-Consistent Gaussian Approximation (SCGA)\cite{Garanin1999PhysRevB.59.443}, the applicability of which for hard spins can be justified by Luttinger-Tisza ideas \cite{Luttinger1946PhysRev.70.954,Luttinger1951PhysRev.81.1015}.
In the SCGA, the hard-spin constraint is enforced only on average,
\[
\label{EQN_S_one_average}
\braket{\mathbf{S}_i^2} = 1 \ .
\]
by using   a Lagrange multiplier that can also viewed as a chemical potential term. Thus, up to this chemical potential shift,  the classification scheme we presented extends to the  SCGA scheme.

In fact, the constrainer Hamiltonian form of Heisenberg spins has been used for analyzing frustrated magnets already. 
Perhaps the most instructive example is that of the pyrochlore spin models \cite{Isakov2004PhysRevLett.93.167204,Isakov2005PhysRevLett.95.217201}, where the Hamiltonian is naturally written as a sum over constrainers on all tetrahedra:  
\[
\begin{split}
  \mathcal{H}_\mathsf{bare} 
  =&  \;\sum_{\braket{i,j}}\mathbf{S}_i \cdot \mathbf{S}_j \\
  =& 
  \;\frac{1}{2}\sum_\text{all tet.}\left(\sum_{i \in \text{tet.}}\mathbf{S}_i\right)^2 + \text{const.}   \\
  \equiv &
  \;\mathcal{H}_\mathsf{constrainer}+ \text{const.}   
\end{split}
\]
The  Hamiltonian $\mathcal{H}_\mathsf{constrainer}$ is actually what SCGA yields at the limit $T \rightarrow 0$. 

A subtlety to be aware of is when the sublattice sites are not equivalent to each other via space group symmetries. 
In such cases, as in e.g. the centered pyrochlore lattice, the bottom flat band from diagonalizing the bare Hamiltonian $\mathcal{H}_\mathsf{bare}$ may not satisfy Eq.~\eqref{EQN_S_one_average} for all sublattice sites. 
In such cases, a generalized Luttinger-Tisza method proposed by Lyon and Kaplan \cite{Lyons1960PhysRev} and improved by Schmidt and Richter \cite{Schmidt_2022} can be used to properly derive the physical ground states by ``renormalizing'' the spins. 
Nevertheless, in the end, the renormalized Hamiltonian still hosts  bottom flat bands, and can be analyzed with our classification scheme. 
A detailed application of this formalism to the centered pyrochlore lattice model \cite{Nutakki2023PhysRevResearch} can be found in Ref.~\onlinecite{nutakki2023arxivCPyro}.
This may also be able to explain the observation of disappearance of pinch points for the square-kagome model with varying exchange parameters  in Ref.~\onlinecite{Gembe2023arxiv}.

\subsection{Survey of known CSLs in literature} 
 
The classification scheme we propose provides a comprehensive view of both FT-CSLs and algebraic CSLs,
and encompasses the majority of classical spin liquid models known in literautre.  
In this section, we provide Table.~\ref{table:summary.of.models} of CSL models found in the literature, as well as those constructed in this work, along with brief comments on their place within our classification scheme.
In this table, we see a variety of models that realize different types of algebraic CSLs and fragile topological CSLs, as well as demonstrate the transitions between different CSLs. They all fit snugly into the landscape of CSLs we propose in Fig.~\ref{fig:Fig_overview}.

\section{Summary and outlook}
\label{sec.summary.outlook}

In this work, we have presented a classification scheme for classical spin liquids. The scheme includes two main categories, namely algebraic CSLs and fragile topological CSLs, with a finer classification within each category based on the emergent Gauss's law and homotopy of eigenvectors. Along with placing known examples from the literature into the landscape of CSLs, we introduce new  models to illustrate the major aspects of the classification scheme. We also make connections to flat band theory to analyze the structure of the ground state degeneracy in real space.

The classification scheme is a useful tool for understanding known and new CSL models and constructing new ones using the constrainer Hamiltonian formalism. 

We do note that the large-$\mathcal{N}$/SCGA treatment may fail for classical magnets, as it treats the non-linear hard spin constraints only 'on average'. There are known examples where hard spin constraints lead to different -- and often interesting -- new physics. 
One such case is the kagome Heisenberg antiferromagnet, which is found to be magnetically ordered, contrary to the large-$\mathcal{N}$ prediction  \cite{Chalker1992PhysRevLett.68.855,Harris1992PhysRevB.45.2899,Ritchey1993PhysRevB.47.15342,Zhitomirsky2008PhysRevB.78.094423,Chern2013PhysRevLett.110.077201}, although our analysis applies for $\mathcal{N}\geq4$-component spins on that lattice \cite{Huse1992PhysRevB.45.7536}.
Another type of cases is when the spins are discrete (Ising spins for example), so that e.g.\ satisfying the constrainers on triangles with their odd number of sites becomes impossible. The discreteness of the spins is a fundamental obstacle here, and the interesting physics of triangular and kagome Ising models lies beyond our present analysis.
Therefore, case-by-case analytical and numerical studies are still necessary for specific models of interest.
While we believe our scheme to be comprehensive for those spin liquids where a soft spin approximation is appropriate, there remains the interesting possibility of spin liquids outside our
scheme in which the non-linear constraints are a crucial element of
the effective description. 
The identification and classification of such cases remains an open question.

It is also interesting to speculate on the fate of CSLs in the presence of quantum dynamics. Although quantum models are usually not solvable, the algebraic CSLs do provide numerous realizations of the electric sector of the generalized rank-2 U(1) electrodynamics, serving as an interesting starting point for constructing quantum spin liquid models that host exotic emergent particles with reduced mobility.

While the constrainer Hamiltonian approach is mathematically convenient for analyzing classical spin liquids, it  relies on fine-tuned interactions between nearest and farther neighbor spins for realising interesting new spin liquids in the $T\rightarrow0$ limit. But even when this fine-tuning is not precisely met, there is a good chance that signatures of the  spin liquid under consideration will be present at moderately low temperatures. This holds the promise that future developments, both in the realm of magnetic materials but also in cold atomic systems, will provide a realization of some of these models.\\

\section*{Acknowledgements}
O.B. and R.M. acknowledge collaborations on related work
with Rafael Flores-Calderon, Adam McRoberts and Benedikt Placke.
H.Y. and A.H.N. were supported by the U.S. National Science Foundation Division of Materials Research under the Award DMR-1917511.
This work was in part supported by the Deutsche Forschungsgemeinschaft under
grant SFB 1143 (project-id 247310070) and the cluster of excellence 
ct.qmat (EXC 2147, project-id 390858490).\\
 
\noindent\textbf{NOTE} During completion of this manuscript, a
preprint by Davier {\it et al.} \cite{davier2023combined} appeared, which independently presents results regarding the classification of CSLs.

\renewcommand{\emph}{\textit}
\bibliography{reference}
   
\end{document}